\documentclass[referee,sn-standardnature]{sn-jnl}

\usepackage[caption=false,font=normalsize,labelfont=bf]{subfig}
\usepackage{placeins}
\begin{document} 
\title[Metastable LLPT in silicon]{Liquid-liquid phase transition in deeply supercooled Stillinger-Weber silicon} 

\author[1]{\fnm{Yagyik} \sur{Goswami}}

\author*[1]{\fnm{Srikanth} \sur{Sastry}}\email{sastry@jncasr.ac.in}

\affil*[1]{\orgdiv{Theoretical Sciences Unit}, \orgname{Jawaharlal Nehru Centre for Advanced Scientific Research}, \orgaddress{\street{Jakkur}, \city{Bengaluru}, \postcode{560064}, \state{Karnataka}, \country{India}}}

\abstract{
The existence of a phase transition between two distinct liquid phases in single-component network-forming liquids (e.g., water, silica, silicon) has elicited considerable scientific interest. The challenge, both for experiments and simulations, is that the liquid-liquid phase transition occurs under deeply supercooled conditions, where crystallization occurs very rapidly. Thus, early evidence from numerical equation of state studies was challenged, with the argument that slow spontaneous crystallization had been misinterpreted as evidence of a second liquid state. Rigorous free energy calculations have subsequently confirmed the existence of a liquid-liquid phase transition in some models of water, and exciting new experimental evidence has since supported these computational results. Similar results have so far not been found for silicon. Here, we present results from free energy calculations performed for silicon modelled with the classical, empirical Stillinger-Weber potential. Through a careful study employing state-of-the-art constrained simulation protocols and numerous checks for thermodynamic consistency, we find that there are two distinct metastable liquid states and a phase transition.  Our results resolve a long-standing debate concerning  the existence of a liquid-liquid transition in supercooled liquid silicon and address key questions regarding the nature of the phase transition and the associated critical point. 
}



\maketitle


The possibility of a phase transition between distinct liquid states in a single-component liquid has been the subject of intense scientific investigation and debate \cite{StanleyPolymorphism}. 
Liquid silicon, which we study here, is one such case where a number of experimental and computational studies have addressed the existence of such a liquid-liquid phase transition (LLPT)\cite{vasisht2013liquid,sastry2003liquid,vasisht2011liquid,beye2010liquid,ganesh2009liquid}. 
The most prominent example of a liquid undergoing a liquid-liquid phase transition is water.
The possibility of a liquid-liquid transition in water was first discussed while attempting to understand the apparent divergence of isothermal compressibility and other quantities in the supercooled state\cite{speedyangell,speedy1982stability}.
Based on molecular dynamics simulations for the ST2 model of water, Poole \emph{et al.} \cite{poole1992phase} proposed the existence of a liquid-liquid critical point under metastable conditions. 
Other scenarios have also been proposed, including those not invoking any singular behaviour\cite{sastry1996singularity}.
The possibility of a liquid-liquid transition has since been explored in a large variety of substances, and efforts made to understand its origins \cite{StanleyPolymorphism,tanaka2020liquid}.
Verifying the existence of a liquid-liquid phase transition experimentally has proved to be immensely challenging both for water\cite{kim2017maxima,kim2020experimental,nilsson2022origin} and for silicon\cite{kim2005situ,beye2010liquid}.
A number of numerical studies have reported the existence of a liquid-liquid phase transition in silicon, both using first-principles or \emph{ab initio} methods\cite{ganesh2009liquid} and using molecular dynamics with a classical empirical potential\cite{sastry2003liquid,vasisht2011liquid}.
However, the body of numerical evidence pointing to a liquid-liquid phase transition was brought into question by the work of Limmer and Chandler\cite{limmer2011putative,limmer2013putative}, who argued that the appearance of a second liquid phase was due to the misinterpretation of slow and spontaneous crystallization. Their study included results on models of water, including the ST2 model, as well as silicon modelled by the classical, empirical Stillinger-Weber (SW) potential\cite{stillinger1985computer}.
The debate around water has since been resolved and the existence of the liquid-liquid phase transition confirmed in a comprehensive study of the free energy surface for the ST2 potential by Palmer~\emph{et al.}\cite{palmer2014metastable} and for other realistic potentials \cite{debenedetti2020second}. Equally compelling results have also been reported for models of silica \cite{chen2017liquid,guo2018fluctuations}. 
For Stillinger-Weber silicon, the question has remained open, with recent investigations with free energy calculations similar to those performed by Limmer and Chandler and Palmer \emph{et al.} showing only one liquid state before the free energy barrier with respect to crystallization vanishes at conditions where the phase transition is expected\cite{ricci2019computational}.
This scenario of spontaneous, barrier-less, crystallization  in silicon has since been ruled out in the work of Goswami \emph{et al.}\cite{goswami2021thermodynamics}. In that work, the choice of a global order parameter $Q_6$ -- typically used in such contexts -- to constrain the system to reversibly sample states with different degrees of crystallinity, was demonstrated to give misleading results.
However, the question of whether there are in fact two metastable liquid states remains an open and challenging one. 
Starr and Sciortino\cite{starr2014crystal} set out to understand the relative propensity of different model network forming liquids to display either a stable or metastable liquid-liquid phase transition based on the angular rigidity of the tetrahedral bonds. This analysis led to the successful design of a patchy colloidal model that exhibits a liquid-liquid phase transition without the intervention of crystallization \cite{smallenburg2014erasing}. The analysis in \cite{starr2014crystal} revealed that the angular rigidity for Stillinger-Weber silicon was the highest among the models considered. Thus, any phase transition between two metastable liquids would be expected in the deeply metastable regime, possibly prevented by the onset of spontaneous crystallization. Silicon thus assumes special significance among the class of network-forming liquids that have been investigated.


Here we present results from numerical free energy calculations for Stillinger-Weber silicon at conditions where the possibility of a liquid-liquid phase transition is discussed.
We use a combination of a constrained sampling protocol and an appropriate order parameter, which has been demonstrated to accurately determine the free energy barriers to crystallization at deep supercooling\cite{saika2006test,goswami2021thermodynamics}.
We develop an extension to that methodology to effectively sample both the liquid state(s) and the transition state with respect to crystallization. Our results show clear evidence of the coexistence between two metastable liquid states, with the characteristic double-well feature in the free energy surface of the liquid state at the relevant conditions.
Through rigorous checks for thermodynamic consistency, including simulations at larger system sizes, we are able to demonstrate that the free energy reconstructions are robust and consistent with expectations concerning a liquid-liquid phase transition\cite{palmer2014metastable}.
We further investigate the behaviour of the relevant order parameter\cite{wilding1997simulation} and characterise the associated critical fluctuations\cite{debenedetti2020second}, which behave in accordance with the 3D Ising universality class.

We perform umbrella sampling Monte Carlo\cite{torrie1977nonphysical} (USMC) simulations of silicon modelled using the $3$-body Stillinger-Weber potential\cite{stillinger1985computer} in the constant pressure, constant temperature (NPT) ensemble. 
The density and the size of the largest crystalline cluster are simultaneously constrained with a harmonic umbrella bias and with a hard-wall bias\cite{saika2006test}
, respectively.
Crystalline atoms and clusters of connected crystalline atoms are identified using the local analogue of the Steinhardt-Nelson bond orientational order parameters\cite{steinhardt1983bond} using the procedure described in \cite{van1992computer,ten1995numerical,romano2011crystallization} with cut-offs specific to SW silicon as used in~\cite{goswami2021thermodynamics}.
Parallel tempering swaps are performed between adjacent bias windows and adjacent temperatures to enhance sampling of different densities (see Methods and SI for details).
Convergence of the simulations is checked by monitoring the decay of the auto-correlation functions of the density and global $Q_6$\cite{steinhardt1983bond}. Further, visit and excursion statistics from the parallel tempering swaps are also monitored to determine the efficacy of sampling.
Free energy estimates from the different bias windows and simulation conditions are reweighted using an in-house implementation of the weighted histogram analysis method\cite{chodera2007use,debenedetti2020second} to both obtain unbiased free energy estimates and reweight across temperature and pressure. Errors are obtained by estimating the number of decorrelated samples, based on the integrated autocorrelation time for the slowly relaxing variables, density and $Q_6$\cite{kumar1992weighted,chodera2007use,palmer2014metastable}.
Finally, this reweighting procedure is used to compare directly obtained free energy profiles with those obtained by reweighting from other conditions, giving identical results (see SI). This is a strong indication of converged, equilibrium sampling. 

\begin{figure*}[htpb!]
    \centering
    \includegraphics[height=4.2cm,width=4.2cm]{1D_barrier.eps}
    \includegraphics[height=4.2cm,width=4.2cm]{bDG_rho_compiled_error_final.eps}
    \includegraphics[trim=40 10 65 10,clip,height=4.4cm,width=4.8cm]{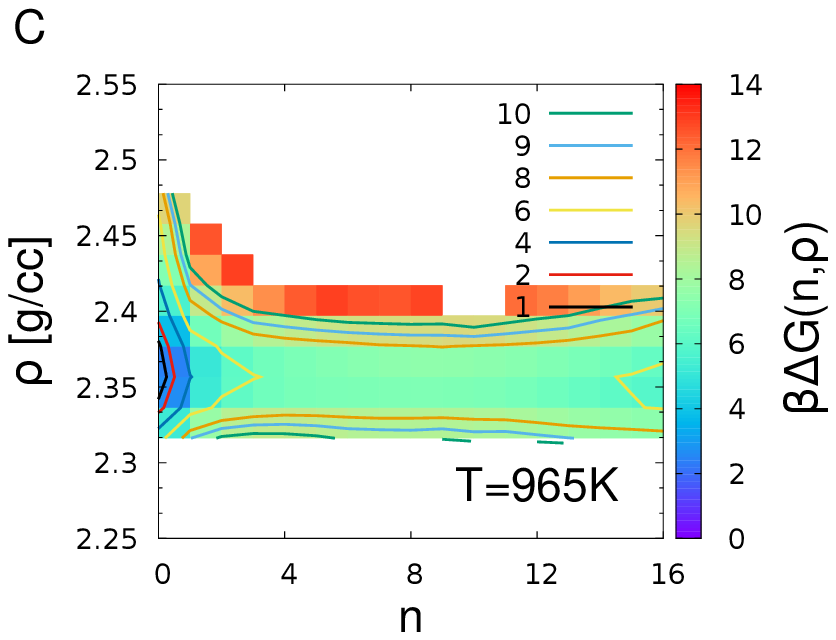}
    \includegraphics[trim=55 10 60 10,clip,height=4.4cm,width=4.8cm]{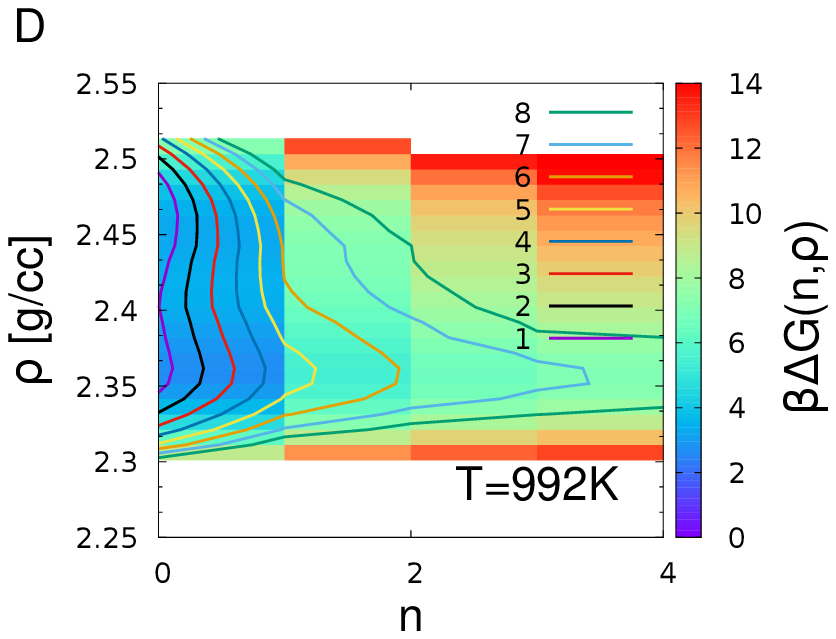}
    \caption{Panel (A) shows the free energy barrier to crystal nucleation from NPT US MC simulations of $N=512$ atoms at $P=0.75~GPa$ at the temperatures shown. The free energy is obtained from the full cluster size distribution using $\beta\Delta G(n)=-ln(P(n))+{\rm const.}$ with the condition $\beta\Delta G(0)=0$ fixing the value of the constant. The free energy barrier is finite at all temperatures, with the height at the lowest temperature being $6-7k_BT$. Panel (B) shows the free energy as a function of density, obtained from the negative log of the contracted distribution defined in Eq.~\ref{eq:Prho_marginal}, from the same simulations. Errors are obtained from estimates of the number of decorrelated samples in each constrained simulation. The double well feature at $T=992K$ is indicative of co-existence. Panel (C) shows the free energy from the joint distribution of density and cluster size at $T=965K$, $P=0.75~GPa$. The liquid has a low density of $2.37 gcc^{-1}$ even when the degree of crystallinity is zero. Contour lines are placed at the values mentioned in the legend.
    Panel (D) shows a two order parameter free energy reconstruction zoomed into the low $n$ region, showing a bi-modal feature along density.}
    \label{fig:fig-1}
\end{figure*}

{\bf \noindent Free energy reconstruction from constrained sampling:}
Free energies are reconstructed, in the first instance, for systems of $N=512$ atoms along the $P=0.75~GPa$ isobar.
The size of the largest crystalline cluster, $n_{max}$, is constrained within overlapping hardwall constraints, $[n^{lo},n^{hi}]$, while the density is constrained with harmonic bias potentials.
Parallel tempering swaps are performed between adjacent windows in $n_{max}$, density and temperature. 
Total simulation run lengths are in excess of $10^8$ MC steps for each case (at all system sizes) and are compared to the autocorrelation time, number of swaps performed along each axis, and the mean duration required for parallel tempering swaps to ``return" to the initial window.
In Fig.~\ref{fig:fig-1}(A), we show the free energy barrier to crystallization with a finite free energy cost to the formation of the critical nucleus 
at each of the temperatures considered.
The choice of temperatures is based on estimates of the LLPT line reported in~\cite{vasisht2011liquid}, where the estimated transition temperature for $P=0.75~{\rm GPa}$ is $T~\sim~990-995 {\rm K}$. We then construct the density distribution subject to the constraint $n_{max}\leq 4$, integrating over the multivariate distribution to get
\begin{equation}
P(\rho) = \sum\limits_{n_{max}=0}^{n_{max}=4} P(n_{max},\rho).
\label{eq:Prho_marginal}
\end{equation}
The corresponding free energies obtained from $\beta\Delta G(\rho)=-ln(P(\rho))$ are shown in Fig.~\ref{fig:fig-1}(B), displaying a jump in the most probable density of the liquid across $T=992K$, and a double-well form at $T=992K$, indicative of coexistence between two liquids.
In Fig.~\ref{fig:fig-1}(C) and (D), we present two order parameter free energy reconstructions as a function of the size of crystalline clusters along $x$ and the density along $y$. The liquid state(s) can be observed by considering the small $n$ region while the transition state (critical cluster for which $\beta\Delta G(n)$ is maximum) and the beginnings of the globally stable crystalline basin are observed by scanning along the $x$ axis. In these reconstructions, we compute the free energy from the relative probability of observing a cluster of size $n$ in the liquid at density $\rho$ (see Methods and SI for more details).
Fig.~\ref{fig:fig-1}(C) shows the free energy at the lowest temperature considered, $T=965~{\rm K}$, where the metastable liquid is purely in the low density state (LDL), with a barrier with respect to the growth of crystalline order centered at $n=8$. In Fig.~\ref{fig:fig-1}(D), for $T=992~{\rm K}$, two basins are visible at high (HDL) and low (LDL) densities respectively, in the low $n$ region. Integrating over $n$ (or $n_{max}$) will yield the contracted surface shown in Fig.~\ref{fig:fig-1}(B).

\begin{figure}[htpb!]
    \centering
    \includegraphics[height=5cm,width=5cm]{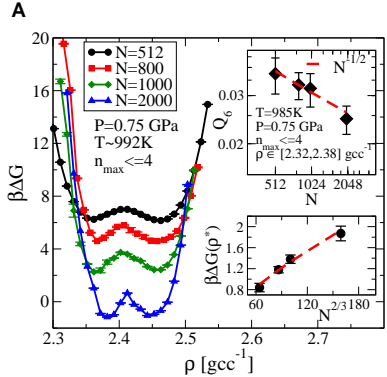}
    \caption{
    $\beta\Delta G(\rho)$ from USMC simulations at $T=992K$ at $N=512,800,1000,2000$ atoms. Density is sampled subject to the constraint on $n_{max}$. Error bars are a measure of the number of uncorrelated samples obtained for the free energy calculation (see SI). Bottom inset shows the height of the barrier as a function of system size. The barrier height scales as $N^{2/3}$, as expected when a stable interface can form between two phases. Error bars indicate the uncertainty arising from measurements at the bottom of the well and the top of the barrier and the variation in depth of the LDL basin and HDL basin. Top inset shows the scaling of average $Q_6$ with system size, when measured in the low density basin. The value of $Q_6$ decreases with $N$ as $N^{-1/2}$, demonstrating that the low density phase is macroscopically disordered. The error bars are the standard deviation obtained from the same samples.
    }
    \label{fig:fig-2}
\end{figure}

{\bf \noindent Free energy reconstructions at larger system sizes:}
In Fig.~\ref{fig:fig-2} we show the free energy profile as a function of density along the $P=0.75~{\rm GPa}$ isobar at different system sizes ranging from $N=512$ to $N=2000$.
Free energy estimates at exact coexistence conditions are obtained by reweighting from the available data directly simulated at $P=0.75~{\rm GPa}$ (see SI).
We note a slight shift (of $< 3 K$) to higher temperatures for the coexistence conditions at larger system sizes, as also noted in \cite{wilding1997simulation}. The coexistence temperature mentioned in Fig.~\ref{fig:fig-2} is for $N = 512$. 

The formation of a stable interface between two liquid phases will result in a scaling of the barrier height with $N$ as $N^{2/3}$\cite{palmer2014metastable,ricci2017free}. Results shown in the bottom inset in Fig.~\ref{fig:fig-2} are consistent with this scaling with system size.
Additionally, for the low density phase to be a disordered phase, the degree of global orientational ordering should scale as $N^{-1/2}$\cite{limmer2013putative,palmer2014metastable}. We find this to be the case from inspection of the top inset of Fig.~\ref{fig:fig-2}, where the error bars are obtained from the standard deviation of $Q_6$ measured under the conditions and constraints specified.

\begin{figure*}
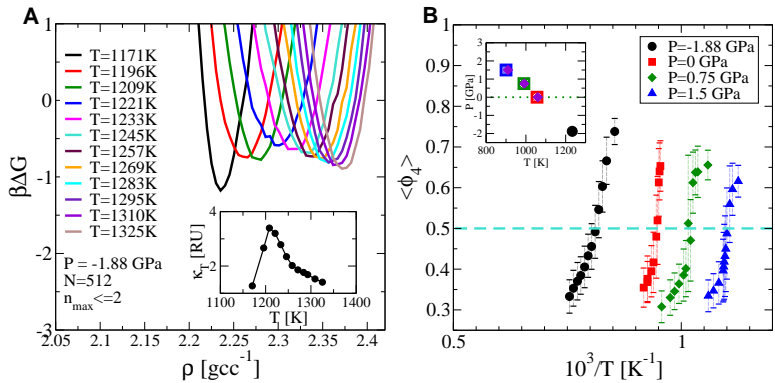

    \centering
    \includegraphics[height=5cm,width=5cm]{Pm05_bDG_rho.eps}
    \includegraphics[height=5cm,width=5cm]{phi4_isobars_final.eps}
    \caption{ Panel (A) Equilibrium sampled density distributions from USMC simulations of $N=512$ atoms along the $P=-1.88~GPa$ isobar. The distributions are unimodal throughout and show no hint of phase separation. {\it Inset} The compressibility measured for different temperatures along the $P=-1.88~GPa$ isobar showing a peak at $T\sim 1230K$, generally consistent with that reported by Vasisht \emph{et al.}
    \cite{vasisht2011liquid}.
    Panel (B) The mean fraction of $4$-coordinated atoms from the equilibrium sampling probability measured subject to the constraint, $n_{max}\leq 4$ shown for $3$ isobars below the critical point from NPT USMC simulations of $N=512$ atoms. $\phi_4$ is $\sim 0.65$ at $T=965K$, $P=0.75~GPa$. Error bars represent the standard deviation of $\phi_4$. ({\it Inset}) The LLPT line obtained by estimating the point of crossing $\langle \phi_4 \rangle=0.5$ for each isobar, shown with symbols of the corresponding colour. The LLPT line obtained from $\beta\Delta G(\rho)$ along each of the $3$ isobars, $P=0~{\rm GPa},~P=0.75~{\rm GPa},~P=1.5~{\rm GPa}$ (see SI for data at $P=0~{\rm GPa}$ and $P=1.5~{\rm GPa}$), is shown with violet triangles. Estimates are found to be consistent with each other and with the equation of state data reported in
    ~\cite{vasisht2011liquid}.
    }
    \label{fig:fig-3}
\end{figure*}

{\bf \noindent Trends across other state points -- absence of bi-modality beyond the critical point:}
We perform USMC simulations constraining only the largest cluster size with a hardwall bias at conditions far from co-existence, where a single liquid phase exists. 
A set of overlapping bias windows is used to constrain $n_{max}$ and the density distribution is measured subject to a constraint of $n_{max}~\leq~4$. 
Parallel tempering swaps across temperatures enhance the sampling of different values of density and away from co-existence this procedure gives quantitatively similar estimates of $\beta\Delta G(\rho)$ as the procedure where both $n_{max}$ and $\rho$ are constrained (see SI for details).
We perform similar USMC computations (both variants) along the $P=0~{\rm GPa}$ and $P=1.5~{\rm GPa}$ isobars for a range of temperatures straddling the LLPT (see SI), as well as at a negative pressure of $P=-1.88~{\rm GPa}$ which is in the supercritical region of the phase diagram reported in~\cite{vasisht2011liquid}.
In this region the extension of the LLPT line corresponds to a locus of maximum compressibility, also known as the Widom line\cite{mishima1998relationship,vasisht2011liquid}. 
No phase separation is expected to occur, though weak bi-modality in the density distribution may be observed at small system sizes when measured in close proximity to the critical point.
One finds no indication of a double-well feature in the free energy reconstructions along the $P=-1.88~{\rm GPa}$ isobar shown in Fig.~\ref{fig:fig-3}(A) suggesting a fully continuous change in the character of the liquid across the Widom line.
From the equilibrium sampling distribution of the fraction of $4$-coordinated (LDL-like) atoms, $\phi_4$, we extract the mean and standard deviation and plot $\langle \phi_4 \rangle$ as a function of temperature along isobars, shown in Fig.~\ref{fig:fig-3}(B).
At coexistence conditions the liquid is composed of equal fractions of LDL-like and HDL-like atoms, enabling an estimate of the LLPT temperature across which the fraction of $4$-coordinated atoms changes sharply, with larger fluctuations around a mean of $0.5$ in the vicinity of the transition temperature, as discussed in \cite{holten2014two}.
The change in the fraction is more gradual across the $P=-1.88~{\rm GPa}$ isobar, indicative of a continuous transformation in the properties of the liquid as seen in Fig.~\ref{fig:fig-3}(A). 


\begin{figure*}[htb!]
\centering
\subfloat{\includegraphics[trim=50 0 65 10,clip,height=4cm,width=4cm]{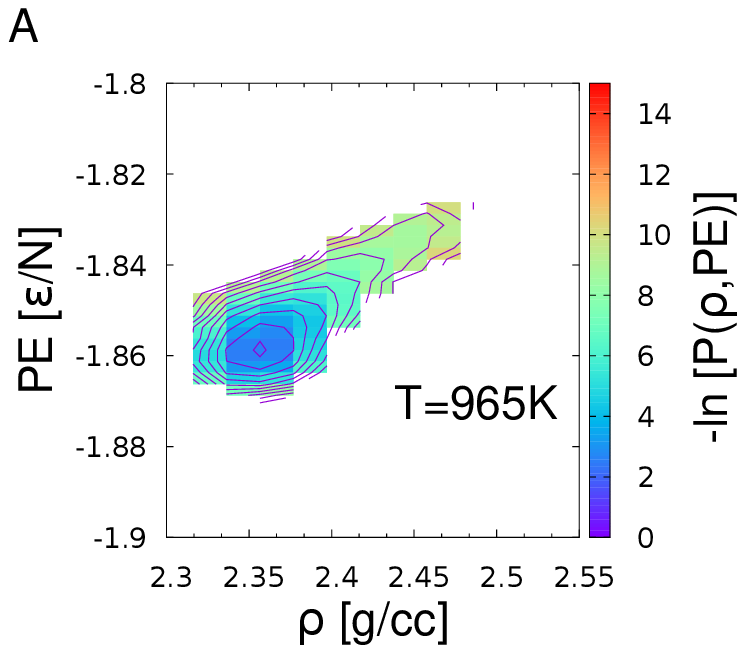}}
\subfloat{\includegraphics[trim=60 0 65 10,clip,height=4cm,width=4cm]{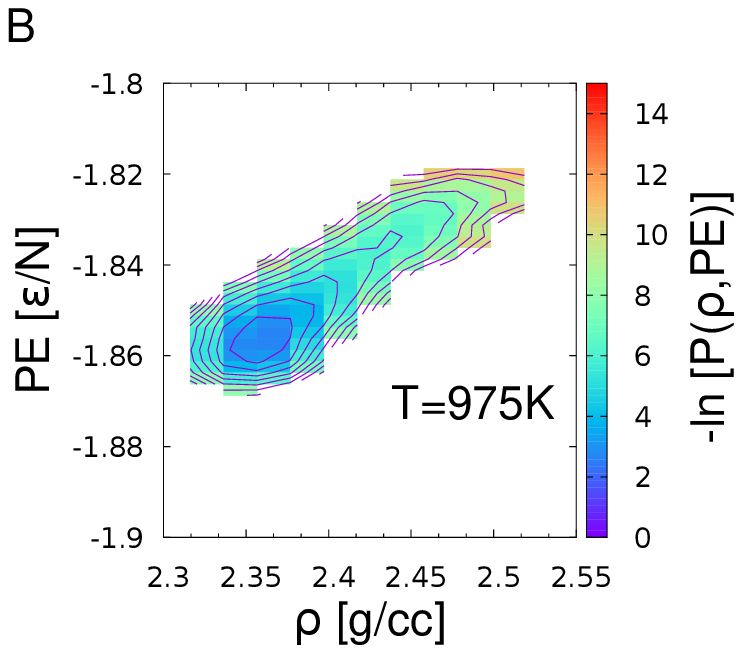}}
\subfloat{\includegraphics[trim=60 0 65 10,clip,height=4cm,width=4cm]{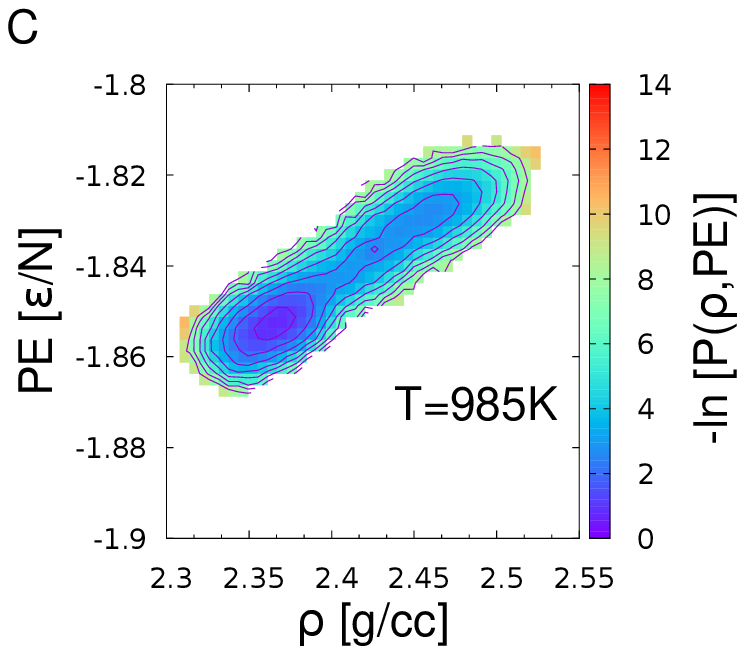}}

\subfloat{\includegraphics[trim=60 0 65 10,clip,height=4cm,width=4cm]{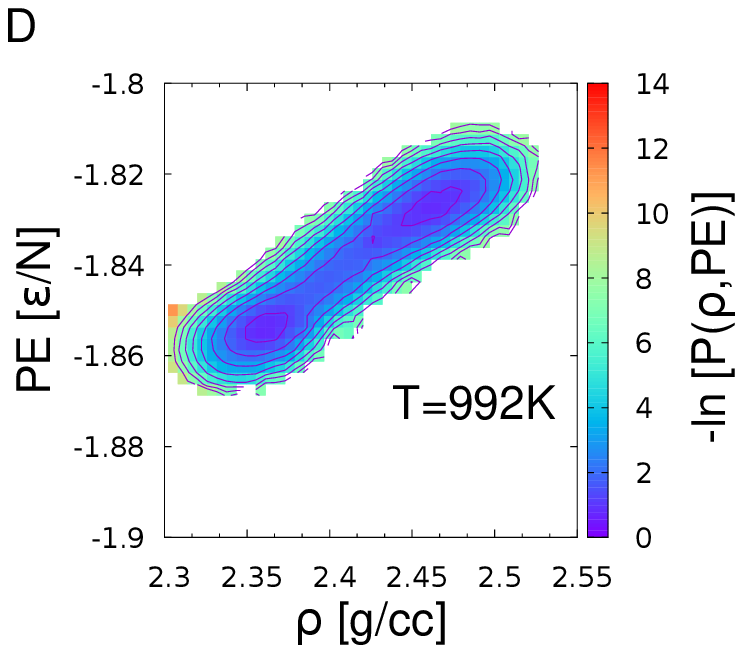}}
\subfloat{\includegraphics[trim=60 0 65 10,clip,height=4cm,width=4cm]{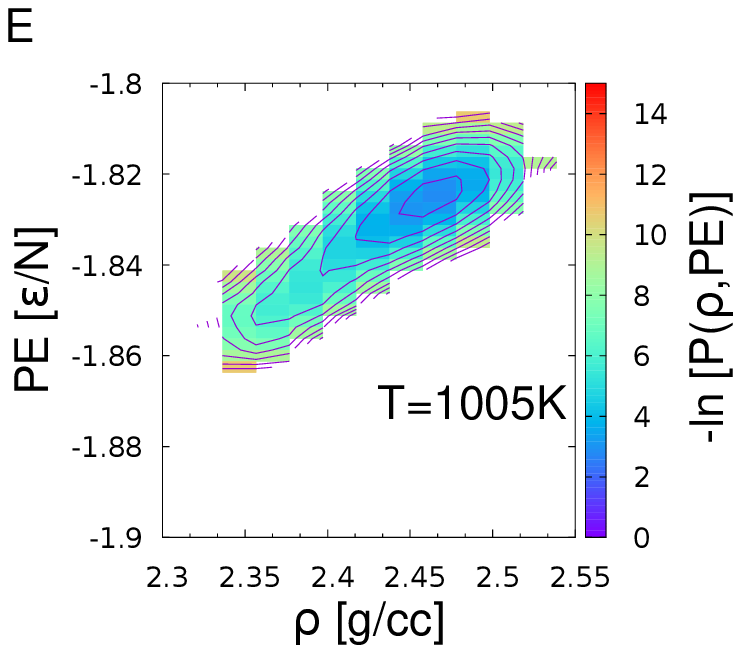}}
\subfloat{\includegraphics[trim=0 -15 -20 0, clip,height=3.5cm,width=3.7cm]{Pm_1085_0p5_final.eps}}
\caption{Negative log of the distribution of density and potential energy per atom obtained subject to the constraint of $n_{max}\leq 4$ along the $P=0.75~{\rm GPa}$ isobar at five temperatures, $T=965{\rm K}$ (A), $T=975{\rm K}$ (B), $T=985{\rm K}$ (C), $T=992{\rm K}$ (D) and $T=1005{\rm K}$ (E). Data is obtained from NPT US MC simulations of $N=512$ atoms. The two liquid phases differ both energetically and in density.
Panel (F) shows a comparison of the distribution of the field-mixing order parameter with the reference Ising 3D distribution at the critical point. The critical point and the field mixing coefficient, $s$, are estimated by iteratively reweighting free energy estimates, obtained directly from umbrella sampling simulations at $P=0~{\rm GPa}$, to different $T,P$ and minimising the difference between $P(M)$ and $P_{ising}(M)$ (see text), where $M=\rho + sE$ is rescaled to have unit variance. The estimate of the critical point obtained from this procedure is $T_c=1085.5{\rm K},~P_c=-0.5~{\rm GPa}$.
}
\label{fig:fig-4}
\end{figure*}

{\bf \noindent Energy and density dependent distributions and critical behavior:}
In Fig.~\ref{fig:fig-4}, we show the multivariate distribution of density and potential energy per atom, subject to the same constraints as in Fig.~\ref{fig:fig-1}(B). Here, one observes basins corresponding to the two liquids, a high energy-high density liquid and a low energy-low density liquid, with a double-well at $T=992~{\rm K}$ as in Fig.~\ref{fig:fig-1}(B). The fact that the HDL has a higher energy and is more stable at the high temperature side of the transition, suggests that the LDL has a lower entropy or fewer favourable configurations. This behaviour is consistent with expectations derived from using the Clausius-Clapeyron equation $(dP/dT)_{LLPT}  = \Delta S/\Delta V$ which relates the slope of the transition line to the difference in entropy and density between the two liquid phases\cite{buldyrev2002models}. The implication of a negative slope for the transition line is that the lower density (higher volume) phase has a lower entropy\cite{holten2012entropy}.

Recent work has investigated the nature of critical fluctuations associated with the LLPT in water, which is generally understood to be of the 3D Ising universality class\cite{debenedetti2020second}.
The critical order parameter has been shown to be a linear combination of the density and potential energy\cite{wilding1997simulation} ($M= \rho + sE$) and its probability distribution at the critical point can be well represented by a standard form\cite{tsypin2000probability} (see Eq.~\ref{eq:ising_ref} in Methods).
In order to identify the field-mixing parameter, $s$, and the critical point, we follow a procedure of iteratively estimating $T_c,P_c$ (to which the histgrams are reweighted) and $s$ for which the order parameter distribution best matches the reference distribution (see Methods and SI), following the approach of Debenedetti \emph{et al.}\cite{debenedetti2020second}.
Fig.~\ref{fig:fig-4}(F) shows the results of this procedure with the estimated critical point and a value of $s=0.6$, which indicates that the distribution of the order parameter agrees closely with the expectation for the 3D Ising universality class. The critical parameters reported in \cite{vasisht2011liquid} ($T_c=1120K,~P_c=-0.6~GPa$)  are in reasonable agreement with the values we obtain in the present analysis. 


{\bf \noindent Comparison with analyses for other models:}
Free energy investigations of the liquid-liquid phase transition for similar tetrahedrally ordered liquids have shown typical barriers of $\sim 1-4~k_BT$. While the barrier height increases with system size, performing constrained simulations at arbitrarily large system sizes is prohibitively expensive.
For ST2 water, Palmer \emph{et al.}\cite{palmer2014metastable} report a barrier height of $\sim 4~k_BT$ with $N=192$ molecules at conditions of coexistence far from the critical point.
Poole \emph{et al.}\cite{poole2013free} have reported similar barrier heights earlier with $N=216$ molecules.
Simulations of silica by Chen \emph{et al}\cite{chen2017liquid} have identified a barrier of less than $4~k_BT$ away from the critical point, but for a system size of $N=1500$ atoms. 
Recent work with two variants of the TIP4P model (which also report analysis of critical fluctuations as belonging to the Ising universality class) have reported density histograms corresponding to a barrier height of less than $2~k_BT$ for $N=300$\cite{debenedetti2020second}. 
The barrier heights we obtain, of $0.8~{\rm k_B T}$ for $N = 512$ and $1.9~{\rm k_B T}$ for $N = 2000$ are thus comparable to these earlier reported values. 


{\bf \noindent Discussion:}
In summary, we find through rigorous free energy calculations and extensive analysis of the consistency of our results, that two well-defined metastable liquid states, with corresponding free energy minima, 
exist in supercooled Stillinger-Weber silicon. 
Co-existence conditions are identified in the sub-critical part of the phase diagram that are in agreement with estimates reported previously from equation of state studies. 
At each of the state points considered, a clear and significant free energy barrier to crystal nucleation is observed, ruling out the possibility that the low density liquid is a transient artefact resulting from slow, spontaneous crystallization. 
At several state points (in particular $T = 965 K$, $P=0.75 GPa$ shown in Fig.~\ref{fig:fig-1}(C)) we observe a free energy minimum corresponding to the low density liquid phase, with a large fraction of tetrahedrally coordinated atoms and zero crystallinity, decisively ruling out the slow crystallization scenario. 
The free energy barrier between the two liquids is found to scale with the size of the simulated system -- an important test of the presence of a first order transition.
Reweighting of free energy profiles across conditions results in identical results, 
providing a strong test of converged equilibrium sampling.
We also find that the same analysis finds no evidence of phase separation when performed along an isobar in the super-critical region of the phase diagram, also consistent with the two-critical point scenario\cite{poole1992phase,vasisht2011liquid}. 
We note that the density difference between the two liquids is small, and remains small as distance from the critical point increases, in contrast to the case of other similar network-forming liquids such as ST2 water\cite{poole2013free,palmer2014metastable} and WAC silica\cite{chen2017liquid,guo2018fluctuations}.
Given the low barriers to crystallization for silicon under these conditions, and the small difference in the densities of the two liquid phases, the barrier separating the two phases is expected to be small.
However, the scaling of barrier height with system size shown here confirms the existence of two well-defined metastable liquid phases.
The barrier heights are of comparable order to these and to other cases such as ST2 water\cite{palmer2014metastable}, TIP4P water\cite{debenedetti2020second} and silica\cite{chen2017liquid}. The two liquid states do not differ in density alone, as shown by a free energy reconstruction along two order parameters, density and potential energy per atom, subject to the constraint of low $n_{max}$. These point to the interplay of energy and entropy in driving the transition, as discussed in the context of other liquids showing a liquid-liquid phase transition. Our work thus provides a comprehensive analysis that resolves the long-standing debate concerning the existence of a liquid-liquid
transition in supercooled Stillinger-Weber silicon. Taken together with the simulation investigations in the case of water and silica, and experimental results concerning water and silicon, there is now a preponderance of evidence in support of liquid-liquid phase transitions in pure substances.

\FloatBarrier
\section{Methods}\label{sec:methods}
\subsection{Interaction potential and simulation protocol}\label{subsec:SW_pot}
We use the classical three-body Stillinger-Weber potential to model silicon\cite{stillinger1985computer}. Monte Carlo simulations are performed in the constant pressure, temperature and particle number (NPT) ensemble. Enhanced reversible sampling is achieved by using the umbrella sampling scheme\cite{torrie1977nonphysical}.
An in-house code with an efficient double-sum implementation\cite{saw2009structural} of the three-body Stillinger Weber potential was used for the umbrella sampling Monte Carlo simulations. 
The bias variables are the size of the largest crystalline cluster, $n_{max}$, and the density $\rho$. A hard wall bias (which is zero within prescribed limits and infinite outside) is used to constrain the $n_{max}$ values as used in\cite{saika2006test,goswami2021thermodynamics} and a harmonic bias constrains the density. Parallel tempering swaps are performed across temperature, $n_{max}$ bias and density bias windows. 
In simulations where only $n_{max}$ is constrained, far from coexistence conditions, parallel tempering swaps are only performed across temperature and $n_{max}$ bias windows.
Swaps are performed between adjacent windows in $n_{max}$, density and adjacent temperatures every $2\times10^2$ MC steps, $10^3$ MC steps and $2\times10^3$ MC steps respectively.
Convergence is determined by monitoring the decay of the time-autocorrelation functions of the density and of the global bond orientational order parameter, $Q_6$\cite{steinhardt1983bond}. These are found to decay in less than $\tau~\sim~10^5-10^6$ MC sweeps at all conditions and system sizes considered. Simulations lengths exceed $10^8$ MC steps at all the conditions studied, with histograms sampled over $\sim~100-200\tau$ for each window.

Statistics of traversal due to parallel tempering swaps are also used to determine adequate sampling. More than $10^4$ parallel tempering swaps are performed in each direction, with observed mean return times being $\sim~ 10^4 - 10^6$ MC steps for simulations in each window. The return time is the number of MC steps before a simulation returns to its initial temperature or bias potential after being swapped out as a result of the replica exchanges.
Further details on the umbrella sampling and parallel tempering scheme are provided in the Supplementary information. 

\subsection{Defining atom types}
Bulk crystalline atoms are identified as those with high degree of local tetrahedral ordering and also surrounded by similarly tetrahedrally ordered atoms. We use the cut-offs described in the SI and also in \cite{romano2011crystallization,vasisht2014nesting,goswami2021thermodynamics}. The local order is identified by using the local bond orientational order for each atom, $q_3(i)$. Neighbouring atoms with correlated neighbourhoods are said to be ``bonded", with the correlation function used being $q_3(i).q_3(j)$. Atoms bonded to $3$ or more neighbours are defined as bulk crystalline atoms. A $4$-coordinated or ``LDL" atom is identified as one with high local $q_3$ but bonded to fewer than $3$ of its neighbours. The fraction of such $4$-coordinated liquid-like atoms, $\phi_4$, is also used to estimate co-existence conditions. At co-existence, the fraction of such $4$-coordinated atoms in LDL-like local structures is expected to be $\sim0.5$\cite{holten2012entropy,holten2014two}. The details of the cut-offs used and the relevant distributions are shown in the SI.

\subsection{Free energy as a function of cluster size and density}
We measure the unbiased probability of observing a cluster of size $n$ in the liquid at density $\rho$ and take the negative log to obtain a free energy as shown below:
\begin{equation}
\Delta G(n,\rho) = -k_BT \ln(P(n,\rho)).
\label{eq:F2D_HW}
\end{equation}
To obtain this, one is required to obtain the following equilibrium probability distribution:
\begin{equation}
P(n,\rho) = \frac{1}{\tau_{s}}\sum_{t=0}^T\frac{N(n,t)}{N(0,t)}\delta(\rho(t) - \rho))\quad if\quad n_{max}^l \leq n \leq n_{max}^u.
\end{equation}
Sampling is performed in the biased ensemble and we use the iterative scheme of the weighted histogram analysis method\cite{kumar1992weighted,chodera2007use} (described in the following section) to obtain the unweighted, normalised distribution, $P(n,\rho)$ from which we obtain the free energy surfaces shown in Fig.~\ref{fig:fig-1}(C) and (D). 
Note that the contracted free energy surface in Fig.~\ref{fig:fig-1}(B) and Fig.~\ref{fig:fig-4}(A-E) are obtained by first constructing the unbiased estimate for $P(n_{max},\rho)$ from free energy reweighting. Then the contracted free energy, $\beta\Delta G(\rho)$, is obtained by summing $P(n_{max},\rho)$ up to the chosen largest value of $n_{max}$ and taking the negative logarithm (see Eq.~\ref{eq:Prho_marginal}).

\subsection{Reweighting and stitching free energy surfaces}\label{subsec:FE_stitch}
For umbrella sampling runs with two order parameters, we employ an in-house code that implements the self-consistent iterative scheme of the WHAM equations\cite{kumar1992weighted,chodera2007use} (see SI ). 
Errors are estimated from the number of decorrelated samples and the integrated autocorrelation times.
Tests for thermodynamic consistency are performed by comparing reweighted estimates of the free energy to directly measured estimates at different conditions (see SI).
For single order parameter umbrella sampling simulations only the largest cluster size is constrained with parallel tempering across temperatures enhancing sampling of density (see SI). The two methods agree quantitatively for conditions far from co-existence, whereas only the full two-order parameter US simulations give reliable results at or near co-existence conditions (see SI for details).

\subsection{Critical fluctuations of the order parameter}
We investigate whether the liquid-liquid critical point belongs to the 3D-Ising universality class by comparing the probability distribution of the relevant order parameter with the reference distribution for the 3D-Ising model. In the case of the Ising model, the magnetisation, $M$, undergoes critical fluctuations in the vicinity of the critical point. In the case of the LLPT, the relevant order parameter is a linear combination of density and the potential energy ($\rho + sE$)\cite{wilding1997simulation}. The following general expression is found to be a good approximation to the distribution of $M$\cite{tsypin2000probability}
\begin{equation}
P_{ising}(M) \propto exp \Biggl \{ - \left (\frac{M^2}{M_0^2} -1 \right )^2\left ( a\frac{M^2}{M_0^2} +c \right )^2    \Biggr \}
\label{eq:ising_ref}
\end{equation}
The appropriate choice of constants yields a distribution of unit variance (see SI for details). The distribution of the order parameter, $M=\rho+sE$, is expected to match the reference distribution at the critical point. The critical point is identified by finding the optimal set of $T_c,P_c,s$ that minimizes the root-mean-squared error of $P(M)$ with respect to $P_{ising}(M)$ (see SI for details). This procedure gives both an estimate of the critical point as well as the field-mixing parameter, $s$.



Refer Supplementary Information.


\bmhead{Acknowledgements}
We gratefully acknowledge C. Austen Angell, Pablo G. Debenedetti, Francesco Sciortino, Francis Starr, Vishwas Vasisht, Daan Frenkel and Peter H. Poole  for discussions and TUE-CMS, SSL, JNCASR, and the National Supercomputing Mission,  (Param Yukti) at the Jawaharlal Nehru Centre for Advanced Scientific Research (JNCASR), for computational resources. SS acknowledges support through the JC Bose Fellowship (JBR/2020/000015) from the Science and Engineering Research Board, Department of Science and Technology, India.



\part*{}   
\setcounter{equation}{0}
\setcounter{figure}{0}
\setcounter{section}{0}
\renewcommand{\theequation}{S\arabic{equation}}%
\renewcommand{\thefigure}{S\arabic{figure}}%
\renewcommand{\thesection}{S\arabic{section}}%

\section*{Supplementary Information}




The supplementary information contains additional details on the model and methods and several sections with details relevant to the results shown in the main text. These are organised as follows:
\begin{enumerate}
    \item Additional details of the model and methods.
    \item Convergence tests for simulations of $N=512$ atoms along the $P=0.75~GPa$ isobar.
    \item Comparison of free energy reconstruction from umbrella sampling with both $n_{max}$ and $\rho$ biased to umbrella sampling simulations where only $n_{max}$ is biased, far from coexistence conditions.
    \item Free energy reconstructions along the $P=0~GPa$ and $P=1.5~GPa$ isobars.
    \item Free energy reconstruction at larger system sizes. 
    \item Fitting order parameter distribution to the Ising universality class.
\end{enumerate}
\section{Additional Details of the Model and Methods}\label{sec:SI_methods}
We construct the free energy landscape here using two methods, umbrella sampling with a hard wall bias potential and a harmonic bias potential. 

Here we describe a prescription to extend these methods to the case of two order parameters and discuss the results obtained.
\subsection*{Model - The Stillinger-Weber potential} \label{subsec:potential}
The Stillinger-Weber potential consists of a two-body term and a three-body term, $U_2$ and $U_3$, respectively.~\cite{stillinger1985computer}
\begin{equation}
U_{SW} = \sum_{j>1}^N U_2(r_{ij}) + \sum_{j>i<k}^N U_3({\bf r_i,r_j,r_k})
\end{equation}
The {\bf $r_i,r_j,r_k$} are position vectors for atoms $i,j,k$. $r_{ij}$ is the distance between the $i^{th}$ and $j^{th}$ atoms. N is the total number of atoms in the system. \\
\begin{equation}
U_2(r_{ij})=
\begin{cases}
\quad \epsilon A \left( \frac{B}{r^4_{ij}} -1 \right)e^{\frac{1}{r_{ij} - r_c} } \quad &if \quad r<r_c\\
\quad 0 \quad &if \quad r \ge r_c\\
\end{cases}
\end{equation}
The three-body interaction term is defined by
\begin{equation}
\begin{split}
U_3({\bf r_i,r_j,r_k}) = h(r_{ij},r_{ik},\theta_{jik})+h(r_{ij},r_{jk},\theta_{ijk})+\\h(r_{ik},r_{jk},\theta_{ikj})
\end{split}
\end{equation}
In turn, 
\begin{equation*}
h(r_{ij},r_{ik},\theta_{jik})=
\begin{cases}
\quad \epsilon \lambda \left[ cos\theta_{jik} + \alpha \right]^2 e^{\frac{\gamma}{r_{ij}-r_c} + \frac{\gamma}{r_{ik}-r_c}} \quad &if \quad r_{ij},r_{ik} < r_c\\
\quad  0 &if \quad r_{ij}\quad or \quad r_{ik} \ge r_c\\
\end{cases}
\end{equation*}
The constants used in the equations above are listed in the table below:
\begin{center}
\begin{tabular}{ c c c c c c c}
\hline \hline 
Symbol & $A$ & $B$ & $r_c$ & $\lambda$ & $\alpha$ & $\gamma$ \\
\hline
Value & $7.04955$ & $0.60222$ & $1.80$ & $21.0$ & $1/3$ & $1.20$ \\
\hline \hline
\end{tabular}
\end{center}
Interconversion factors between standard units and reduced units are listed in Table~\ref{tab:Conversion} below:
\begin{table}[ht]
     \centering
     \begin{tabular}{| c | c |}
          \hline
          Observable & Factor (Unit)\\ 
          \hline
          \hline
	  Length & $r^* \times 2.0951 $ (\AA) \\
          \hline
          Temperature & $T^* \times 25173 $ (K) \\
          \hline
          Energy & $E^* \times 209.5 $ (kJ/mol) \\
          \hline
	  Mass & $m^* \times 28.0855 $ (gm/mol) \\
          \hline
          Time & $t^* \times 76.6 $ (fs) \\
          \hline
          Pressure & $P^* \times 37.776 $ (GPa) \\
          \hline
	  Density & $\rho^* \times 5.0571 $ (gm/$cm^3$) \\
          \hline
	  Diffusivity & $D^* \times 0.005730345 $ ($cm^2$/s) \\
          \hline
	  Viscosity & $\eta^* \times 0.029060146 $ (poise) \\
          \hline
          \hline 
     \end{tabular}
     \caption{Conversion factor for various observables calculated from
     the Stillinger-Weber model potential for silicon.}  
     \label{tab:Conversion}
\end{table}
\subsection*{Order Parameters}\label{subsec:OP}
The bond orientational order parameters of Steinhardt, Nelson and Ronchetti \cite{steinhardt1983bond} are used to distinguish bulk crystalline atoms from liquid-like atoms and further to distinguish LDL-like liquid atoms from HDL-like atoms. Specifically, the local analogue of this order parameter can be used to distinguish the neighbourhoods of individual atoms and classify them as being ordered or disordered.
\begin{equation}
q_{lm}(i)=\frac{1}{n_b(i)}\sum_{j=1}^{n_b(i)}Y_{lm}[\theta(r_{ij}),\phi(r_{ij})]
\end{equation}
The corresponding order parameter, summed over $m's$ is
\begin{equation}
q_l(i)=[\frac{4\pi}{(2l+1)}\sum_{m=-l}^l|q_{lm}(i)|^2]^{1/2}
\end{equation}
Here, we use $q_3(i)$, noting that using $q_6(i)$ is equivalent and gives very similar results~\cite{vasisht2014nesting}.
The number of neighbours, $n_{b}(i)$, is taken to be the number of atoms within the first coordination shell of the pair-correlation function, i.e., atoms within a cut-off of $2.95~\AA$ from the reference atom.
Other works have considered other definitions, such as considering only the four nearest neighbours. However, when there are more than four atoms at similar distances from the reference atom, certain artefacts arise such as the apparent decrease of tetrahedral ordering with density or an increase with pressure~\cite{vasisht2014nesting}.
We therefore employ a distance-based cut-off to specify nearest neighbours. To identify crystalline atoms, we compute the correlations in the local orientational order of neighbouring atoms, following the prescription described in the literature\cite{van1992computer,ten1995numerical,wolde1996simulation}.
Atoms with correlated neighbourhoods of high local orientational order are classified solid-like atoms.

\begin{figure*}[htpb!]
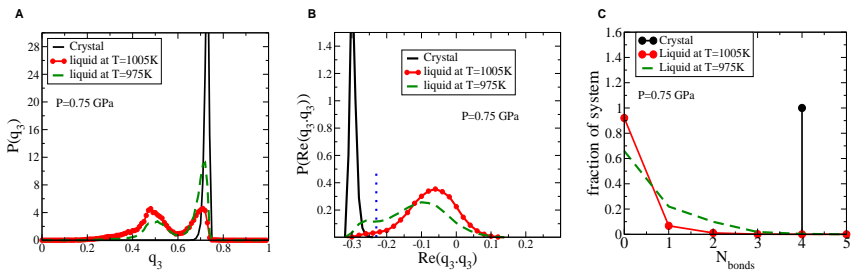

\centering
\subfloat{\includegraphics[height=3.5cm,width=3.5cm]{q3_compare.eps}}\hspace{2mm}
\subfloat{\includegraphics[height=3.5cm,width=3.5cm]{q3q3.eps}}\hspace{2mm}
\subfloat{\includegraphics[height=3.5cm,width=3.5cm]{bonded_neigh_filtered.eps}}
\caption{Panel A shows the $q_3$ distributions for the three types of atoms, panel B shows the distribution of the correlation function and panel C shows the distribution of neighbours that are ``bonded" according to the correlation function for the three phases.}
\label{fig:fig-s1}
\end{figure*}

Quantitatively, this correlation is given by the quantity,~\cite{ten1995numerical,romano2011crystallization,kesselring2013finite}
\begin{equation}
Re \left( q_3(i).q_3(j) \right)=Re\left (\sum_{-3}^{3} q_{3m}(i)q_{3m}^*(j) \right)
\end{equation}
An atom $i$ and an atom $j$ are considered to be ``bonded" if $Re(q_3(i).q_3(j))<-0.23$. We note here the significance of the the cut-off value of $-0.23$ which demands that the crystal structure formed is diamond cubic, to the exclusion of the  hexagonal crystal structure which also has local tetrahedral ordering~\cite{romano2011crystallization,goswami2021thermodynamics}.
Crystalline atoms have a $q_3>0.6$ and are ``bonded" to at least 3 neighbours. Further, crystalline atoms within the SW-cutoff distance of each other belong to the same cluster. In this study we consider both the size of the largest cluster, $n_{max}$ and the full distribution of cluster sizes $P(n)$.
We observe that using $q_6(i).q_6(j)$ to identify crystalline atoms gives nearly identical results when the appropriate cut-off is chosen. The choice of cut-off will depend on whether a normalisation factor is included in the definition~\cite{ricci2019computational,goswami2021thermodynamics}. LDL-like atoms have a high $q_3(i)>0.6$, showing high tetrahedral ordering, but have fewer than $3$ neighbours with similar ordering. Finally, HDL-like atoms have disordered neighbourhoods with $5$ or more neighbours. Fig. \ref{fig:fig-s1} shows the distributions of $q_3$, $Re \left( q_3(i).q_3(j) \right)$ and the number of bonded neighbors for typical crystalline, LDL and HDL configurations. 

\subsection*{Umbrella sampling}\label{subsec:US}
Umbrella sampling Monte Carlo (USMC) simulations were performed at the state points mentioned with a hard wall bias applied that strictly constrains the size of the largest crystalline cluster to be between $n_{max}^l$ and $n_{max}^u$ as described in Eq.~\ref{eq:W_HW}. The full cluster size distribution is used to calculate the free energy upto an additive constant using Eq.~\ref{eq:F_HW}.\\
\begin{equation}
    \beta\Delta G(n) = -\ln[P(n)] + \mathrm{const.}
    \label{eq:F_HW}
\end{equation}
Parallel tempering swaps between adjacent windows are carried out to enhance sampling and speed up equilibration.
The general expression for the Hamiltonian under application of bias is given by:
\begin{equation}
H_C = H + W_1(n_{max}) + W_2(\rho)
\label{eq:bias_full}
\end{equation}
where $H$ is the original Hamiltonian, $W_1(n_{max})$ represents the bias potential on $n_{max}$, and $W_2(\rho)$ is the bias potential on $\rho$.\\
Here, $W_1$ is defined by
\begin{align}
W_1 &=
\begin{cases}
0 & n_{lo} \leq n_{max} < n_{hi}\\
\infty & otherwise
\end{cases}
\label{eq:W_HW}
\end{align}
For $W_2$, a harmonic bias of the form 
\begin{equation}
W_2(\rho; \rho_{0},k_{\rho}) = \frac{1}{2} k_{\rho} (\rho - \rho_{0})^2
\label{eq:W_harmonic}
\end{equation}
is used to enhance sampling around a desired value of $\rho$, labelled $\rho^{0}$.
We can write the constrained Hamiltonian as:
\begin{equation}
H_C = H + W_1(n_{max};n_{lo},n_{hi}) + W_2(W_2(\rho; \rho_{0},k_{\rho})
\label{eq:bias_full_new}
\end{equation}

The unbiased expectation value of some system property such as the density, $\rho$ can be written as (in general for a bias applied on any combination of collective variables such as $(n_{max},\rho)$):
\begin{equation}
\left < \rho \right >  = \frac{\left < \rho e^{\beta W_1+W_2}\right >_C }{\left < e^{\beta W_1+W_2)}\right >_C}
\label{eq:unbias_full}
\end{equation}
The expectation subscript $C$ is the sampled probability from the simulation under the modified Hamiltonian. Likewise, $W_C$ is the bias potential that describes the constrained ensemble.
\subsection*{Parallel tempering}\label{subsec:PT}
The general expression for probability of acceptance of parallel tempering swaps in the NPT ensemble between simulations indexed $i$ and $j$ is given by
\begin{equation}
\begin{split}
P_{accept} = min\Biggl(1, exp\Bigl[\bigl[ (E_i -E_j) + P(V_i - V_j)\bigr] (\beta_i - \beta_j)\Bigr]\Biggr. \\
\Biggl. exp\bigl[-\beta_j W_i(n_{max_j}) - \beta_i W_j(n_{max_i})\bigr]\Biggr.\\
\Biggl. exp\bigl[\beta_i W_i(n_{max_i}) + \beta_j W_j(n_{max_j}) \bigr]\Biggr )
\end{split}
\end{equation}
The details of parallel tempering are as follows:
\begin{itemize}
\item Consider $N$ independent simulations run in parallel - different temperatures or different bias potentials.
\item To ensure better sampling of the configuration space 
and consequently of the order parameter, we swap adjacent configurations periodically.
\item Two types of swaps are performed, one type where simulations with different temperatures but the same bias potential exchange configurations and one type where simulations at the same temperature but different bias potentials exchange configurations.
\item A swap between adjacent simulations indexed $i$ and $j$, at different temperatures, $1/\beta_i$ and $1/\beta_j$, but with the same bias potential is executed with a probability of $min\Biggl ( 1,exp\bigl[ (E_i -E_j) + P(V_i - V_j)\bigr] (\beta_i - \beta_j) \Biggr )$
\item For cases where $\beta$ is the same but the bias potential varies, the probability is $min\left ( 1,exp\left[ \beta(W_N - W_O) \right] \right )$
\item Here, the term $W_N - W_O$ represents the sum of the bias potentials after the swap minus the sum of the bias potentials before the swap (the sum being over the bias applied on the two runs in consideration.
\begin{eqnarray}
W_N &=& W_j(n_{max_i}) + W_i(n_{max_j}) \nonumber \\
W_O &=& W_i(n_{max_i}) + W_j(n_{max_j}) \nonumber
\end{eqnarray}
In all simulations replica exchanges are attempted across adjacent temperatures and bias windows. Thus, in simulations where both $n_{max}$ and $\rho$ are constrained, parallel tempering swaps are performed across $T$, $[n_{max}^l:n_{max}^u]$ and $\rho_0$.
For the hard wall bias, the swap is accepted with probability $1$ if the $n_{max_i}$ and $n_{max_j}$ are both within the new constraints after the swap and rejected otherwise.
\end{itemize}
\subsection*{Unbiasing and stitching free energies with WHAM}\label{subsubsec:WHAM}
The weighted histogram analysis scheme\cite{kumar1992weighted,chodera2007use} is used to unbias and stitch together free energy estimates from the different independent umbrella sampling simulations, as well as to reweigh the unbiased distributions to other values of $T,P$. We describe the procedure generally, before describing the exact details of implementation in each of the cases where free energy stitching and/or reweighting is performed.
In what follows, we begin by considering a general case where our goal is to obtain the unbiased equilibrium distributions of $E,\rho (1/V)$ and any other order parameter(s). In our case, the additional order parameter is $n_{max}$. 



The value of the bias potential $W_i(E,\rho,n_{max})$ in a simulation indexed $i$ refers to the total bias that includes a bias potential on $n_{max}$ and the density $\rho$ (but not on $E$ in our simulations),  shown in Eq.~\ref{eq:bias_full_new}: 


\begin{equation}
    W_i(E,\rho,n_{max}) = W_1(n_{max};n_{lo}^i,n_{hi}^i) + W_2(\rho; \rho_{0}^i,k_{\rho}^i),
\end{equation}
Given $R$ NPT simulations performed under different conditions (different temperature, pressure and/or bias), one obtains equilibrium unbiased estimates of $E,\rho,n_{max}$, i.e., the internal energy, volume and order parameter which can be reweighted to nearby temperatures and pressures. 
The density of states, $\Omega(E,\rho,n_{max})$, is given by iteratively solving the following self-consistent equations\cite{kumar1992weighted,chodera2007use,debenedetti2020second}
\begin{align}
\Omega(E,\rho,n_{max}) &= \frac{\sum\limits_{i=1}^{R}H^b_i(E,\rho,n_{max})}{\sum\limits_{i=1}^{R} N_i e^{-\beta_i E}e^{-\beta_i \frac{P_iN}{\rho}}e^{-\beta W_i(E,\rho,n_{max})}e^{F_i}} \nonumber \\
e^{-F_i} &= \sum_{\{E,V,n_{max}\}}\Omega(E,\rho,n_{max})e^{-\beta_i E}e^{-\beta_i \frac{P_iN}{\rho}}e^{-\beta W_i(E,\rho,n_{max})}
\end{align}
where $H^b_{i}(E,\rho,n_{max})$ is the histogram obtained in simulation $i$, $N_i$ is the total number of entries from simulation $i$, and $F_i$ are the shifts applied to each window. 
These equations are solved iteratively to self-consistency to obtain the shifts corresponding to each simulation window and the density of states, $\Omega(E,\rho,n_{max})$.


{\bf \noindent On-the-fly unbiasing}, i.e., factoring out the Boltzmann weight associated with the bias potential for each configuration sampled, is useful to obtain the unbiased distributions of all quantities, under the given conditions of $P,\beta$.\\
For this, the histograms for each simulation of length $\tau_i$ are computed as  
\begin{equation}
H_i(E,\rho,n_{max}) = \frac{1}{\tau_i}\sum_{t=1}^{\tau_i}\delta (E(t) - E)\delta (\rho(t) - \rho)\delta (n_{max}(t) - n) e^{\beta_i W_i(E,\rho,n_{max})},
\end{equation}
as opposed to a flat histogram count without the inverse of the Boltzmann factor for the bias potential. The iterative equations employing $H_i$ are: 

\begin{align}
\Omega(E,\rho,n_{max}) &= \frac{\sum\limits_{i=1}^{R}H_i(E,\rho,n_{max})}{\sum\limits_{i=1}^{R} N_i e^{-\beta_i E}e^{-\beta_i \frac{P_iN}{\rho}} e^{F_i}} \nonumber \\
e^{-F_i} &= \sum_{\{E,V,n_{max}\}}\Omega(E,\rho,n_{max})e^{-\beta_i E}e^{-\beta_i \frac{P_iN}{\rho}}.
\end{align}

We can write an unnormalised distribution from the density of states, reweighted to some target $\beta,P$ as:

\begin{align}
N_{ub}(E,\rho,n;_{max}\beta,P) &= \Omega(E,\rho,n_{max})e^{-\beta E}e^{-\beta \frac{PN}{\rho}} \nonumber \\
N_{ub}(E,\rho,n_{max};\beta,P) &=  e^{-\beta E}e^{-\beta \frac{PN}{\rho}}\frac{\sum\limits_{i=1}^{R}H_i(E,\rho,n_{max})}{\sum\limits_{i=1}^{R} N_i e^{-\beta_i E}e^{-\beta_i \frac{P_iN}{\rho}}e^{F_i}}
\end{align}
This can be explicitly normalised to obtain the probability distributions, 
\begin{equation}
P_{ub}(E,\rho,n_{max};\beta,P) = \frac{\Omega(E,\rho,n_{max})e^{-\beta E}e^{-\beta \frac{PN}{\rho}}}{\sum\limits_{\{E,\rho,n_{max}\}}\Omega(E,\rho,n_{max})e^{-\beta E}e^{-\beta \frac{PN}{\rho}}}
\end{equation}
which are written in terms of $H_i$ as 
\begin{align}
P_{ub}(E,\rho,n_{max};\beta,P) &= \frac{\sum\limits_{i=1}^{R}H_i(E,\rho,n_{max})}{\sum\limits_{i=1}^{R} N_i e^{(\beta -\beta_i) E}e^{(\beta P -\beta_i P_i)\frac{N}{\rho}}e^{F_i}} \nonumber \\
e^{-F_i} &=\sum_{\{E,\rho,n_{max}\}}P_{ub}(E,\rho,n_{max})e^{(\beta - \beta_i) E}e^{(\beta P - \beta_i P_i)\frac{N}{\rho}}
\label{eq:wham_big}
\end{align}
from which we can obtain the free energies.
The procedure is as follows:
\begin{enumerate}
\item The $F_i$ are initialised to arbitrary non-zero values.

\item $P_{ub}(E,\rho,n;\beta,P)$ in  Eq. \ref{eq:wham_big} (a) is computed using histograms $H_i$.

\item $F_i$ are computed from $P_{ub}(E,\rho,n;\beta,P)$ using  Eq. \ref{eq:wham_big} (b)

\item Steps $2$ and $3$ are repeated till a tolerance value of $10^{-4}$ is reached for the quantity, $\sqrt{\frac{1}{R}\sum\limits_{i=1}^{R}(F_i - F_i^{old})^2}$.
\end{enumerate}


\noindent {\bf Integrated auto-correlation time and errors} are computed from the auto-correlation function of the order parameter (we have used $\rho$),
\begin{equation}
C_{\rho}(t) = \frac{\langle \rho(t)  \rho(0)\rangle - \langle \rho \rangle^2}{\langle \rho^2 \rangle - \langle \rho \rangle^2}
\end{equation}
The integrated auto-correlation time is obtained from the self auto-correlation as 
\begin{equation}
g = 1 + 2\sum\limits_{t=1}^{T-1} \left ( 1 - \frac{t}{T}\right ) C_{\rho}(t).
\label{eq:Integ_AC}
\end{equation}
We desire the integrated auto-correlation time for the slowly varying density, $\rho$, and the resultant measure of the error in our estimates of $\beta\Delta G(\rho)$. We thus weight sampling according to the number of decorrelated samples obtained as a function of $\rho$.
The error as a function of $\rho$ is 
\begin{equation}
\omega(\rho) = \left ( \sum\limits_{i=1}^{N_{sim}} \frac{g_i^{\rho}}{\langle H_i(\rho) \rangle} \right )^{1/2}
\label{eq:wham_error}
\end{equation}
The WHAM equations are then modified as:
\begin{align}
P_{ub}(E,\rho,n_{max};\beta,P) &= \frac{\sum\limits_{i=1}^{R}g_i^{-1}H_i(E,\rho,n_{max})}{\sum\limits_{i=1}^{R} g_i^{-1}N_i e^{(\beta -\beta_i) E}e^{(\beta P -\beta_i P_i)\frac{N}{\rho}}e^{F_i}} \nonumber \\
e^{-F_i} &=\sum_{\{E,\rho,n_{max}\}}P_{ub}(E,\rho,n)e^{(\beta - \beta_i) E}e^{(\beta P - \beta_i P_i)\frac{N}{\rho}}
\label{eq:wham_main}
\end{align}
In the subsequent discussion, we adapt the WHAM equations described above to the specific cases of:
\begin{enumerate}
    \item Obtaining the free energy barrier as a function of cluster size, $\beta\Delta G(n)$
    \item Stitching and reweighted the distribution as a function of $E,\rho$ in order to obtain $\beta\Delta G(\rho)$
    \item Constructing the two order parameter free energy $\beta\Delta G(n,\rho)$
\end{enumerate}
\noindent {\bf Case 1: Stitching free energy as a function of cluster size, $\beta\Delta G(n)$}
Here, we obtain the unbiased distribution of all cluster sizes, $P_{ub}(n)$, from which we write the free energy, $\beta\Delta G(n)$.
The WHAM equations are used to obtain $P_{ub}(n)$ (at the same temperature and pressure at which the simulations are performed):
\begin{align}
P_{ub}(n) &= \frac{\sum\limits_{i=1}^{R} H_i(n)}{\sum\limits_{i=1}^{R} N_i e^{-F_i}} \nonumber \\
e^{-F_i} &= \sum_{n}P_{ub}(n)
\label{eq:wham_bDG_n}
\end{align}
Histogram entries $H_i(n)$ are given by:
\begin{align}
H_i(n) &= \frac{1}{\tau_i}\sum\limits_{t=1}^{\tau_i}
\begin{cases}
\delta(n(t)-n) &\text{if} \quad n_{lo} \leq n \leq n_{hi}\\
0 & otherwise
\end{cases}
\label{eq:Hn_update}
\end{align}
While the bias constraints are applied on $n_{max}$, we track the full cluster size distribution because the approximation $P(n_{max})\approx P(n)$ does not hold at deep supercooling when the size of the critical cluster is small\cite{wolde1996simulation,goswami2021thermodynamics}.
In obtaining the estimates for $H_i(n)$ and consequently for $P_{ub}(n)$ ($\beta\Delta G(n)=-\ln[P_{ub}(n)]$), we only consider data for $n$ values within the bounds $[n_{lo}^i,n_{hi}^i]$ for each simulation, even though all cluster sizes $n~\leq n_{hi}^i$ are sampled.
This is done because the frequency of occurrence of clusters of size $n_{lo}\leq n\leq n_{hi}$ satisfies the requirement that at least one cluster in the specified size range must be present at any point in time\cite{goswami2021thermodynamics}. 
The normalisation factor for $n$ within the bounds is therefore not meaningfully applicable to values of $n$ outside the bounds; it is therefore simpler to discard data for $n$ outside the bounds.

\begin{figure}[htpb!]
\centering
\includegraphics[scale=0.35]{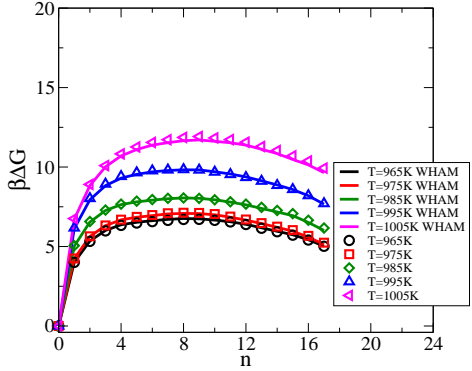}
\caption{A comparison of free energy stitching procedures using the WHAM equations in Eq.~\ref{eq:wham_bDG_n} and the minimisation of $\chi_{HW}$ in Eq.~\ref{eq:HW_stitch_n} showing identical results from simulations of $N=512$ atoms at $P=0.75~GPa$.}
\label{fig:wham_n_test}
\end{figure}

The stitching using the WHAM procedure can be compared with a procedure where the free energies are stitched by determining the appropriate additive constant, $b_d$, which minimizes the discrepancy between free energy estimates from different, but overlapping, bias windows, in the overlapping regions.
From a set of independent simulations, each indexed by $d$ and having distinct but adjacent bounds for $n_{max}$, one obtains the free energy differences $\beta\Delta G_d(n)$ up to an undetermined constant, $b_d$. The constants, $b_d$, are obtained by minimising the error described in Eq.~\ref{eq:HW_stitch_n}, $\chi_{HW}$, sequentially between overlapping data points from simulations with adjacent bounds.  
\begin{equation}
\chi_{HW} = \sum_{d=1}^{N_{sim}} \sum_{n=n_{lo}^d}^{n_{hi}^d} \left [\beta\Delta G_d(n) - \beta\Delta G_{d+1}(n) - b_d \right ]^2
\label{eq:HW_stitch_n}
\end{equation}
This is done subject to the constraint
\begin{equation}
 \beta\Delta G(0)= 0 \quad \text{if}\quad n_{lo}^{d}=0.
 \label{eq:HW_fix}
\end{equation}
A comparison of results is shown in Fig.~\ref{fig:wham_n_test} where we find that the two procedures give quantitatively identical results.

\noindent {\bf Case 2: Stitching and reweighting $P_{ub}(E,\rho)$ and obtaining $\beta\Delta G(\rho)$}
The following equations are solved self-consistently, with a condition imposed that the largest cluster size, $n_{max}\le n_{lim}$, when updating the histogram $H_i(E,\rho)$.

\begin{align}
P_{ub}(E,\rho;\beta,P) &= \frac{\sum\limits_{i=1}^{R}g_i^{-1}H_i(E,\rho)}{\sum\limits_{i=1}^{R} g_i^{-1}N_i e^{(\beta -\beta_i) E}e^{(\beta P -\beta_i P_i)N/\rho}e^{F_i}} \nonumber \\
e^{-F_i} &=\sum_{\{E,\rho\}}P_{ub}(E,\rho,n)e^{(\beta - \beta_i) E}e^{(\beta P - \beta_i P_i)N/\rho}
\label{eq:wham_rho_PE}
\end{align}
The rule for updating the histogram $H_i(E,\rho)$ is given by:
\begin{align}
H_i(E,\rho) &= \frac{1}{\tau_i}\sum\limits_{t=1}^{\tau_i}
\begin{cases}
\delta(E(t)-E)\delta(\rho(t)-\rho) &\text{if} \quad n_{max}(t)\leq n_{lim}\\
0 & otherwise
\end{cases}
\label{eq:H_Erho_update}
\end{align}
$P_{ub}(E,\rho;\beta,P)$ is the unbiased sampling probability of energy and density, measured subject to a constraint on $n_{max}$, weighted on $\beta$ and $P$. The density histogram can be obtained by summing over all values of $E$.
The term $g_i$ is the integrated auto-correlation time for each simulation window, shown in Eq.~\ref{eq:Integ_AC}. The results from this procedure are shown in Fig.~\ref{fig:reweight_match_nmax_2}, Fig.~\ref{fig:reweight_match_nmax_4} and Fig.~\ref{fig:rho_PE_reweight_992}.

{\bf \noindent Case 3: Constructing the two order parameter free energy $\beta\Delta G(n,\rho)$}
$\beta\Delta G(n,\rho)$ shows the degree of crystallinity along one axis and the density along the other so that the two liquids can be characterised in relation to the crystalline phase.
We write the WHAM equations as in Eq.~\ref{eq:wham_big}, where we are interested now in $P_{ub}(n,\rho)$. As discussed in the main text and Methods, this is the probability of observing a cluster of size $n$ when the liquid has density, $\rho$.
We begin by noting that the histogram $H(n,\rho)$ is updated as follows:
\begin{equation}
H_i(n,\rho) = \frac{1}{\tau_{i}}\sum\limits_{t=0}^{\tau_i}\frac{N(n,t)}{N(0,t)}\delta(\rho(t) - \rho)\quad if\quad n_{lo}^i \leq n \leq n_{hi}^i
\end{equation}
We then write the unbiased probability $P_{ub}(n,\rho)$ as
\begin{align}
P_{ub}(n,\rho;\beta,P) &= \frac{\sum\limits_{i=1}^{R}g_i^{-1}H_i(n,\rho)}{\sum\limits_{i=1}^{R} g_i^{-1}N_ie^{F_i}} \nonumber \\
e^{-F_i} &=\sum_{\{E,\rho\}}P_{ub}(n,\rho)
\label{eq:wham_n_rho}
\end{align}
The bi-variate distribution, $P(n,\rho)$ yields the full two-order parameter free energy $\beta\Delta G(n,\rho)$.
\begin{equation}
\Delta G(n,\rho) = -k_BT \ln(P(n,\rho)).
\label{eq:F2D_HW_si}
\end{equation}
One may also consider the free energy $\beta\Delta G(n_{max},\rho)$ however, this would lead to the appearance of an artificial minimum at small $n_{max}$\cite{maibaum2008comment,chakrabarty2008chakrabarty,goswami2021thermodynamics}.
Considering $\beta\Delta G(n,\rho)$ allows us to clarify features of the free energy landscape at deep supercooling, as well as to verify the presence of the LDL phase in the absence of any crystalline ordering.
\section{Convergence tests for $N=512$ atoms along the $P=0.75~GPa$ isobar}\label{subsec:2DUS_convergence}

\begin{figure}[htpb!]
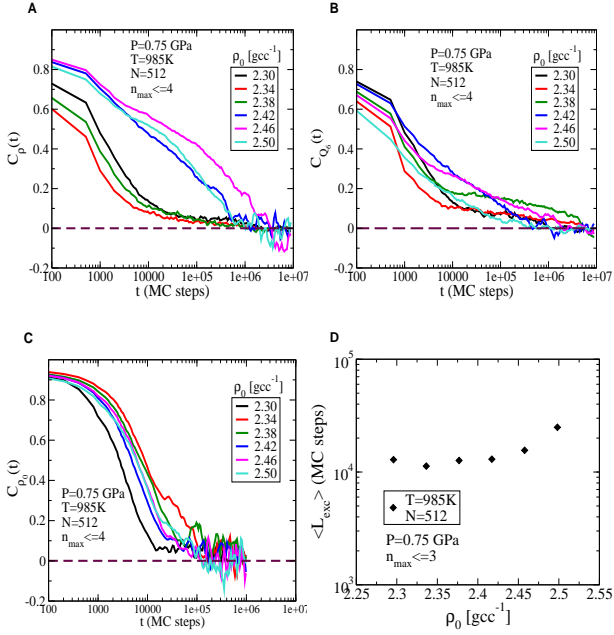

    \centering
    \subfloat{\includegraphics[height=4cm,width=4cm]{2DAC_rho_985.eps}}
    \subfloat{\includegraphics[height=4cm,width=4cm]{2DAC_q6_985.eps}}
    
    \subfloat{\includegraphics[height=4cm,width=4cm]{2DAC_PT_985.eps}}
    \subfloat{\includegraphics[height=4cm,width=4cm]{2DPT_exc_len_985.eps}}
    \caption{Decay of time auto-correlation function for density (Panel A), $Q_6$ (Panel B), and for density window index, $c_{id}$, (Panel C) for each of the density bias windows, $\rho_0$, subject to the constraint of $n_{max}~\leq 4$ at $T=985K$, $P=0.75~GPa$. Panel D shows the mean excursion length or return time as a function of $\rho_0$ subject to the constraint on $n_{max}$ at $T=985K$. 
    In each case, the different curves are labelled according to the initial reference density, $\rho_0$, for the given independent simulation.}
    \label{fig:2DUS_stats_985}
\end{figure}

\begin{figure}[htpb!]
    \centering
    \subfloat{\includegraphics[height=4cm,width=4cm]{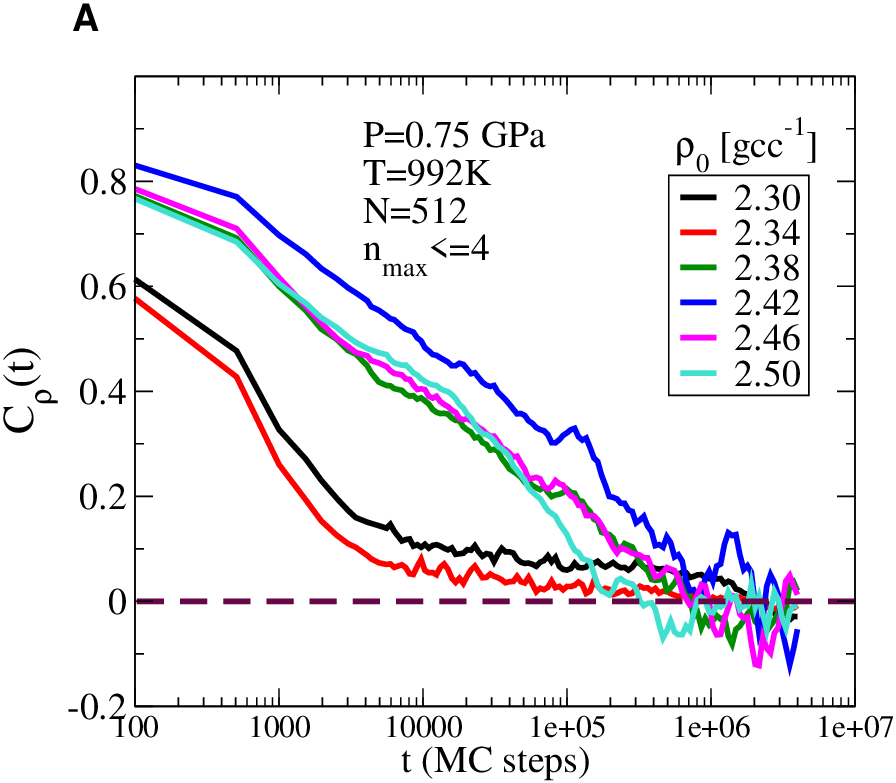}}
    \subfloat{\includegraphics[height=4cm,width=4cm]{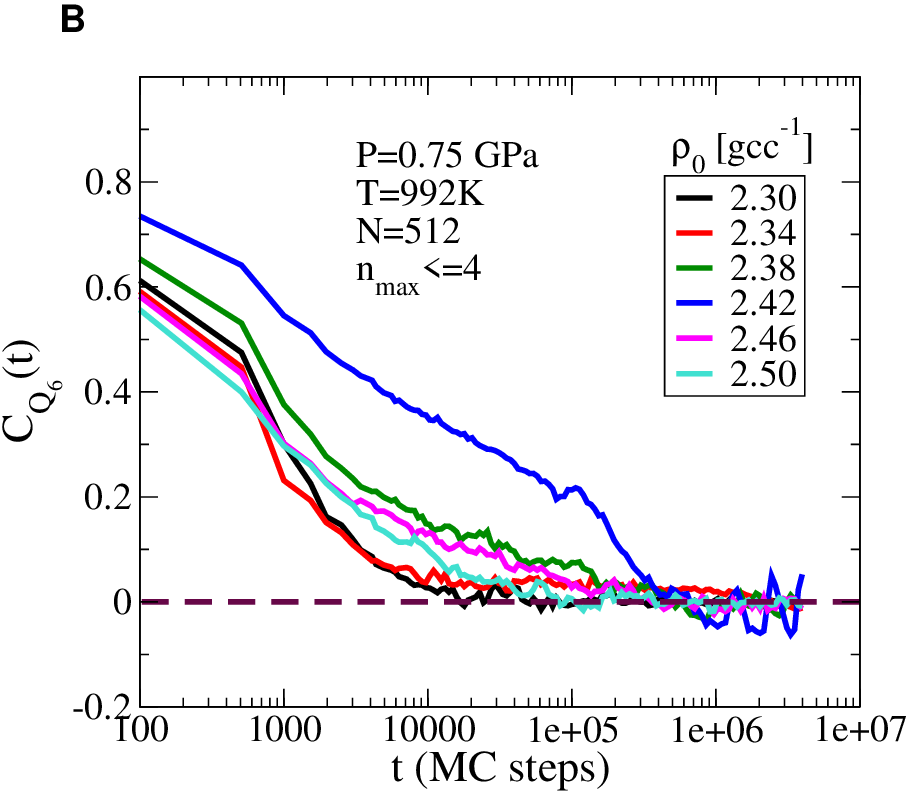}}
    
    \subfloat{\includegraphics[height=4cm,width=4cm]{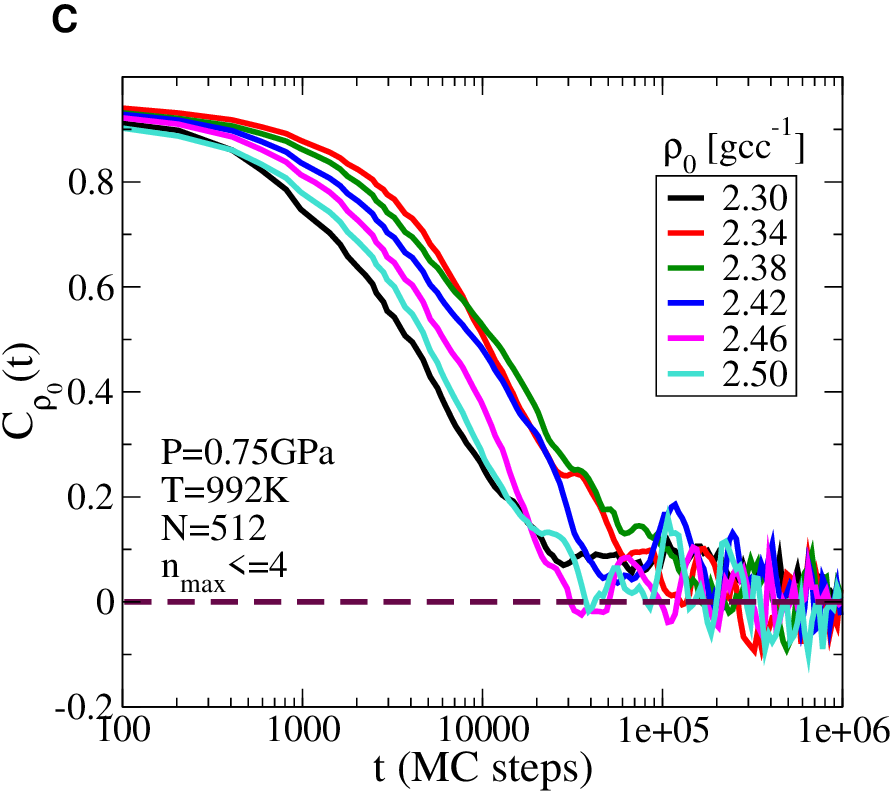}}
    \subfloat{\includegraphics[height=4cm,width=4cm]{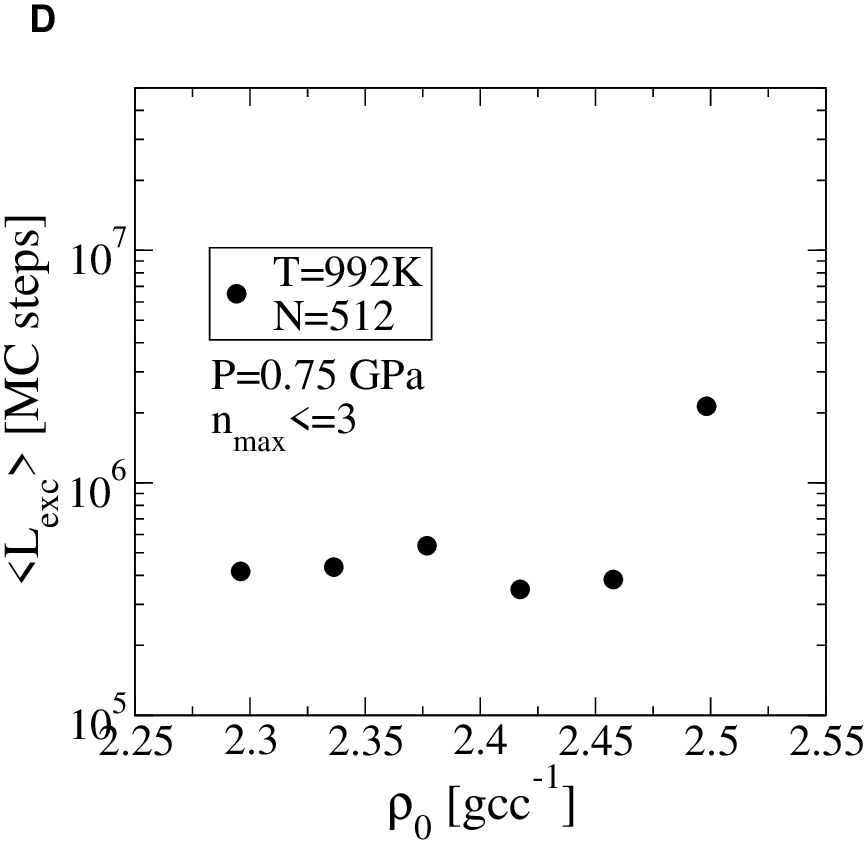}}
    \caption[Decay of time auto-correlation function for density, $Q_6$, and for density window index for each of the density bias windows, $\rho_0$, subject to the constraint of $n_{max}~\leq 4$ at $T=992K$, $P=0.75~GPa$.]{Decay of time auto-correlation function for density (Panel A), $Q_6$ (Panel B), and for density window index (Panel C) for each of the density bias windows, $\rho_0$, subject to the constraint of $n_{max}~\leq 4$ at $T=992K$, $P=0.75~GPa$. Panel D shows the mean excursion length or return time as a function of $\rho_0$ subject to the constraint on $n_{max}$. 
    In each case, the different curves are labelled according to the initial reference density, $\rho_0$, for the given independent simulation.}
    \label{fig:2DUS_stats_992}
\end{figure}

We measure decorrelation times for key quantities in our umbrella sampling simulations in order to test for the convergence of the free energy estimate, $\beta\Delta G(\rho)$, subject to a constraint on $n_{max}$. In generating the corresponding autocorrelation functions, we consider two types of time series.
The first type is a time series of configurations subject to a given bias potential labelled by $\rho_0$, the reference density. With this, we compute the autocorrelation of the $\rho$ and $Q_6$ shown in panels A and B respectively of Fig.~\ref{fig:2DUS_stats_985} and Fig.~\ref{fig:2DUS_stats_992}
({\bf Note:} This time series exhibits discontinuities when a swap occurs). 
The second type is a time series of each trajectory initialised in a given bias window. These are labelled by the bias potential applied at the initial time and are subjected to different bias potentials over time when parallel tempering swaps are performed. These are continuous trajectories, but the bias potential changes over time, giving a corresponding time series of the reference $\rho_0$ values. We compute the autocorrelation function of the reference $\rho_0$, shown in panel C of Fig.~\ref{fig:2DUS_stats_985} and Fig.~\ref{fig:2DUS_stats_992}).
For the first type of time series, the bias potential remains the same, whereas for the second type, the bias potential changes with time.

In Panel D of the same figures, the mean excursion length away from the initial reference density, or mean return time, is shown for different density windows, indexed $\rho_0$, subject to the constraint of $n_{max}~\leq 4$.

\FloatBarrier
\subsection*{Histogram reweighting at $P=0.75~GPa$ to test for equilibrium sampling}
\begin{figure}[htpb!]
\centering
\includegraphics[scale=0.5]{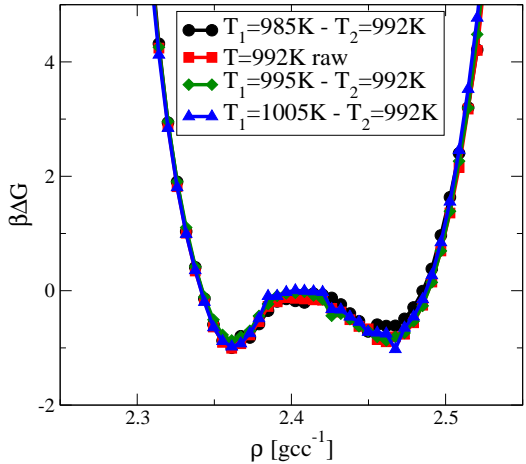}
\caption{Raw and re-weighted free energies at three temperatures, for $P=0.75~GPa$, $N=512$, with a constraint on $n_{max}$ at $n_{max} \le 4$. Note that for this constraint, the coexistence temperature is $T=992K$.}
\label{fig:reweight_match_nmax_4}
\end{figure}

\begin{figure}[ht!]
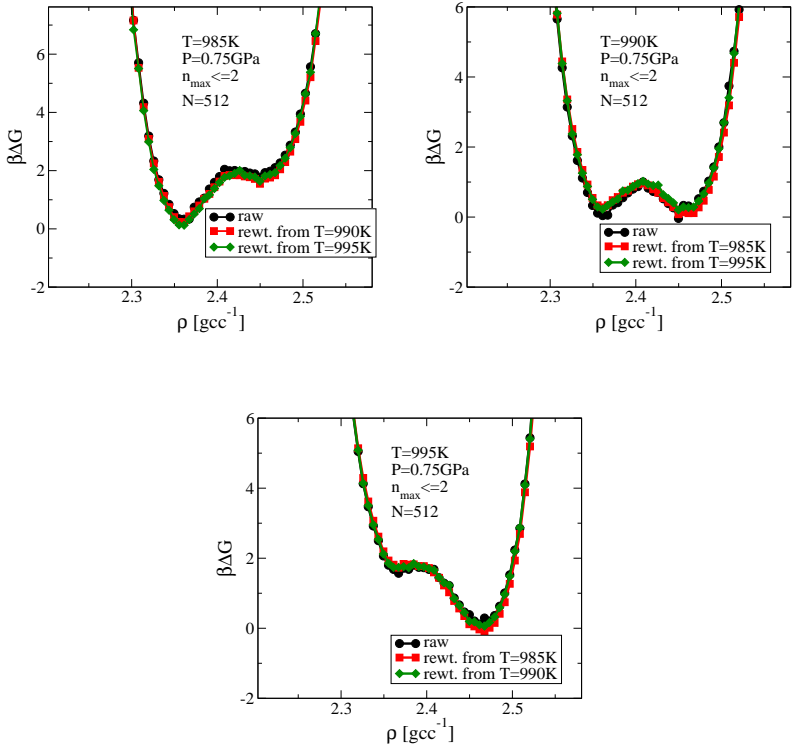

\centering
\vspace{10mm}
\includegraphics[scale=0.35]{T985_rw_WHAM_Prho_512.eps}\hspace{5mm}
\vspace{10mm}
\includegraphics[scale=0.35]{T990_rw_WHAM_Prho_512.eps}\hspace{3mm}
\vspace{8mm}
\includegraphics[scale=0.35]{T995_rw_WHAM_Prho_512.eps}
\caption{Raw and re-weighted free energies at three temperatures, for $P=0.75~GPa$, $N=512$, with a constraint on $n_{max}$ at $n_{max} \le 2$. For this constraint of smaller $n_{max}$ (compared to Fig.~\ref{fig:reweight_match_nmax_4}), the coexistence temperature shifts to a lower temperature of $T=990K$.}
\label{fig:reweight_match_nmax_2}
\end{figure}

The histogram reweighting procedure in Eq.~\ref{eq:wham_rho_PE} is used to obtain the unbiased, reweighted, bivariated distribution $P_{ub}(E,\rho;\beta,P)$ at the target conditions of $T=1/\beta$, $P$.
The free energy, shown in Fig.~\ref{fig:reweight_match_nmax_4} and in Fig.~\ref{fig:reweight_match_nmax_2} is given by $\beta\Delta G(\rho;\beta,P)=-ln(P_{ub}(\rho;\beta,P)$. Note that changing the constraint on $n_{max}$ alters the coexistence temperature at a given isobar (compare Fig.~\ref{fig:reweight_match_nmax_2} and Fig.~\ref{fig:reweight_match_nmax_4}), however, the feature of coexistence remains and is robust to changes in the choice of the upper bound in $n_{max}$ less than the critical cluster size.

\subsection*{Histogram reweighting of $\beta\Delta G(\rho,E)$}
\begin{figure}[htp!]
\centering
\includegraphics[trim=60 30 80 20,clip,scale=0.7]{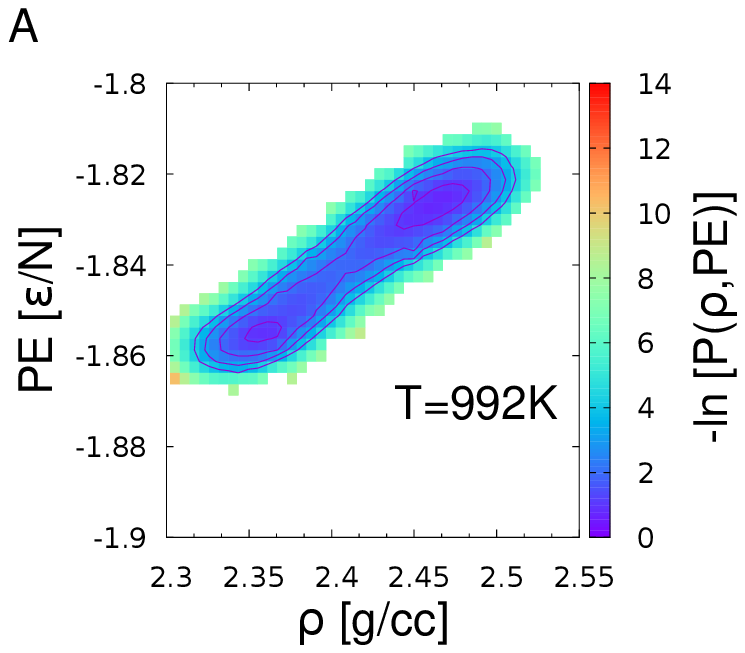}
\vspace{4mm}
\includegraphics[trim=60 30 80 20,clip,scale=0.7]{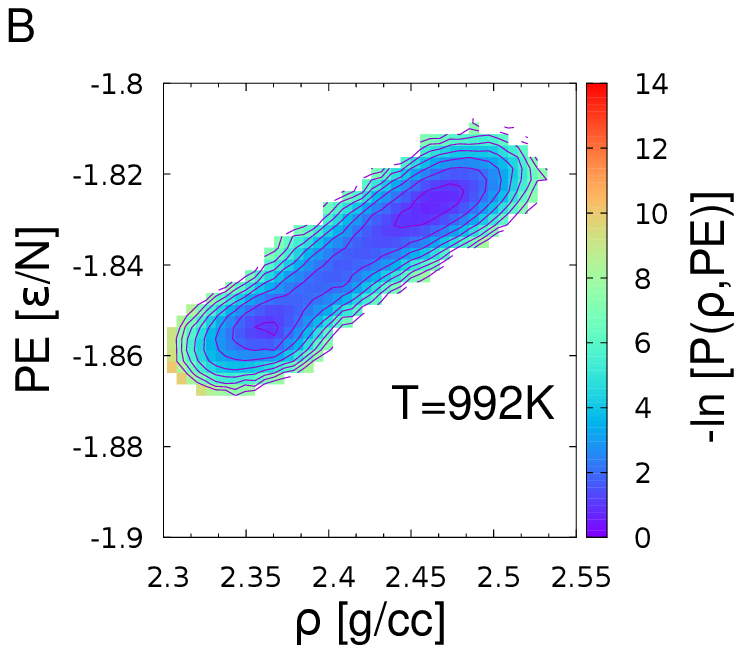}
\vspace{4mm}
\includegraphics[trim=60 30 80 20,clip,scale=0.7]{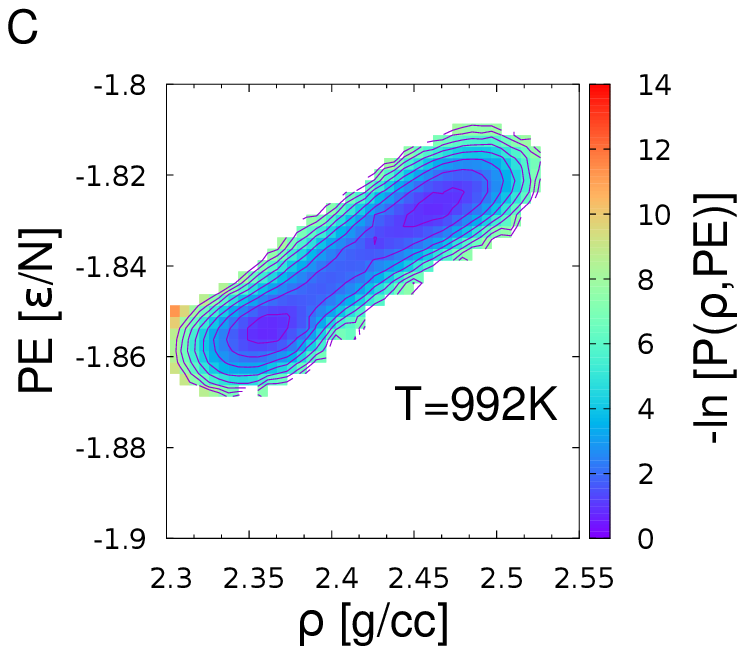}
\vspace{4mm}
\caption{({\it Left:}) Re-weighted from $T_1=985K$ to $T_2=992K$, ({\it Centre:}) computed directly from umbrella sampling simulations, ({\it Right:}) re-weighted from $T_1=995K$ to $T_2=992K$.}
\label{fig:rho_PE_reweight_992}
\end{figure}

In Fig.~\ref{fig:rho_PE_reweight_992}, we show results from applying the histogram reweighting procedure described in Eq.~\ref{eq:wham_rho_PE} to the bivariate distribution of $\rho$ and the potential energy. By reweighting across temperatures along the $P=0.75~GPa$ isobar, we find that directly measured free energy estimates are identical to those obtained by $T,~P$ reweighting, which is a strong indication of converged, equilibrium sampling.

\FloatBarrier
\section{Comparison of methods far from co-existence}\label{sec:compare_2d}
Away from the state points where two liquids co-exist, the two umbrella sampling schemes are expected to give the same results. Close to co-existence, the scheme of only performing parallel tempering swaps across temperature, without constraining the density, may or may not give converged estimates of the free energy on reasonable simulation timescales. This is because the temperature parallel tempering needs to effect a barrier crossing. The comparison is made either side of co-existence along the $P=0.75~GPa$ isobar.
\subsection{Convergence and sampling tests for umbrella sampling runs constraining $n_{max}$ only}\label{sec:SI_convergence}
\begin{figure}[h!]
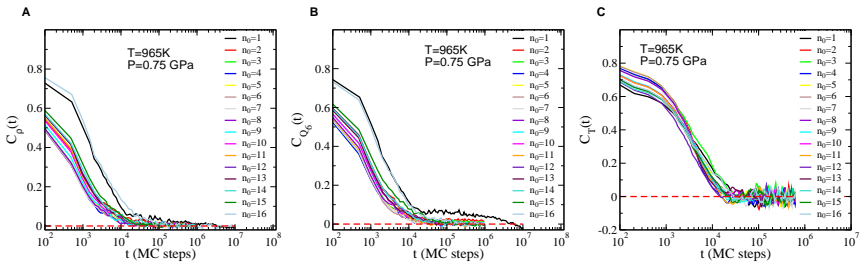

    \centering
    \includegraphics[scale=0.25]{Ac_rho_965.eps}
    \includegraphics[scale=0.25]{Ac_Q6_965.eps}
    \includegraphics[scale=0.25]{Ac_PT_965.eps}
    \caption{Autocorrelation of density, $C_{\rho}(t)$ (Panel A), $Q_6$ , $C_{Q_6}(t)$ (Panel B) and simulation temperature $C_{T}(t)$ (Panel C) at $T=965K$, $P=0.75~GPa$ with $N=512$. The initial temperature for each simulation is as shown in the panel, the different curves are labelled according to the midpoint of the $n_{max}$ bias window, $n_0$.}
    \label{fig:AC_TPT_965}
\end{figure}

\begin{figure}[h!]
    \centering
    \includegraphics[scale=0.25]{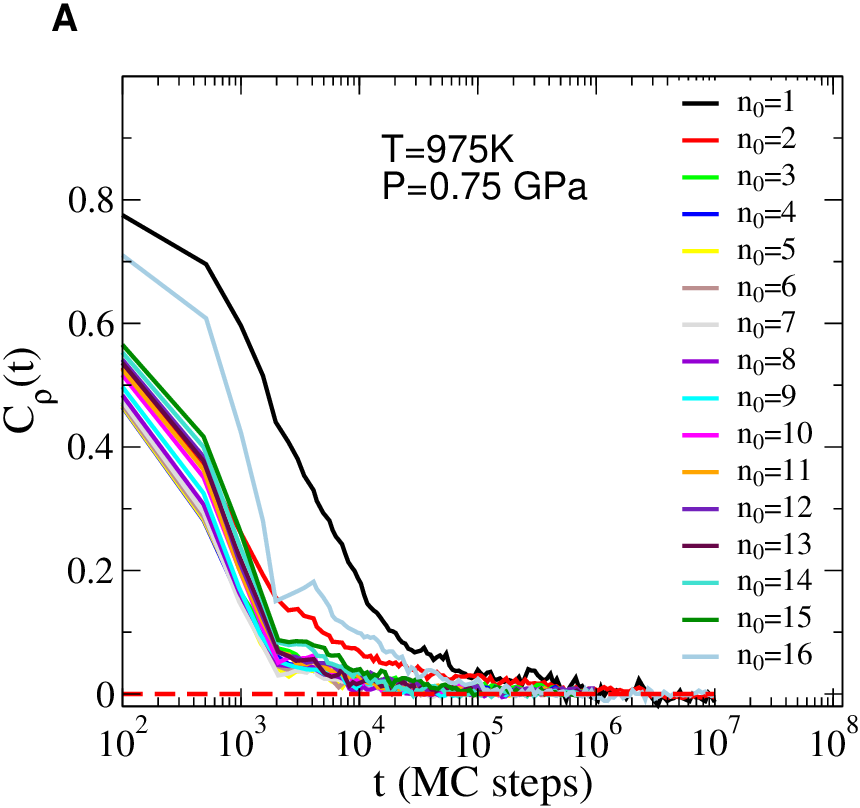}
    \includegraphics[scale=0.25]{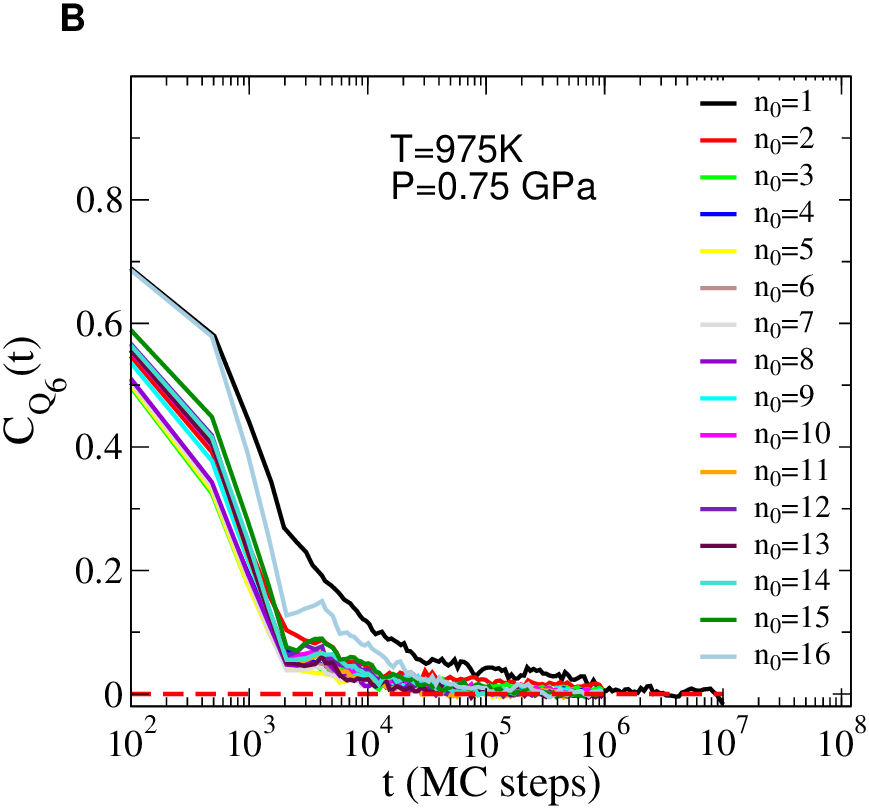}
    \includegraphics[scale=0.25]{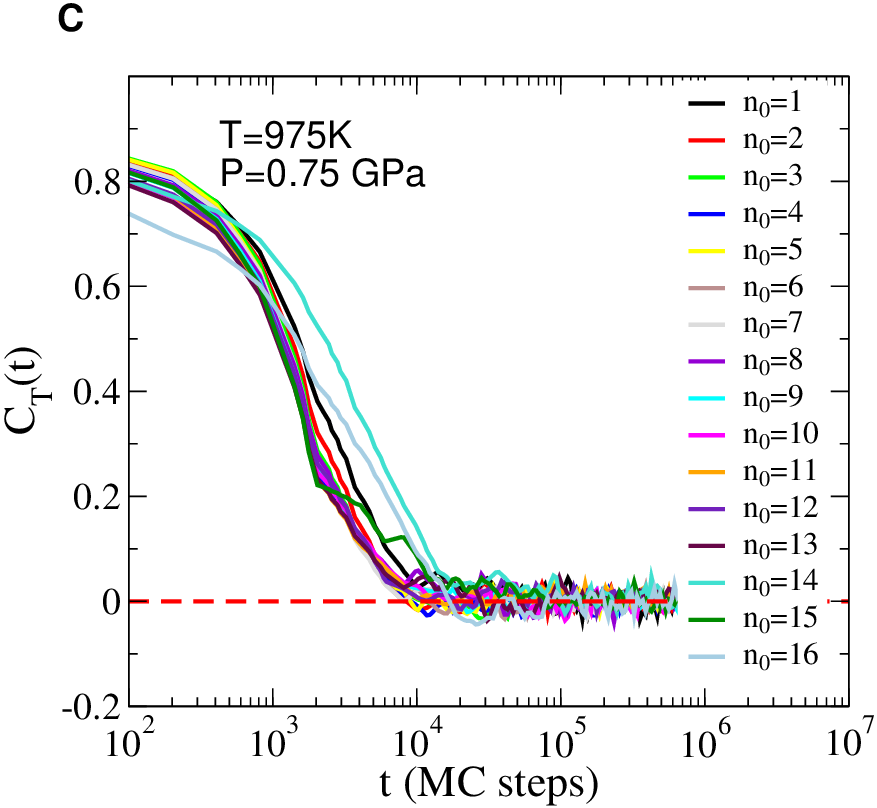}
    \caption{Autocorrelation of density, $C_{\rho}(t)$ (Panel A), $Q_6$ , $C_{Q_6}(t)$ (Panel B) and simulation temperature $C_{T}(t)$ (Panel C) at $T=975K$, $P=0.75~GPa$ with $N=512$. The initial temperature for each simulation is as shown in the panel, the different curves are labelled according to the midpoint of the $n_{max}$ bias window, $n_0$.}
    \label{fig:AC_TPT_975}
\end{figure}

\begin{figure}[h!]
    \centering
    \includegraphics[scale=0.25]{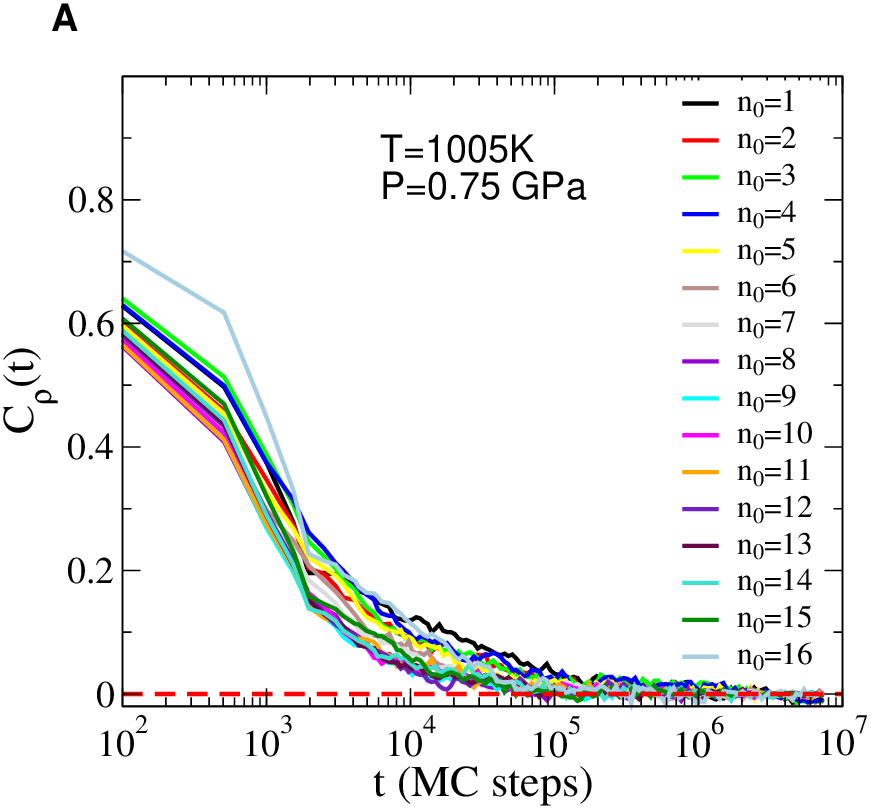}
    \includegraphics[scale=0.25]{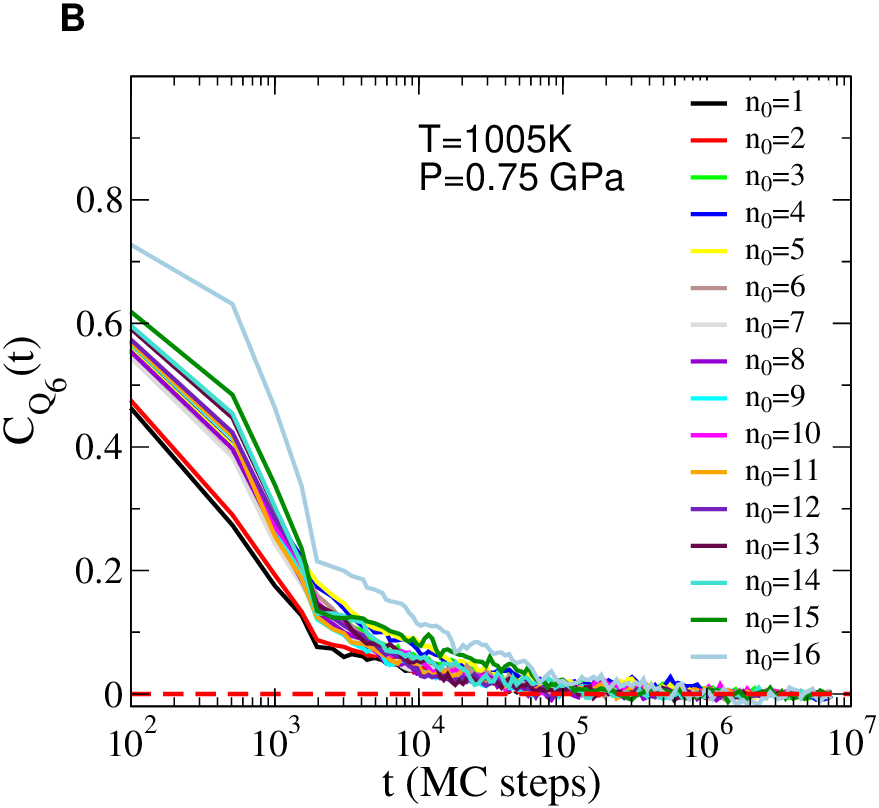}
    \includegraphics[scale=0.25]{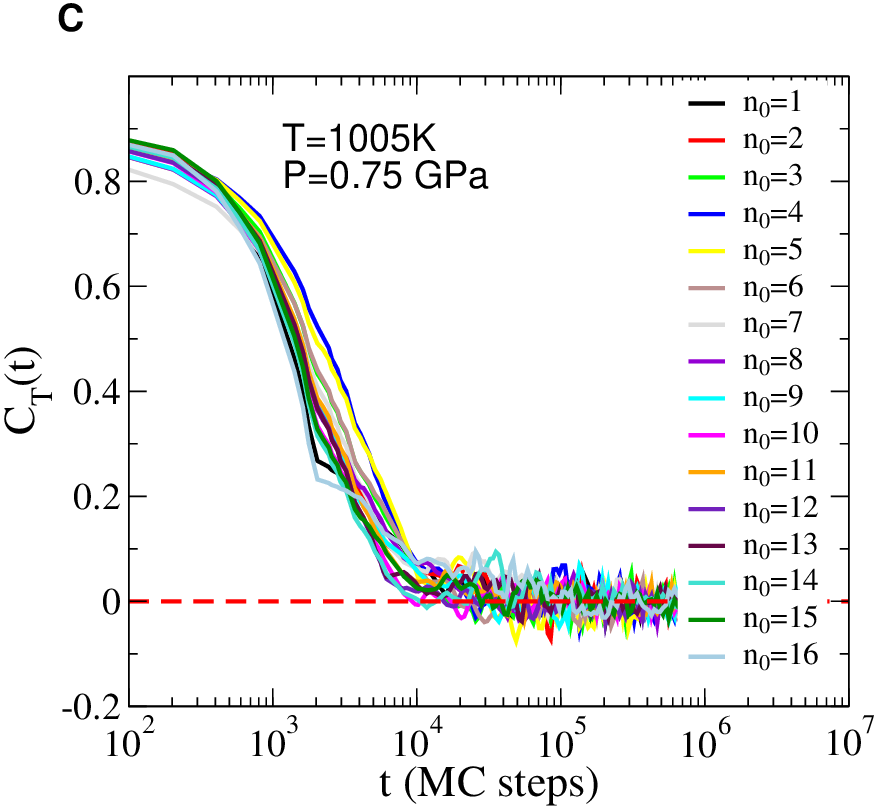}
    \caption{Autocorrelation of density, $C_{\rho}(t)$ (Panel A), $Q_6$ , $C_{Q_6}(t)$ (Panel B) and simulation temperature $C_{T}(t)$ (Panel C) at $T=1005K$, $P=0.75~GPa$ with $N=512$. The initial temperature for each simulation is as shown in the panel, the different curves are labelled according to the midpoint of the $n_{max}$ bias window, $n_0$.}
    \label{fig:AC_TPT_1005}
\end{figure}

\begin{figure}[h!]
    \centering
    \includegraphics[scale=0.25]{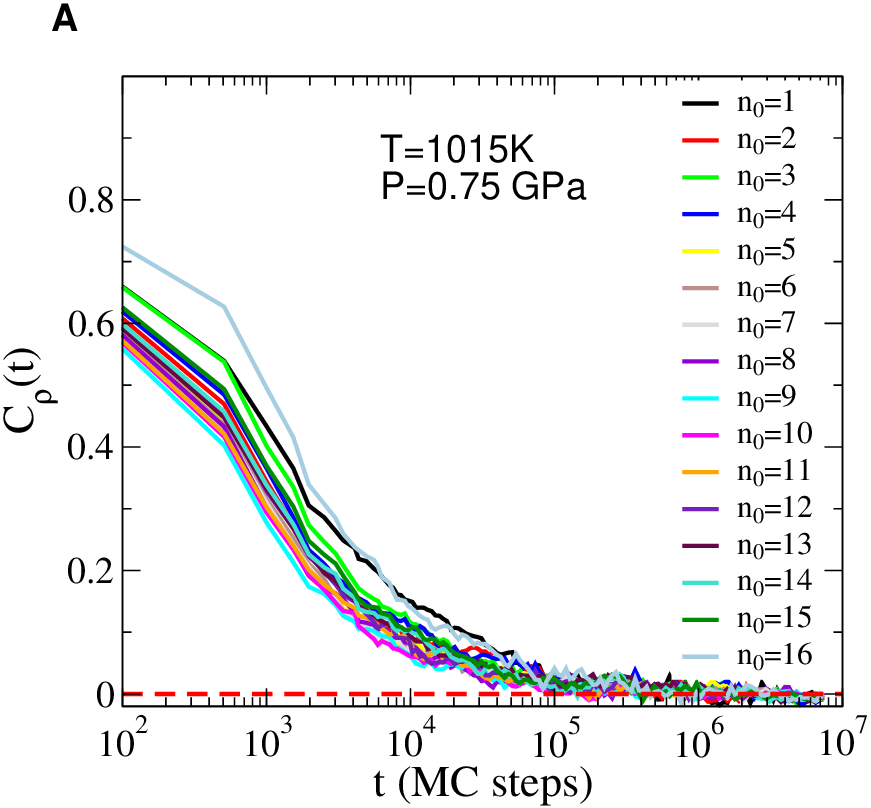}
    \includegraphics[scale=0.25]{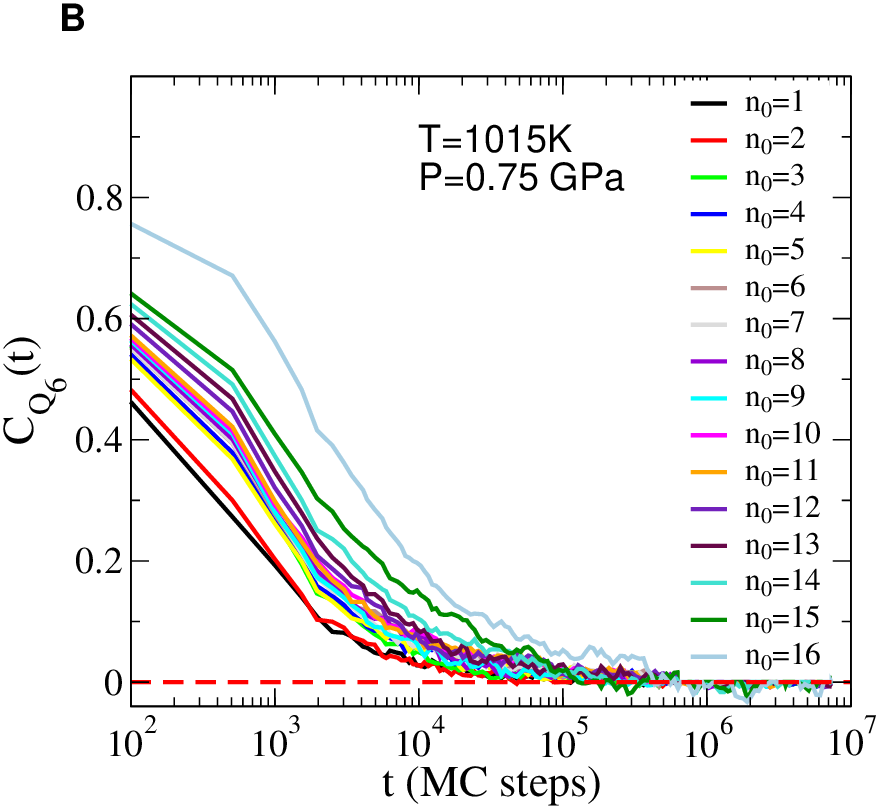}
     \includegraphics[scale=0.25]{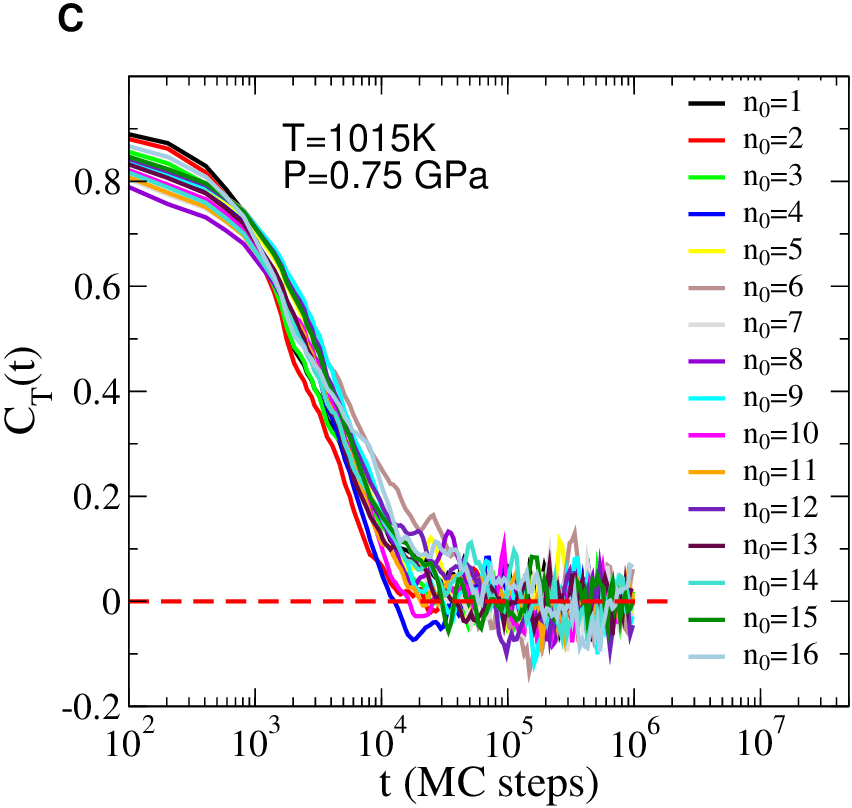}
    \caption{Autocorrelation of density, $C_{\rho}(t)$ (Panel A), $Q_6$ , $C_{Q_6}(t)$ (Panel B) and simulation temperature $C_{T}(t)$ (Panel C) at $T=1005K$, $P=0.75~GPa$ with $N=512$. The initial temperature for each simulation is as shown in the panel, the different curves are labelled according to the midpoint of the $n_{max}$ bias window, $n_0$.}
    \label{fig:AC_TPT_1015}
\end{figure}

\begin{figure}[htp!]
    \centering
    \includegraphics[scale=0.5]{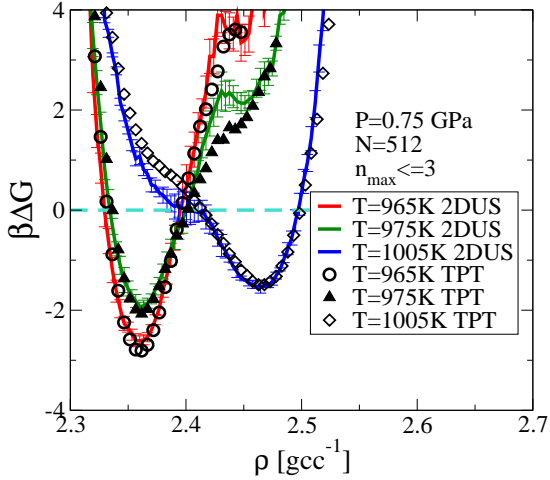}
    \caption{Comparison of free energy along density, $\rho$, with two order parameter umbrella sampling (solid symbols) and one order parameter umbrella sampling along $n_{max}$ and enhanced sampling of density with parallel tempering across temperatures (hollow black symbols). For the two order parameter US simulations, sampling of different densities is enhanced by including a bias potential along $\rho$ for $6$ different $\rho_0$ values. A comparison is made for the density distribution subject to the constraint on $n_{max}$ to test for consistency across methods. Free energy is obtained for the unbiased distributions for simulations of $N=512$ atoms. Convergence is tested by monitoring the distribution of residence times resulting from parallel tempering and from monitoring the decay of the auto-correlation functions for $Q_6$ and $\rho$.}
    \label{fig:SI_method_far_from_coex}
\end{figure}

The decay of the auto-correlation functions for $\rho$, $Q_6$ from the time series of configurations simulated at a given temperature are shown in Fig.~\ref{fig:AC_TPT_965}, Fig.~\ref{fig:AC_TPT_975}, Fig.~\ref{fig:AC_TPT_1005} and Fig.~\ref{fig:AC_TPT_1015}, panels A and B. The temperatures chosen are those outside the co-existence region of LDL and HDL.
We also construct a time series of trajectories initialised at a given temperature, $T$, where each trajectory is subject to different temperatures over time as parallel tempering swaps are performed. This time series is then used to construct the autocorrelation function, $C_T(t)$, shown in panel C of Fig.~\ref{fig:AC_TPT_965}, Fig.~\ref{fig:AC_TPT_975}, Fig.~\ref{fig:AC_TPT_1005} and Fig.~\ref{fig:AC_TPT_1015}.
\FloatBarrier
\FloatBarrier
Results from full two-order parameter umbrella sampling, constraining $n_{max}$ and $\rho$, are compared with results from simulations where only $n_{max}$ is constrained. In the latter case, parallel tempering across temperatures enhances sampling of density. This procedure does not work close to LDL-HDL co-existence conditions since swaps across the density range occur infrequently, affecting estimates of barrier height and basin depth. Away from co-existence conditions, the two methods give the same results, as shown in Fig.~\ref{fig:SI_method_far_from_coex}.

\section{Free energy at $P=0 GPa$ and $P=1.5 GPa$}
\begin{figure}[h!]
    \centering
    \subfloat{\includegraphics[height=4cm,width=4cm]{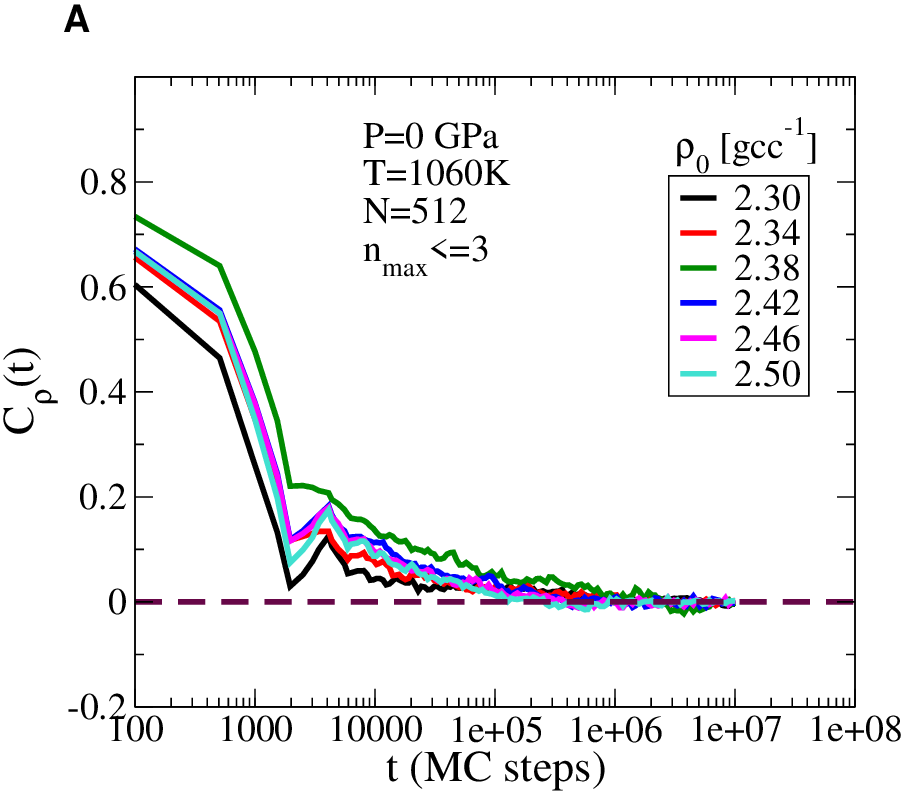}}
    \subfloat{\includegraphics[height=4cm,width=4cm]{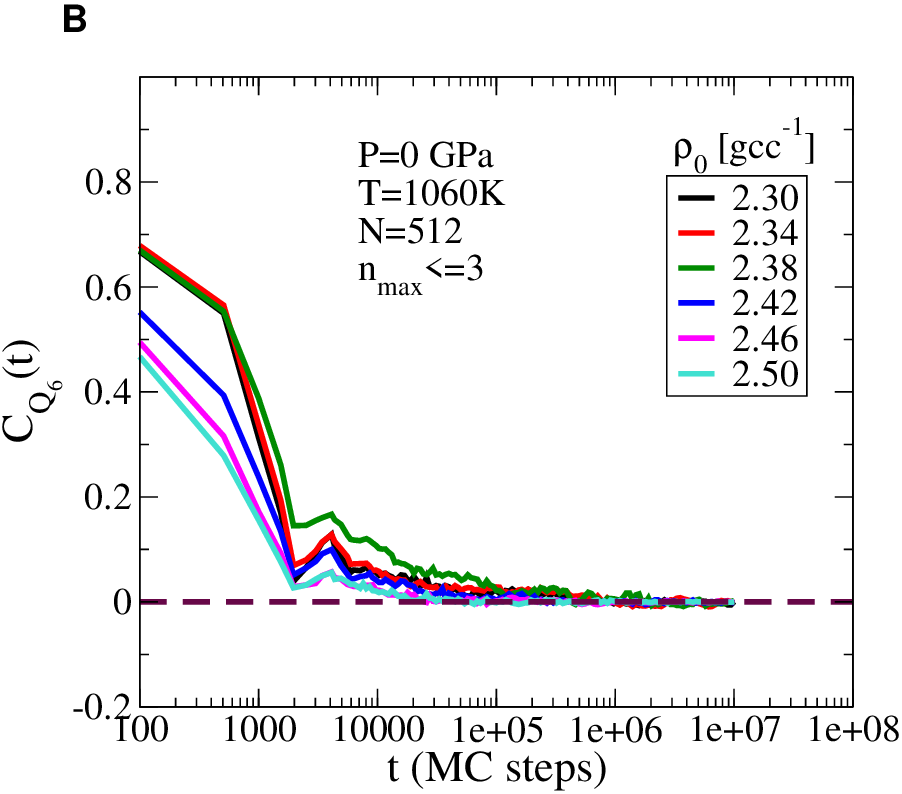}}
    
    \subfloat{\includegraphics[height=4cm,width=4cm]{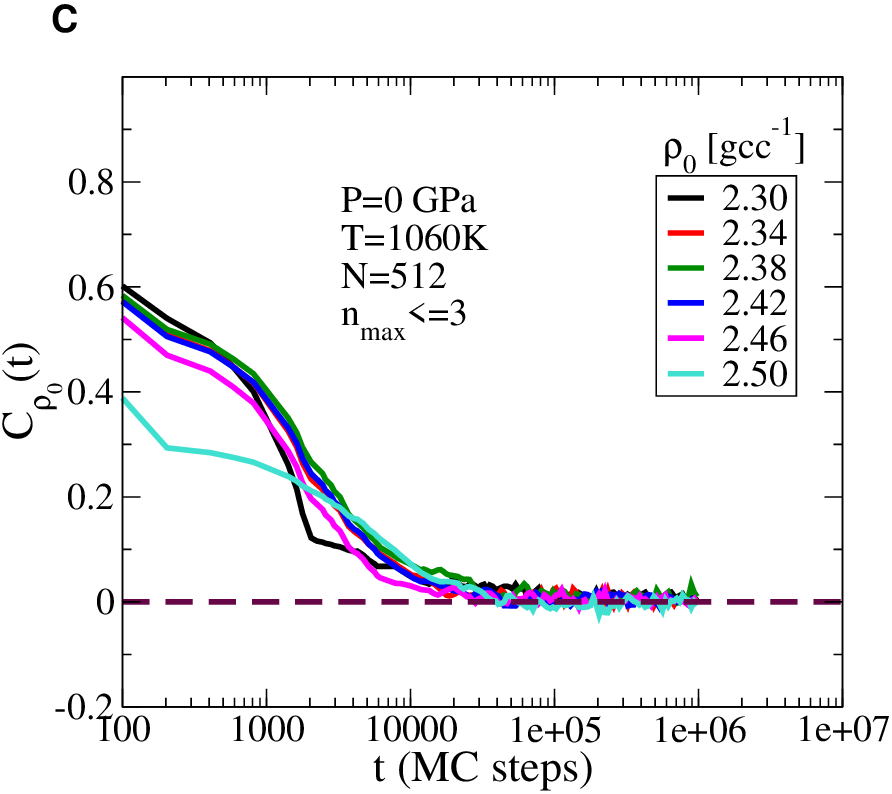}}
    \subfloat{\includegraphics[height=4cm,width=4cm]{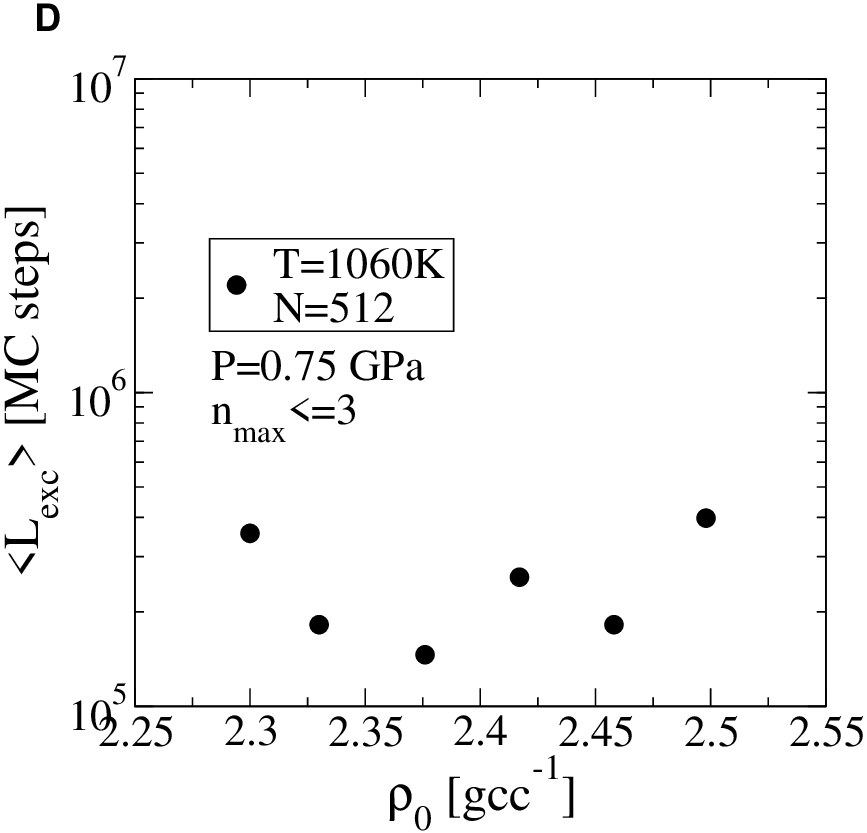}}
    \caption{Decay of time auto-correlation function for density (Panel A), $Q_6$ (Panel B), and for density window index (Panel C) for each of the density bias windows, $\rho_0$, subject to the constraint of $n_{max}~\leq 3$ at $T=1060K$, $P=0~GPa$. Panel D shows the mean excursion length or return time as a function of $\rho_0$ subject to the constraint on $n_{max}$.  
    In each case, the different curves are labelled according to the initial reference density, $\rho_0$, for the given independent simulation.
    }
    \label{fig:2DUS_stats_P0}
\end{figure}

\begin{figure}[h!]
    \centering
    \subfloat{\includegraphics[height=4cm,width=4cm]{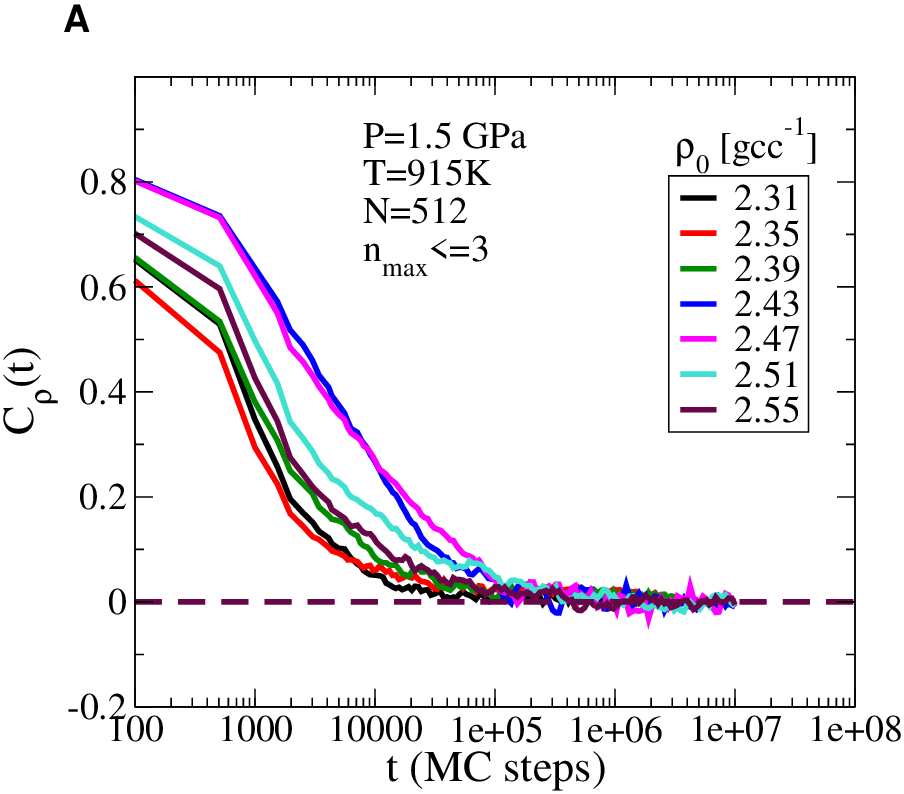}}
    \subfloat{\includegraphics[height=4cm,width=4cm]{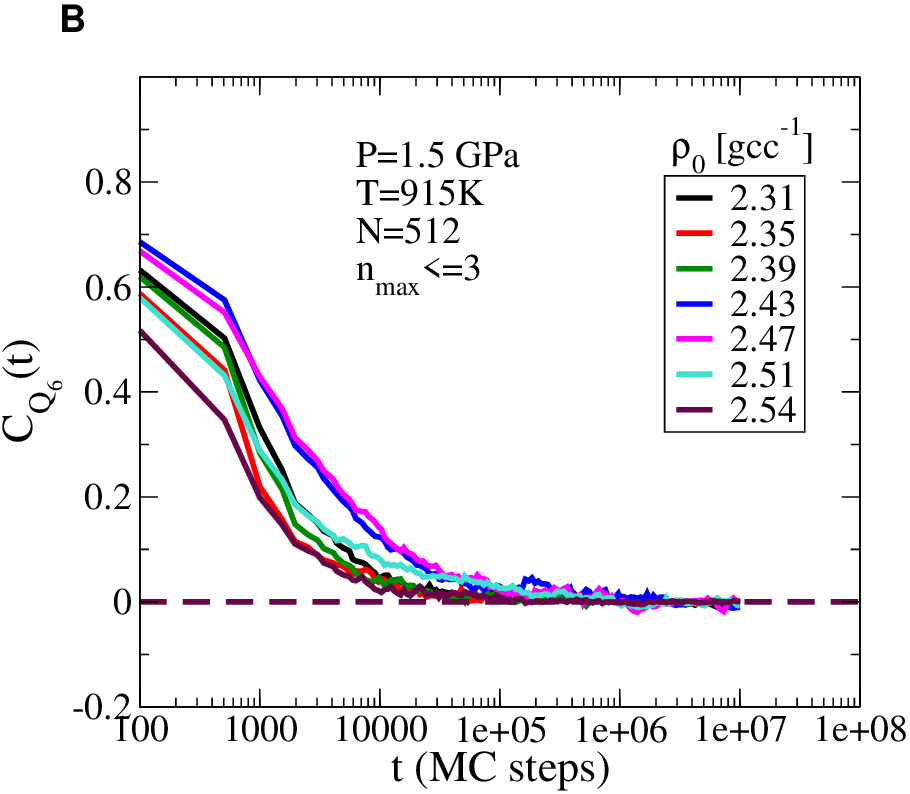}}
    
    \subfloat{\includegraphics[height=4cm,width=4cm]{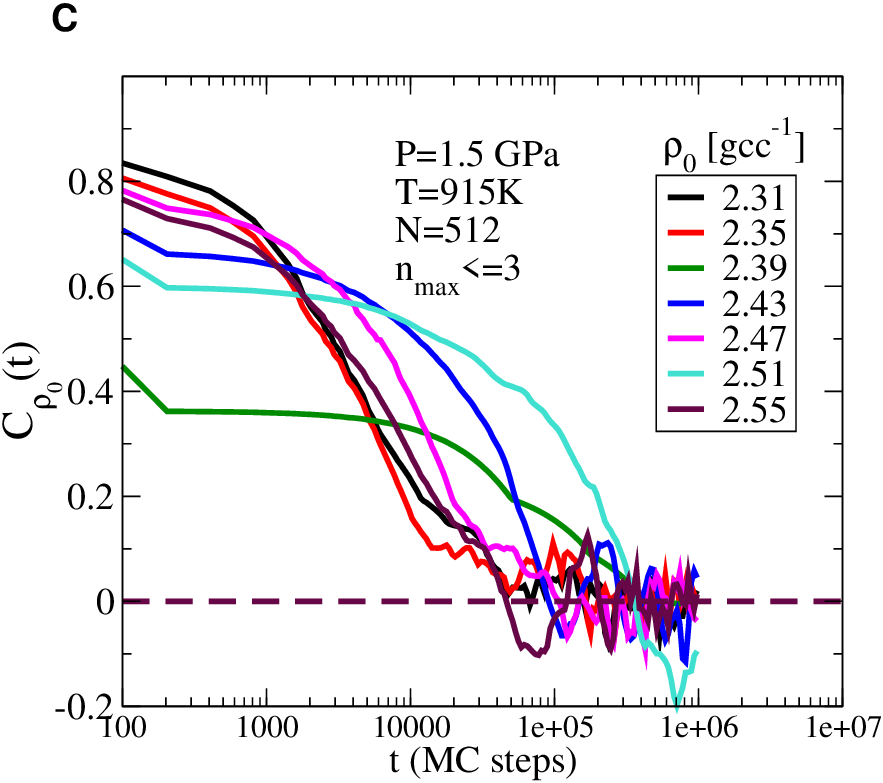}}
    \subfloat{\includegraphics[height=4cm,width=4cm]{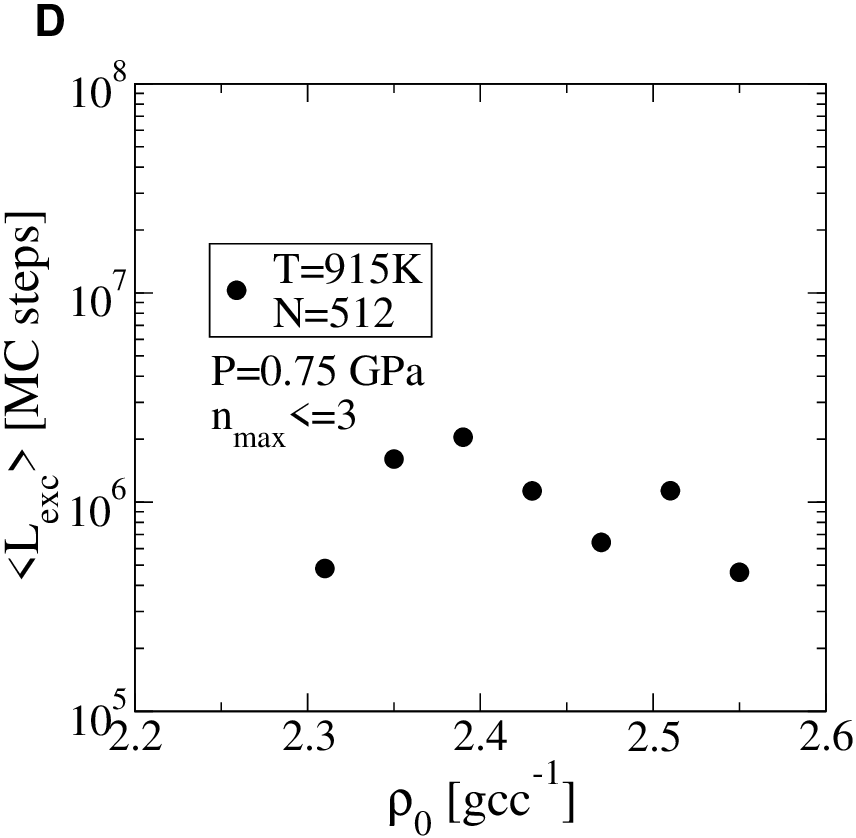}}
    \caption{Decay of time auto-correlation function for density (Panel A), $Q_6$ (Panel B), and for density window index (Panel C) for each of the density bias windows, $\rho_0$, subject to the constraint of $n_{max}~\leq 3$ at $T=915K$, $P=1.5~GPa$. Panel D shows the mean excursion length or return time as a function of $\rho_0$ subject to the constraint on $n_{max}$.  In each case, the different curves are labelled according to the initial reference density, $\rho_0$, for the given independent simulation.
    }
    \label{fig:2DUS_stats_P04}
\end{figure}

We perform a similar investigation along two other isobars in the sub-critical regime, finding that the liquid-liquid phase transition occurs at the expected state points in each case. Along one super-critical isobar, we find no evidence of a discontinuous change in the nature of the liquid. This is expected since the two liquids are expected to be indistinguishable in the supercritical regime.
Fig.~\ref{fig:2DUS_stats_P0} and Fig.~\ref{fig:2DUS_stats_P04} show the autocorrelation functions and mean parallel tempering excursion lengths at state points close to co-existence for $P=0~GPa$ and $P=1.5~GPa$ respectively. 

\begin{figure}[h]
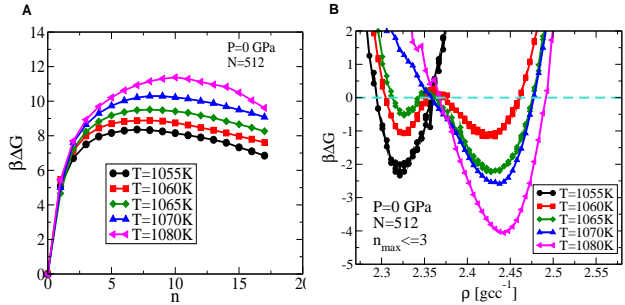

    \centering
    \includegraphics[height=4cm,width=4cm]{1D_barrier_P0.eps}
    \includegraphics[height=4cm,width=4cm]{P0_bDG_rho.eps}
    \caption{Free energy barrier to crystallisation along the $P=0~GPa$ isobar (Panel A) and free energy as a function of density along the $P=0~GPa$ isobar (Panel B). The free energy along density is obtained from the unweighted density distributions measured subject to constraint on $n_{max}$.}
    \label{fig:FE_P0}
\end{figure}

\begin{figure}[h!]
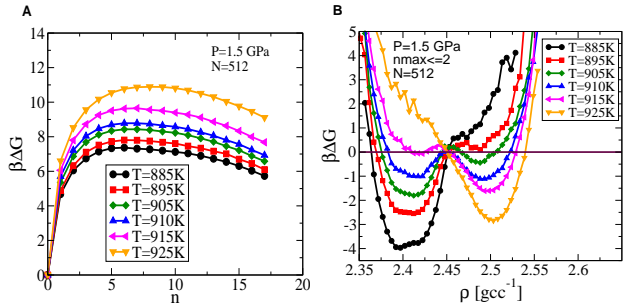

    \centering
    \includegraphics[height=4cm,width=4cm]{1D_barrier_P04.eps}
    \includegraphics[height=4cm,width=4cm]{P04_bDG_rho.eps}
    \caption{Free energy barrier to crystallisation along the $P=1.5~GPa$ isobar (Panel A) and free energy as a function of density along the $P=1.5~GPa$ isobar (Panel B). The free energy along density is obtained from the unweighted density distributions measured subject to constraint on $n_{max}$.}
    \label{fig:FE_P04}
\end{figure}

\FloatBarrier
Co-existence conditions are also identified along other isobars in Fig.~\ref{fig:FE_P0} and Fig.~\ref{fig:FE_P04}. $\beta \Delta G(\rho)$ is  shown, subject to constraint on $n_{max}$, demonstrating the shift in the typical density of the liquid from high to low temperature and the region where the distributions are bi-modal. Results are shown along the $P=0~GPa$ isobar in Fig.~\ref{fig:FE_P0} and along the $P=1.5~GPa$ isobar in Fig.~\ref{fig:FE_P04}.

\section{Free energy reconstructions at larger system sizes }
\begin{figure}[htpb!]
    \centering
    \subfloat{\includegraphics[height=4cm,width=4cm]{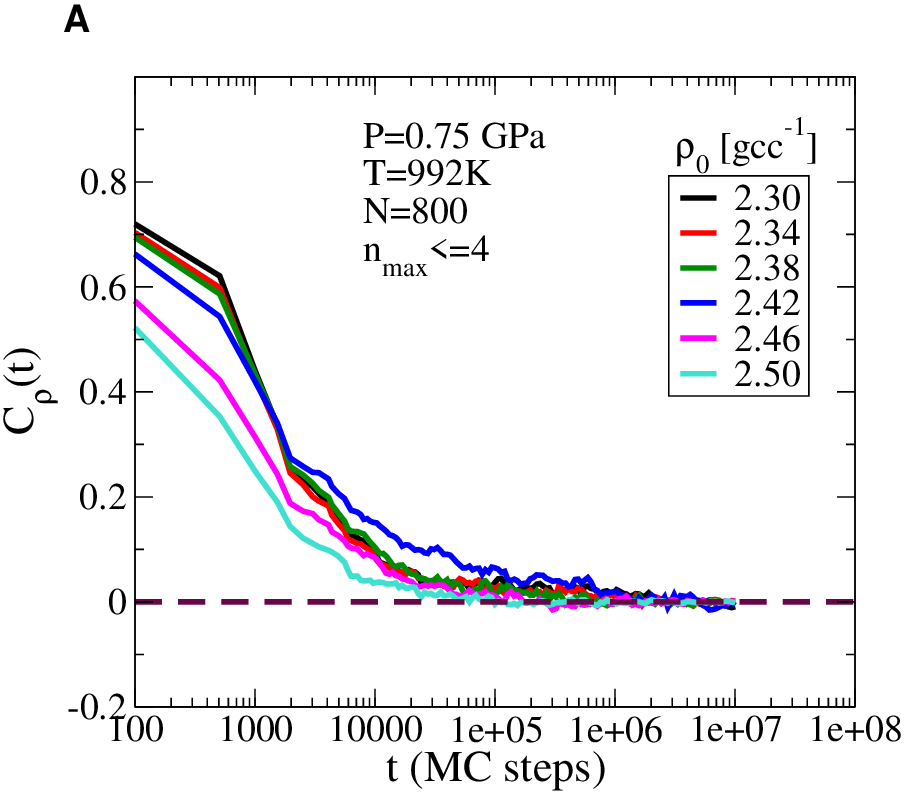}}
    \subfloat{\includegraphics[height=4cm,width=4cm]{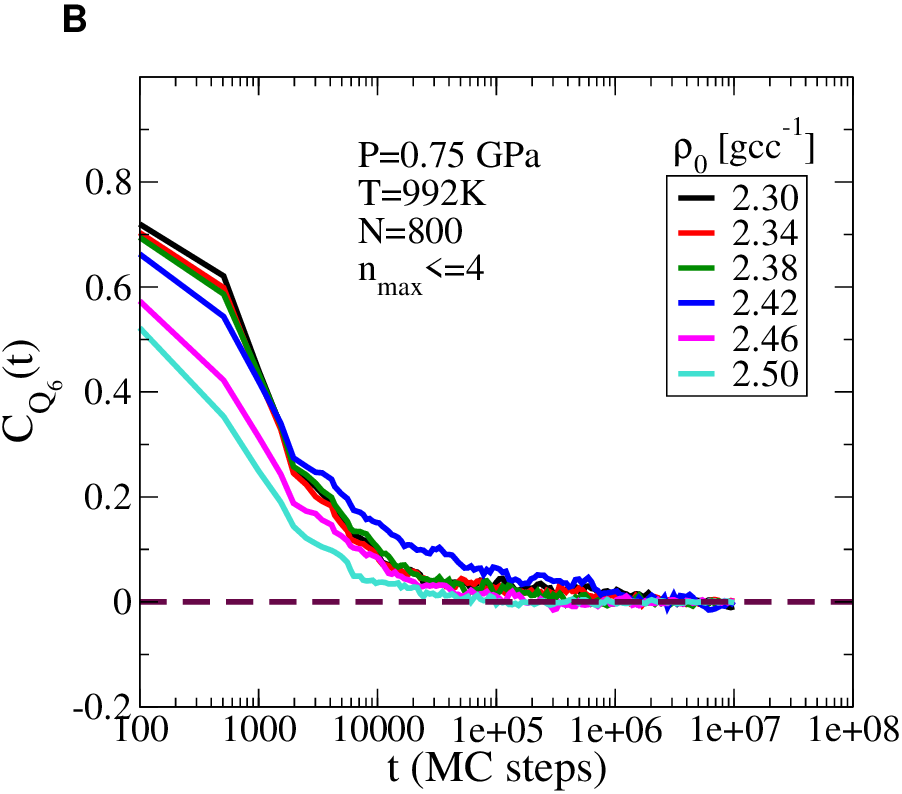}}
    
    \subfloat{\includegraphics[height=4cm,width=4cm]{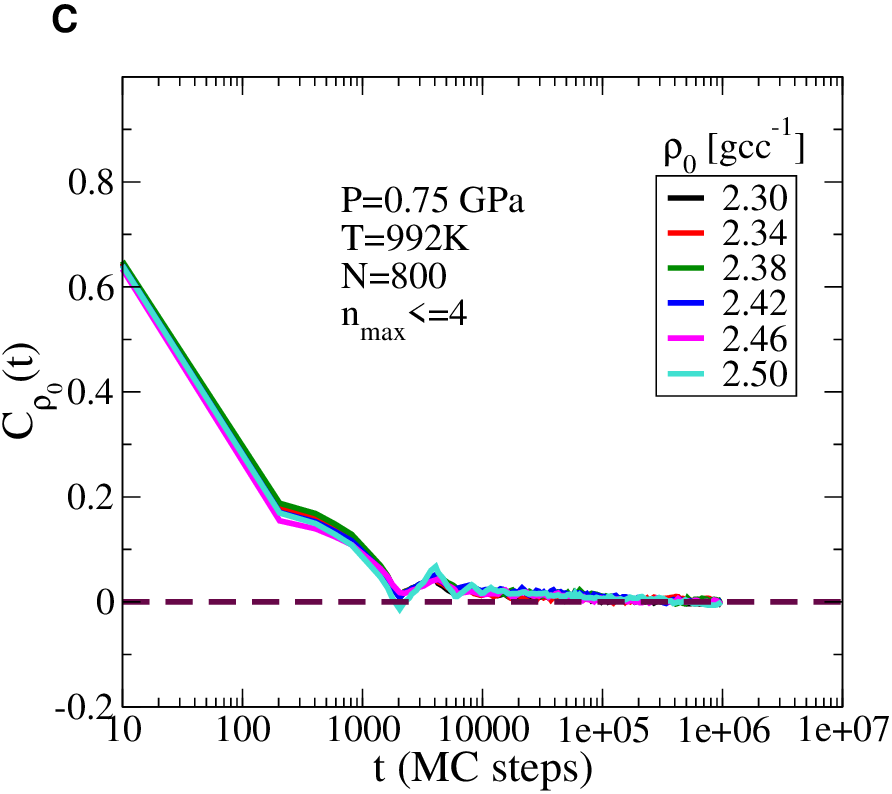}}
    \subfloat{\includegraphics[height=4cm,width=4cm]{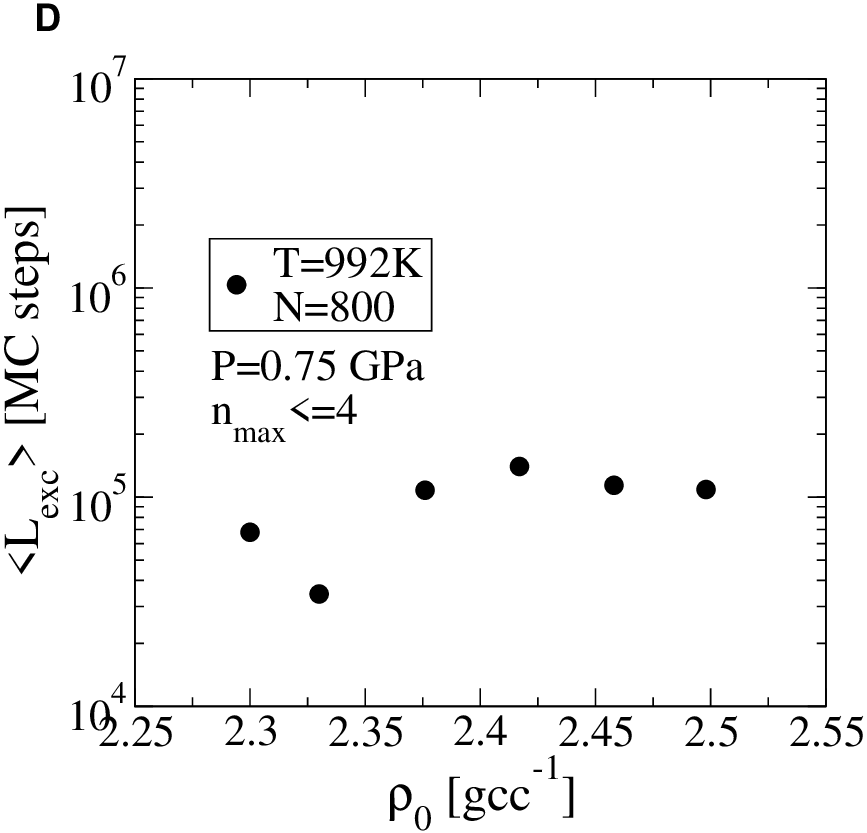}}
    \caption{Decay of time auto-correlation function for density (Panel A), $Q_6$ (Panel B), and for density window index (Panel C) for each of the density bias windows, $\rho_0$, subject to the constraint of $n_{max}~\leq 3$ at $T=992K$, $P=0.75~GPa$. Panel D shows the mean excursion length or return time as a function of $\rho_0$ subject to the constraint on $n_{max}$.  In each case, the different curves are labelled according to the initial reference density, $\rho_0$, for the given independent simulation.
    }
    \label{fig:800_2DUS_stats_992}
\end{figure}

\begin{figure}[htpb!]
    \centering
    \subfloat{\includegraphics[height=4cm,width=4cm]{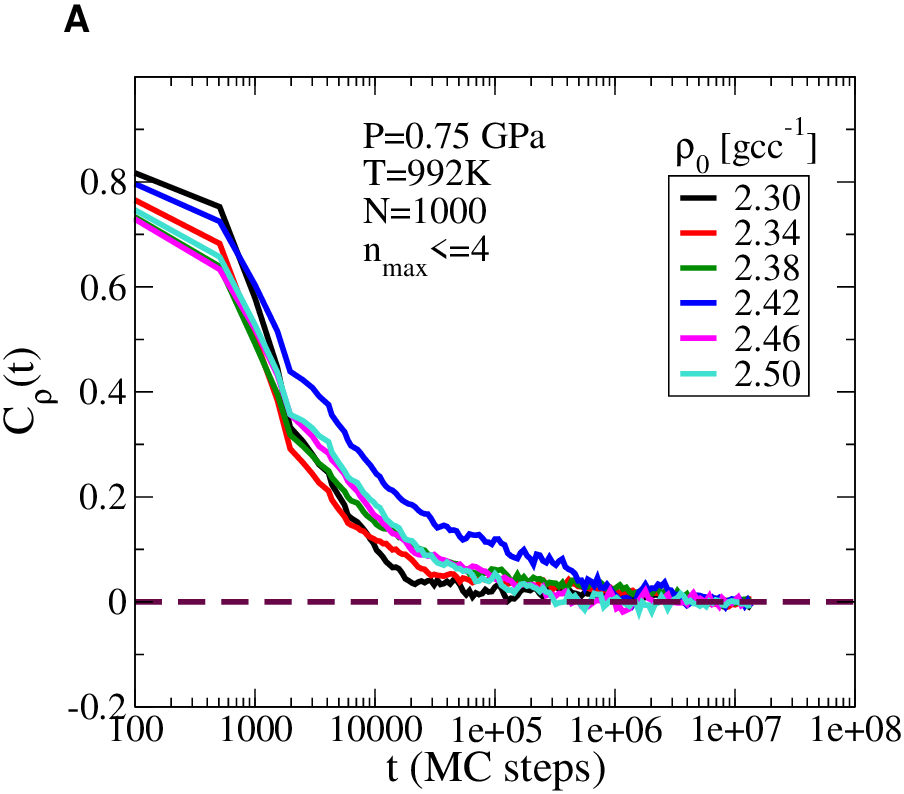}}
    \subfloat{\includegraphics[height=4cm,width=4cm]{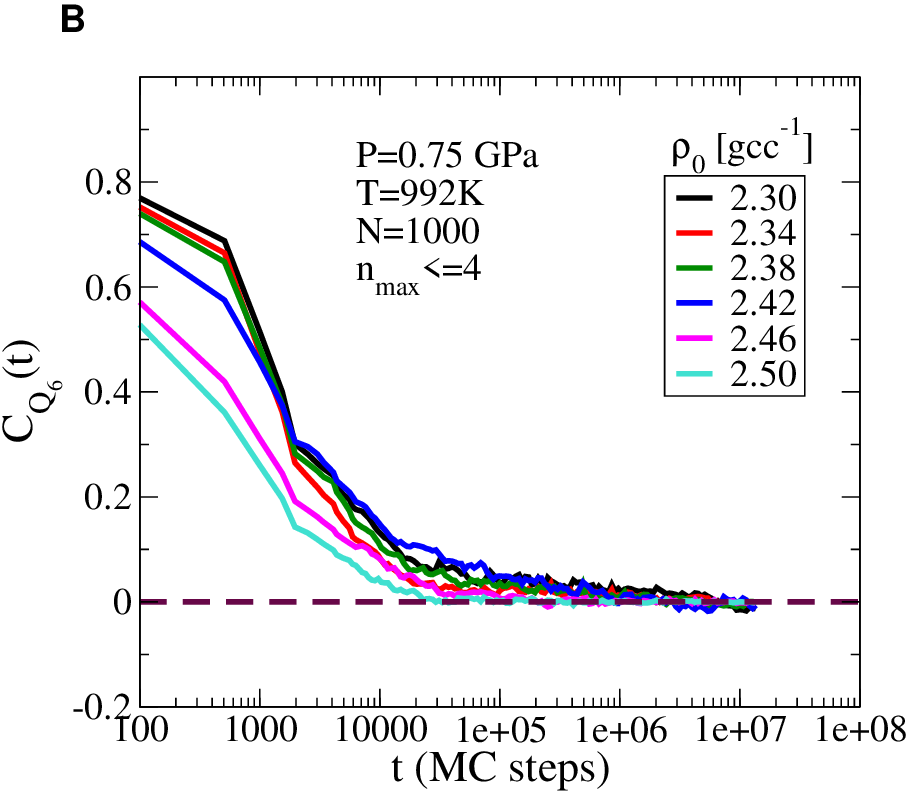}}
    
    \subfloat{\includegraphics[height=4cm,width=4cm]{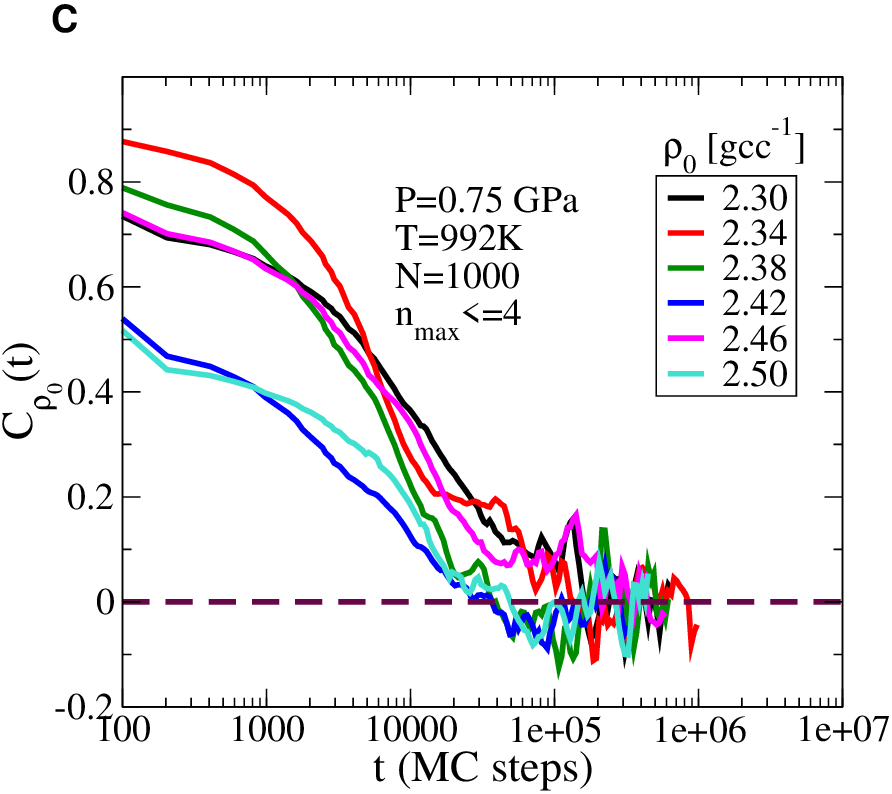}}
    \subfloat{\includegraphics[height=4cm,width=4cm]{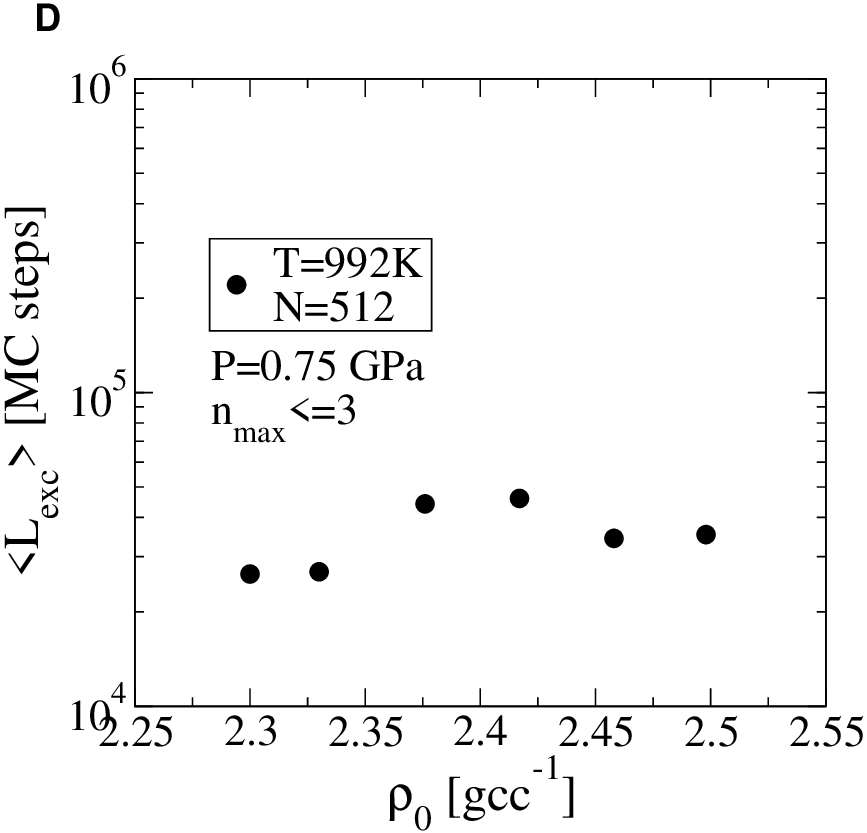}}
    \caption{Decay of time auto-correlation function for density (Panel A), $Q_6$ (Panel B), and for density window index (Panel C) for each of the density bias windows, $\rho_0$, subject to the constraint of $n_{max}~\leq 4$ at $T=992K$, $P=0.75~GPa$. Panel D shows the mean excursion length or return time as a function of $\rho_0$ subject to the constraint on $n_{max}$. In each case, the different curves are labelled according to the initial reference density, $\rho_0$, for the given independent simulation.
    }
    \label{fig:1k_2DUS_stats_992}
\end{figure}

\begin{figure}[htpb!]
    \centering
    \subfloat{\includegraphics[height=4cm,width=4cm]{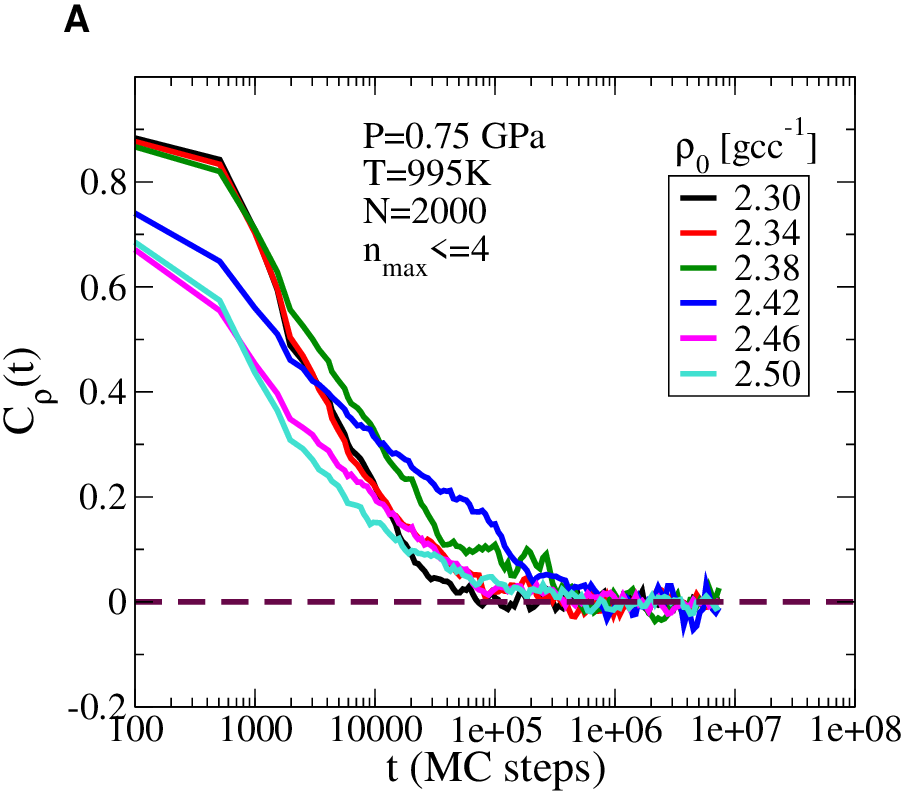}}
    \subfloat{\includegraphics[height=4cm,width=4cm]{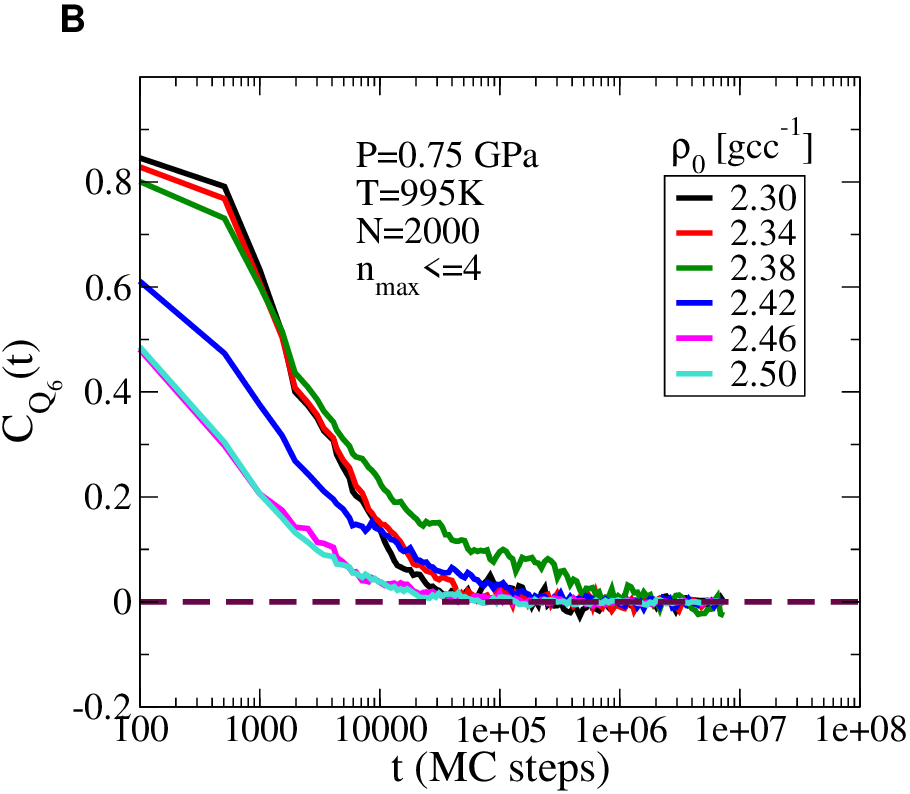}}
    
    \subfloat{\includegraphics[height=4cm,width=4cm]{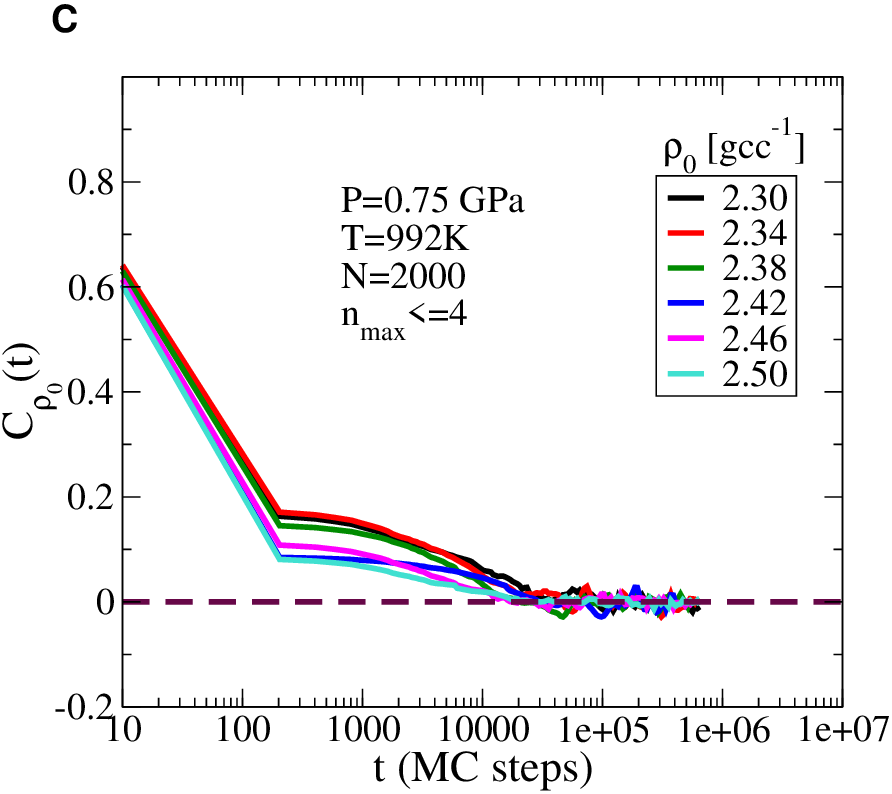}}
    \subfloat{\includegraphics[height=4cm,width=4cm]{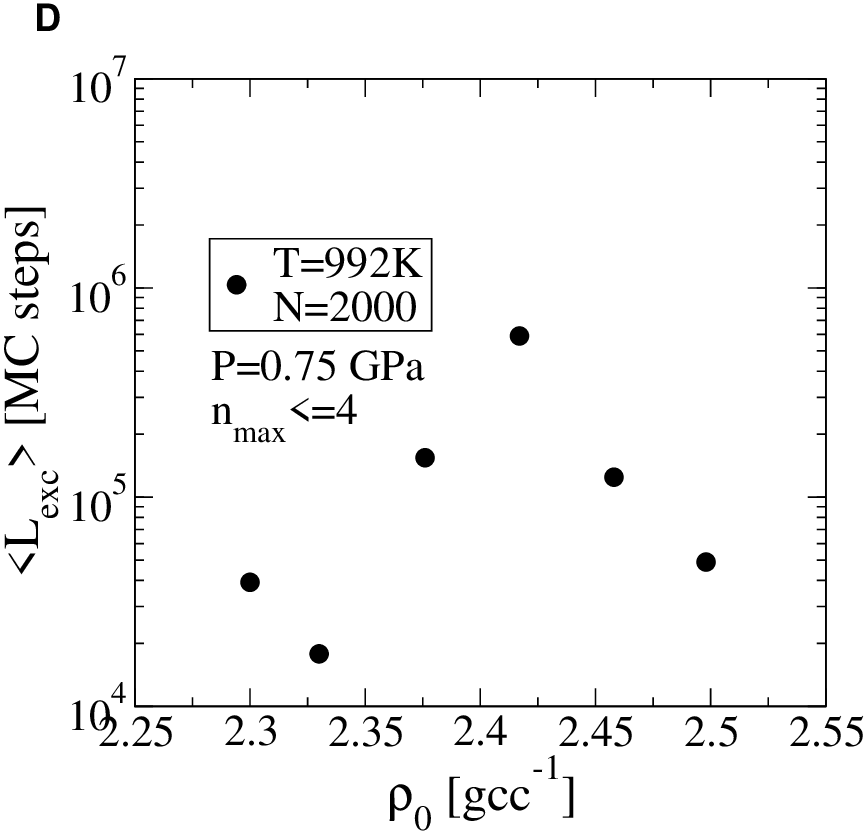}}
    \caption{Decay of time auto-correlation function for density (Panel A), $Q_6$ (Panel B), and for density window index (Panel C) for each of the density bias windows, $\rho_0$, subject to the constraint of $n_{max}~\leq 3$ at $T=992K$, $P=0.75~GPa$. Panel D shows the mean excursion length or return time as a function of $\rho_0$ subject to the constraint on $n_{max}$.  In each case, the different curves are labelled according to the initial reference density, $\rho_0$, for the given independent simulation.
    }
    \label{fig:2k_2DUS_stats_992}
\end{figure}

In Fig.~2 main manuscript, we show the free energy profile as a function of density along the $P=0.75~GPa$ isobar at four system sizes, $N=512$, $N=800$, $N=1000$ and $N=2000$. For the case where the two liquids can have a stable interface between them, the barrier height between the two liquids is expected to scale as $N^{2/3}$, which is shown in Fig.~2 in the main text. The low density phase is a disordered phase, as shown by checking the scaling of $Q_6$ in the low density basin with $N$, which goes as $N^{-1/2}$ as shown in Fig.~\ref{fig:barr_Q6_scale}. This is indicative of a disordered phase, whereas ordered phases would have constant $Q_6$ for all $N$. The sections that follow show the autocorrelation functions and parallel tempering statistics for the corresponding sets of simulations at larger system sizes.
Convergence of simulations for $N=800$ near coexistence conditions are shown in Fig.~\ref{fig:800_2DUS_stats_992}, for $N=1000$ in Fig.~\ref{fig:1k_2DUS_stats_992}, and for $N=2000$ in Fig.~\ref{fig:2k_2DUS_stats_992}.

\clearpage
\subsubsection*{Scaling of LDL basin depth with system size}
\begin{figure}[htp!]
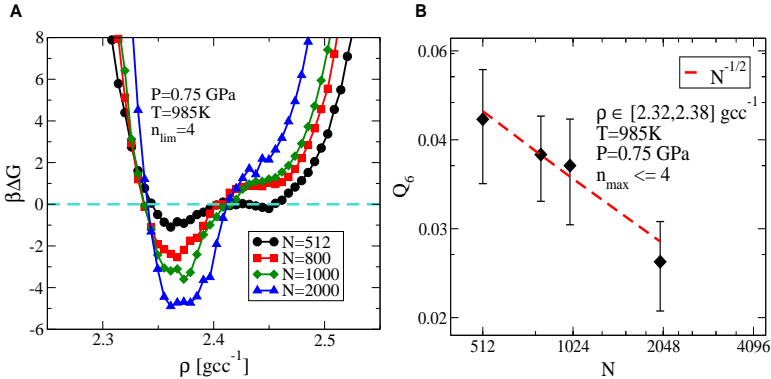

\centering
\vspace{10mm}
\includegraphics[height=5cm,width=5cm]{bDG_rho_N_compare_T985.eps}
\includegraphics[height=5cm,width=5cm]{Q6_scaling.eps}
\caption{$\beta\Delta G(\rho)$ at $P=0.75$ GPa and $T=985K$ from simulations and reweighting at $4$ system sizes, $N=512,~800,~1000,~2000$ (panel A). These conditions correspond to a single stable LDL phase. The scaling of the $Q_6$, measured subject to constraint on $\rho$, is shown in panel B.}
\label{fig:barr_Q6_scale}
\end{figure}

Fig.~\ref{fig:barr_Q6_scale} shows the decrease in basin depth as system size is increased, at $P=0.75~GPa,~T=985K$. The average $Q_6$ subject to constraint on $\rho$ and $n_{max}$ is shown to scale as $N^{-1/2}$ in the main manuscript Fig. 2.
\section{Fit to the Ising Universality class}
\begin{figure}[htpb!]
\centering
\includegraphics[trim=60 0 60 0,clip,scale=1.2]{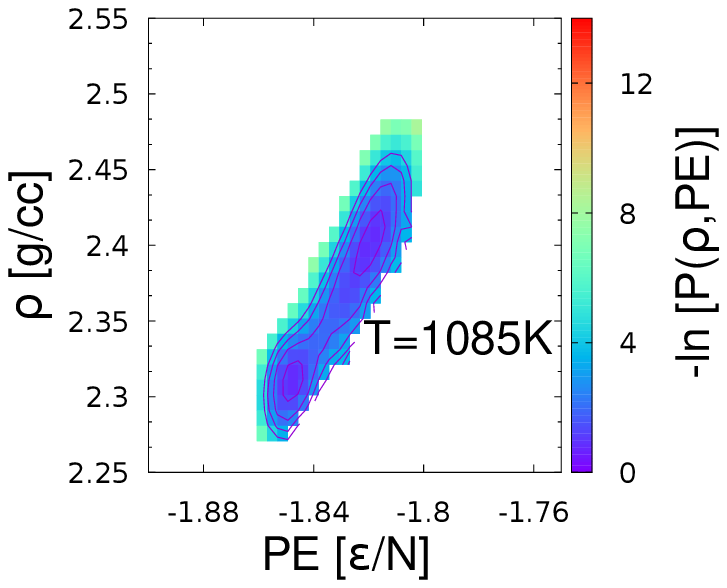}
\caption{Bivariate free energy as a function of $\rho$ and the potential energy, $E$, at the identified $T_c,P_c$ of $T=1085.5K,~P=-0.5~GPa$ showing the weakly double well feature characteristic of critical fluctuations. 
}
\label{fig:reweight_rho_PE}
\end{figure}

The analysis of the critical fluctuations of the order parameter is carried out by comparing the distribution of the order parameter, $M=\rho+sE$ with the reference distribution:
\begin{equation}
P_{ising}(M) \propto exp \Biggl \{ - \left (\frac{M^2}{M_0^2} -1 \right )^2\left ( a\frac{M^2}{M_0^2} +c \right )^2    \Biggr \}
\label{eq:ising_ref_si}
\end{equation}
For $M_0=1.1341665$, $a=0.158$, $c=0.776$, one obtains a universal distribution of unit variance. By identifying the parameter for the LLPT, $r=\rho + sE$, and shifting and scaling it as $M = \frac{r - \langle r \rangle}{\sigma_r}$, one obtains a distribution of unit variance, which is thus system-size independent.
We compare the difference between the distributions, $P(M)$, and $P_{ising}(M)$, and identify the set of $T_c,P_c,s$ that minimise the difference between them. The minimisation is performed using an implementation of the Nelder-Mead optimisation scheme available with the SciPy optimization library\cite{gao2012implementing}.
The critical point can be estimated using this optimisation procedure as the  $T_c,P_c$ for which the distribution best matches the reference distribution. In Fig.~\ref{fig:reweight_rho_PE}, we show the bivariate free energy as a function of $\rho$ and the potential energy, $E$, at the identified critical point conditions. 
Histogram data is reweighted from the $P=0~GPa$ isobar at different temperatures to produce estimates in the vicinity of the critical point.


\begin{thebibliography}{10}
\expandafter\ifx\csname url\endcsname\relax
  \def\url#1{\burl{#1}}\fi
\expandafter\ifx\csname urlprefix\endcsname\relax\def\urlprefix{URL }\fi
\providecommand{\bibinfo}[2]{#2}
\providecommand{\eprint}[2][]{\url{#2}}
\providecommand{\doi}[1]{\url{https://doi.org/#1}}
\bibcommenthead

\bibitem{StanleyPolymorphism}
\bibinfo{author}{Stanley, H.~E.}
\newblock \emph{\bibinfo{title}{Liquid Polymorphism}} Vol.
  \bibinfo{volume}{152} (\bibinfo{publisher}{Wiley Online Library},
  \bibinfo{year}{2013}).

\bibitem{vasisht2013liquid}
\bibinfo{author}{Vasisht, V.~V.} \& \bibinfo{author}{Sastry, S.}
\newblock \bibinfo{title}{Liquid--liquid phase transition in supercooled
  silicon}.
\newblock \emph{\bibinfo{journal}{Liquid Polymorphism}}
  \textbf{\bibinfo{volume}{152}}, \bibinfo{pages}{463--517}
  (\bibinfo{year}{2013}) .

\bibitem{sastry2003liquid}
\bibinfo{author}{Sastry, S.} \& \bibinfo{author}{Angell, C.~A.}
\newblock \bibinfo{title}{Liquid--liquid phase transition in supercooled
  silicon}.
\newblock \emph{\bibinfo{journal}{Nature materials}}
  \textbf{\bibinfo{volume}{2}}~(11), \bibinfo{pages}{739}
  (\bibinfo{year}{2003}) .

\bibitem{vasisht2011liquid}
\bibinfo{author}{Vasisht, V.~V.}, \bibinfo{author}{Saw, S.} \&
  \bibinfo{author}{Sastry, S.}
\newblock \bibinfo{title}{Liquid--liquid critical point in supercooled
  silicon}.
\newblock \emph{\bibinfo{journal}{Nature Physics}}
  \textbf{\bibinfo{volume}{7}}~(7), \bibinfo{pages}{549} (\bibinfo{year}{2011})
  .

\bibitem{beye2010liquid}
\bibinfo{author}{Beye, M.}, \bibinfo{author}{Sorgenfrei, F.},
  \bibinfo{author}{Schlotter, W.~F.}, \bibinfo{author}{Wurth, W.} \&
  \bibinfo{author}{F{\"o}hlisch, A.}
\newblock \bibinfo{title}{The liquid-liquid phase transition in silicon
  revealed by snapshots of valence electrons}.
\newblock \emph{\bibinfo{journal}{Proceedings of the National Academy of
  Sciences}} \textbf{\bibinfo{volume}{107}}~(39), \bibinfo{pages}{16772--16776}
  (\bibinfo{year}{2010}) .

\bibitem{ganesh2009liquid}
\bibinfo{author}{Ganesh, P.} \& \bibinfo{author}{Widom, M.}
\newblock \bibinfo{title}{Liquid-liquid transition in supercooled silicon
  determined by first-principles simulation}.
\newblock \emph{\bibinfo{journal}{Physical review letters}}
  \textbf{\bibinfo{volume}{102}}~(7), \bibinfo{pages}{075701}
  (\bibinfo{year}{2009}) .

\bibitem{speedyangell}
\bibinfo{author}{Speedy, R.~J.} \& \bibinfo{author}{Angell, C.~A.}
\newblock \bibinfo{title}{Isothermal compressibility of supercooled water and
  evidence for a thermodynamic singularity at $-45{}^\circ$c}.
\newblock \emph{\bibinfo{journal}{The Journal of Chemical Physics}}
  \textbf{\bibinfo{volume}{65}}~(3), \bibinfo{pages}{851--858}
  (\bibinfo{year}{1976}).
\newblock \urlprefix\url{https://doi.org/10.1063/1.433153}.
\newblock \doi{10.1063/1.433153},
  \bibinfo{eprint}{{\href{https://arxiv.org/abs/https://doi.org/10.1063/1.433153}{{https://doi.org/10.1063/1.433153}}}}
  .

\bibitem{speedy1982stability}
\bibinfo{author}{Speedy, R.~J.}
\newblock \bibinfo{title}{Stability-limit conjecture. an interpretation of the
  properties of water}.
\newblock \emph{\bibinfo{journal}{The Journal of Physical Chemistry}}
  \textbf{\bibinfo{volume}{86}}~(6), \bibinfo{pages}{982--991}
  (\bibinfo{year}{1982}) .

\bibitem{poole1992phase}
\bibinfo{author}{Poole, P.~H.}, \bibinfo{author}{Sciortino, F.},
  \bibinfo{author}{Essmann, U.} \& \bibinfo{author}{Stanley, H.~E.}
\newblock \bibinfo{title}{Phase behaviour of metastable water}.
\newblock \emph{\bibinfo{journal}{Nature}}
  \textbf{\bibinfo{volume}{360}}~(6402), \bibinfo{pages}{324--328}
  (\bibinfo{year}{1992}) .

\bibitem{sastry1996singularity}
\bibinfo{author}{Sastry, S.}, \bibinfo{author}{Debenedetti, P.~G.},
  \bibinfo{author}{Sciortino, F.} \& \bibinfo{author}{Stanley, H.~E.}
\newblock \bibinfo{title}{Singularity-free interpretation of the thermodynamics
  of supercooled water}.
\newblock \emph{\bibinfo{journal}{Physical Review E}}
  \textbf{\bibinfo{volume}{53}}~(6), \bibinfo{pages}{6144}
  (\bibinfo{year}{1996}) .

\bibitem{tanaka2020liquid}
\bibinfo{author}{Tanaka, H.}
\newblock \bibinfo{title}{Liquid--liquid transition and polyamorphism}.
\newblock \emph{\bibinfo{journal}{The Journal of Chemical Physics}}
  \textbf{\bibinfo{volume}{153}}~(13), \bibinfo{pages}{130901}
  (\bibinfo{year}{2020}) .

\bibitem{kim2017maxima}
\bibinfo{author}{Kim, K.~H.} \emph{et~al.}
\newblock \bibinfo{title}{Maxima in the thermodynamic response and correlation
  functions of deeply supercooled water}.
\newblock \emph{\bibinfo{journal}{Science}}
  \textbf{\bibinfo{volume}{358}}~(6370), \bibinfo{pages}{1589--1593}
  (\bibinfo{year}{2017}) .

\bibitem{kim2020experimental}
\bibinfo{author}{Kim, K.~H.} \emph{et~al.}
\newblock \bibinfo{title}{Experimental observation of the liquid-liquid
  transition in bulk supercooled water under pressure}.
\newblock \emph{\bibinfo{journal}{Science}}
  \textbf{\bibinfo{volume}{370}}~(6519), \bibinfo{pages}{978--982}
  (\bibinfo{year}{2020}) .

\bibitem{nilsson2022origin}
\bibinfo{author}{Nilsson, A.}
\newblock \bibinfo{title}{Origin of the anomalous properties in supercooled
  water based on experimental probing inside “no-man's land”}.
\newblock \emph{\bibinfo{journal}{Journal of Non-Crystalline Solids: X}}
  \bibinfo{pages}{100095} (\bibinfo{year}{2022}) .

\bibitem{kim2005situ}
\bibinfo{author}{Kim, T.} \emph{et~al.}
\newblock \bibinfo{title}{In situ high-energy x-ray diffraction study of the
  local structure of supercooled liquid si}.
\newblock \emph{\bibinfo{journal}{Physical review letters}}
  \textbf{\bibinfo{volume}{95}}~(8), \bibinfo{pages}{085501}
  (\bibinfo{year}{2005}) .

\bibitem{limmer2011putative}
\bibinfo{author}{Limmer, D.~T.} \& \bibinfo{author}{Chandler, D.}
\newblock \bibinfo{title}{The putative liquid-liquid transition is a
  liquid-solid transition in atomistic models of water}.
\newblock \emph{\bibinfo{journal}{The Journal of chemical physics}}
  \textbf{\bibinfo{volume}{135}}~(13), \bibinfo{pages}{134503}
  (\bibinfo{year}{2011}) .

\bibitem{limmer2013putative}
\bibinfo{author}{Limmer, D.~T.} \& \bibinfo{author}{Chandler, D.}
\newblock \bibinfo{title}{The putative liquid-liquid transition is a
  liquid-solid transition in atomistic models of water. ii}.
\newblock \emph{\bibinfo{journal}{The Journal of chemical physics}}
  \textbf{\bibinfo{volume}{138}}~(21), \bibinfo{pages}{214504}
  (\bibinfo{year}{2013}) .

\bibitem{stillinger1985computer}
\bibinfo{author}{Stillinger, F.~H.} \& \bibinfo{author}{Weber, T.~A.}
\newblock \bibinfo{title}{Computer simulation of local order in condensed
  phases of silicon}.
\newblock \emph{\bibinfo{journal}{Physical review B}}
  \textbf{\bibinfo{volume}{31}}~(8), \bibinfo{pages}{5262}
  (\bibinfo{year}{1985}) .

\bibitem{palmer2014metastable}
\bibinfo{author}{Palmer, J.~C.} \emph{et~al.}
\newblock \bibinfo{title}{Metastable liquid--liquid transition in a molecular
  model of water}.
\newblock \emph{\bibinfo{journal}{Nature}}
  \textbf{\bibinfo{volume}{510}}~(7505), \bibinfo{pages}{385}
  (\bibinfo{year}{2014}) .

\bibitem{debenedetti2020second}
\bibinfo{author}{Debenedetti, P.~G.}, \bibinfo{author}{Sciortino, F.} \&
  \bibinfo{author}{Zerze, G.~H.}
\newblock \bibinfo{title}{Second critical point in two realistic models of
  water}.
\newblock \emph{\bibinfo{journal}{Science}}
  \textbf{\bibinfo{volume}{369}}~(6501), \bibinfo{pages}{289--292}
  (\bibinfo{year}{2020}) .

\bibitem{chen2017liquid}
\bibinfo{author}{Chen, R.}, \bibinfo{author}{Lascaris, E.} \&
  \bibinfo{author}{Palmer, J.~C.}
\newblock \bibinfo{title}{Liquid--liquid phase transition in an ionic model of
  silica}.
\newblock \emph{\bibinfo{journal}{The Journal of chemical physics}}
  \textbf{\bibinfo{volume}{146}}~(23), \bibinfo{pages}{234503}
  (\bibinfo{year}{2017}) .

\bibitem{guo2018fluctuations}
\bibinfo{author}{Guo, J.} \& \bibinfo{author}{Palmer, J.~C.}
\newblock \bibinfo{title}{Fluctuations near the liquid--liquid transition in a
  model of silica}.
\newblock \emph{\bibinfo{journal}{Physical Chemistry Chemical Physics}}
  \textbf{\bibinfo{volume}{20}}~(39), \bibinfo{pages}{25195--25202}
  (\bibinfo{year}{2018}) .

\bibitem{ricci2019computational}
\bibinfo{author}{Ricci, F.} \emph{et~al.}
\newblock \bibinfo{title}{A computational investigation of the thermodynamics
  of the stillinger-weber family of models at supercooled conditions}.
\newblock \emph{\bibinfo{journal}{Molecular Physics}} \bibinfo{pages}{1--15}
  (\bibinfo{year}{2019}) .

\bibitem{goswami2021thermodynamics}
\bibinfo{author}{Goswami, Y.}, \bibinfo{author}{Vasisht, V.~V.},
  \bibinfo{author}{Frenkel, D.}, \bibinfo{author}{Debenedetti, P.~G.} \&
  \bibinfo{author}{Sastry, S.}
\newblock \bibinfo{title}{Thermodynamics and kinetics of crystallization in
  deeply supercooled stillinger--weber silicon}.
\newblock \emph{\bibinfo{journal}{The Journal of Chemical Physics}}
  \textbf{\bibinfo{volume}{155}}~(19), \bibinfo{pages}{194502}
  (\bibinfo{year}{2021}) .

\bibitem{starr2014crystal}
\bibinfo{author}{Starr, F.~W.} \& \bibinfo{author}{Sciortino, F.}
\newblock \bibinfo{title}{“crystal-clear” liquid--liquid transition in a
  tetrahedral fluid}.
\newblock \emph{\bibinfo{journal}{Soft Matter}}
  \textbf{\bibinfo{volume}{10}}~(47), \bibinfo{pages}{9413--9422}
  (\bibinfo{year}{2014}) .

\bibitem{smallenburg2014erasing}
\bibinfo{author}{Smallenburg, F.}, \bibinfo{author}{Filion, L.} \&
  \bibinfo{author}{Sciortino, F.}
\newblock \bibinfo{title}{Erasing no-man’s land by thermodynamically
  stabilizing the liquid--liquid transition in tetrahedral particles}.
\newblock \emph{\bibinfo{journal}{Nature physics}}
  \textbf{\bibinfo{volume}{10}}~(9), \bibinfo{pages}{653--657}
  (\bibinfo{year}{2014}) .

\bibitem{saika2006test}
\bibinfo{author}{Saika-Voivod, I.}, \bibinfo{author}{Poole, P.~H.} \&
  \bibinfo{author}{Bowles, R.~K.}
\newblock \bibinfo{title}{Test of classical nucleation theory on deeply
  supercooled high-pressure simulated silica}.
\newblock \emph{\bibinfo{journal}{The Journal of chemical physics}}
  \textbf{\bibinfo{volume}{124}}~(22), \bibinfo{pages}{224709}
  (\bibinfo{year}{2006}) .

\bibitem{wilding1997simulation}
\bibinfo{author}{Wilding, N.~B.}
\newblock \bibinfo{title}{Simulation studies of fluid critical behaviour}.
\newblock \emph{\bibinfo{journal}{Journal of Physics: Condensed Matter}}
  \textbf{\bibinfo{volume}{9}}~(3), \bibinfo{pages}{585} (\bibinfo{year}{1997})
  .

\bibitem{torrie1977nonphysical}
\bibinfo{author}{Torrie, G.~M.} \& \bibinfo{author}{Valleau, J.~P.}
\newblock \bibinfo{title}{Nonphysical sampling distributions in monte carlo
  free-energy estimation: Umbrella sampling}.
\newblock \emph{\bibinfo{journal}{Journal of Computational Physics}}
  \textbf{\bibinfo{volume}{23}}~(2), \bibinfo{pages}{187--199}
  (\bibinfo{year}{1977}) .

\bibitem{steinhardt1983bond}
\bibinfo{author}{Steinhardt, P.~J.}, \bibinfo{author}{Nelson, D.~R.} \&
  \bibinfo{author}{Ronchetti, M.}
\newblock \bibinfo{title}{Bond-orientational order in liquids and glasses}.
\newblock \emph{\bibinfo{journal}{Physical Review B}}
  \textbf{\bibinfo{volume}{28}}~(2), \bibinfo{pages}{784}
  (\bibinfo{year}{1983}) .

\bibitem{van1992computer}
\bibinfo{author}{Van~Duijneveldt, J.} \& \bibinfo{author}{Frenkel, D.}
\newblock \bibinfo{title}{Computer simulation study of free energy barriers in
  crystal nucleation}.
\newblock \emph{\bibinfo{journal}{The Journal of chemical physics}}
  \textbf{\bibinfo{volume}{96}}~(6), \bibinfo{pages}{4655--4668}
  (\bibinfo{year}{1992}) .

\bibitem{ten1995numerical}
\bibinfo{author}{Ten~Wolde, P.~R.}, \bibinfo{author}{Ruiz-Montero, M.~J.} \&
  \bibinfo{author}{Frenkel, D.}
\newblock \bibinfo{title}{Numerical evidence for bcc ordering at the surface of
  a critical fcc nucleus}.
\newblock \emph{\bibinfo{journal}{Physical review letters}}
  \textbf{\bibinfo{volume}{75}}~(14), \bibinfo{pages}{2714}
  (\bibinfo{year}{1995}) .

\bibitem{romano2011crystallization}
\bibinfo{author}{Romano, F.}, \bibinfo{author}{Sanz, E.} \&
  \bibinfo{author}{Sciortino, F.}
\newblock \bibinfo{title}{Crystallization of tetrahedral patchy particles in
  silico}.
\newblock \emph{\bibinfo{journal}{The Journal of chemical physics}}
  \textbf{\bibinfo{volume}{134}}~(17), \bibinfo{pages}{174502}
  (\bibinfo{year}{2011}) .

\bibitem{chodera2007use}
\bibinfo{author}{Chodera, J.~D.}, \bibinfo{author}{Swope, W.~C.},
  \bibinfo{author}{Pitera, J.~W.}, \bibinfo{author}{Seok, C.} \&
  \bibinfo{author}{Dill, K.~A.}
\newblock \bibinfo{title}{Use of the weighted histogram analysis method for the
  analysis of simulated and parallel tempering simulations}.
\newblock \emph{\bibinfo{journal}{Journal of Chemical Theory and Computation}}
  \textbf{\bibinfo{volume}{3}}~(1), \bibinfo{pages}{26--41}
  (\bibinfo{year}{2007}) .

\bibitem{kumar1992weighted}
\bibinfo{author}{Kumar, S.}, \bibinfo{author}{Rosenberg, J.~M.},
  \bibinfo{author}{Bouzida, D.}, \bibinfo{author}{Swendsen, R.~H.} \&
  \bibinfo{author}{Kollman, P.~A.}
\newblock \bibinfo{title}{The weighted histogram analysis method for
  free-energy calculations on biomolecules. i. the method}.
\newblock \emph{\bibinfo{journal}{Journal of computational chemistry}}
  \textbf{\bibinfo{volume}{13}}~(8), \bibinfo{pages}{1011--1021}
  (\bibinfo{year}{1992}) .

\bibitem{ricci2017free}
\bibinfo{author}{Ricci, F.} \& \bibinfo{author}{Debenedetti, P.~G.}
\newblock \bibinfo{title}{A free energy study of the liquid-liquid phase
  transition of the jagla two-scale potential}.
\newblock \emph{\bibinfo{journal}{Journal of Chemical Sciences}}
  \textbf{\bibinfo{volume}{129}}~(7), \bibinfo{pages}{801--823}
  (\bibinfo{year}{2017}) .

\bibitem{mishima1998relationship}
\bibinfo{author}{Mishima, O.} \& \bibinfo{author}{Stanley, H.~E.}
\newblock \bibinfo{title}{The relationship between liquid, supercooled and
  glassy water}.
\newblock \emph{\bibinfo{journal}{Nature}}
  \textbf{\bibinfo{volume}{396}}~(6709), \bibinfo{pages}{329--335}
  (\bibinfo{year}{1998}) .

\bibitem{holten2014two}
\bibinfo{author}{Holten, V.}, \bibinfo{author}{Palmer, J.~C.},
  \bibinfo{author}{Poole, P.~H.}, \bibinfo{author}{Debenedetti, P.~G.} \&
  \bibinfo{author}{Anisimov, M.~A.}
\newblock \bibinfo{title}{Two-state thermodynamics of the st2 model for
  supercooled water}.
\newblock \emph{\bibinfo{journal}{The Journal of chemical physics}}
  \textbf{\bibinfo{volume}{140}}~(10), \bibinfo{pages}{104502}
  (\bibinfo{year}{2014}) .

\bibitem{buldyrev2002models}
\bibinfo{author}{Buldyrev, S.} \emph{et~al.}
\newblock \bibinfo{title}{Models for a liquid--liquid phase transition}.
\newblock \emph{\bibinfo{journal}{Physica A: Statistical Mechanics and its
  Applications}} \textbf{\bibinfo{volume}{304}}~(1-2), \bibinfo{pages}{23--42}
  (\bibinfo{year}{2002}) .

\bibitem{holten2012entropy}
\bibinfo{author}{Holten, V.} \& \bibinfo{author}{Anisimov, M.}
\newblock \bibinfo{title}{Entropy-driven liquid--liquid separation in
  supercooled water}.
\newblock \emph{\bibinfo{journal}{Scientific reports}}
  \textbf{\bibinfo{volume}{2}}~(1), \bibinfo{pages}{1--7}
  (\bibinfo{year}{2012}) .

\bibitem{tsypin2000probability}
\bibinfo{author}{Tsypin, M.} \& \bibinfo{author}{Bl{\"o}te, H.}
\newblock \bibinfo{title}{Probability distribution of the order parameter for
  the three-dimensional ising-model universality class: A high-precision monte
  carlo study}.
\newblock \emph{\bibinfo{journal}{Physical Review E}}
  \textbf{\bibinfo{volume}{62}}~(1), \bibinfo{pages}{73} (\bibinfo{year}{2000})
  .

\bibitem{poole2013free}
\bibinfo{author}{Poole, P.~H.}, \bibinfo{author}{Bowles, R.~K.},
  \bibinfo{author}{Saika-Voivod, I.} \& \bibinfo{author}{Sciortino, F.}
\newblock \bibinfo{title}{Free energy surface of st2 water near the
  liquid-liquid phase transition}.
\newblock \emph{\bibinfo{journal}{The Journal of chemical physics}}
  \textbf{\bibinfo{volume}{138}}~(3), \bibinfo{pages}{034505}
  (\bibinfo{year}{2013}) .

\bibitem{saw2009structural}
\bibinfo{author}{Saw, S.}, \bibinfo{author}{Ellegaard, N.~L.},
  \bibinfo{author}{Kob, W.} \& \bibinfo{author}{Sastry, S.}
\newblock \bibinfo{title}{Structural relaxation of a gel modeled by three body
  interactions}.
\newblock \emph{\bibinfo{journal}{Physical review letters}}
  \textbf{\bibinfo{volume}{103}}~(24), \bibinfo{pages}{248305}
  (\bibinfo{year}{2009}) .

\bibitem{vasisht2014nesting}
\bibinfo{author}{Vasisht, V.~V.}, \bibinfo{author}{Mathew, J.},
  \bibinfo{author}{Sengupta, S.} \& \bibinfo{author}{Sastry, S.}
\newblock \bibinfo{title}{Nesting of thermodynamic, structural, and dynamic
  anomalies in liquid silicon}.
\newblock \emph{\bibinfo{journal}{The Journal of chemical physics}}
  \textbf{\bibinfo{volume}{141}}~(12), \bibinfo{pages}{124501}
  (\bibinfo{year}{2014}) .

\end{thebibliography}

\begin{thebibliography}{10}

\bibitem{stillinger1985computer}
Frank~H Stillinger and Thomas~A Weber.
\newblock Computer simulation of local order in condensed phases of silicon.
\newblock {\em Physical review B}, 31(8):5262, 1985.

\bibitem{steinhardt1983bond}
Paul~J Steinhardt, David~R Nelson, and Marco Ronchetti.
\newblock Bond-orientational order in liquids and glasses.
\newblock {\em Physical Review B}, 28(2):784, 1983.

\bibitem{vasisht2014nesting}
Vishwas~V Vasisht, John Mathew, Shiladitya Sengupta, and Srikanth Sastry.
\newblock Nesting of thermodynamic, structural, and dynamic anomalies in liquid
  silicon.
\newblock {\em The Journal of chemical physics}, 141(12):124501, 2014.

\bibitem{van1992computer}
JS~Van~Duijneveldt and D~Frenkel.
\newblock Computer simulation study of free energy barriers in crystal
  nucleation.
\newblock {\em The Journal of chemical physics}, 96(6):4655--4668, 1992.

\bibitem{ten1995numerical}
Pieter~Rein Ten~Wolde, Maria~J Ruiz-Montero, and Daan Frenkel.
\newblock Numerical evidence for bcc ordering at the surface of a critical fcc
  nucleus.
\newblock {\em Physical review letters}, 75(14):2714, 1995.

\bibitem{wolde1996simulation}
Pieter-Rein{\'a}ten Wolde et~al.
\newblock Simulation of homogeneous crystal nucleation close to coexistence.
\newblock {\em Faraday discussions}, 104:93--110, 1996.

\bibitem{romano2011crystallization}
Flavio Romano, Eduardo Sanz, and Francesco Sciortino.
\newblock Crystallization of tetrahedral patchy particles in silico.
\newblock {\em The Journal of chemical physics}, 134(17):174502, 2011.

\bibitem{kesselring2013finite}
Tobias~A Kesselring, Erik Lascaris, Giancarlo Franzese, Sergey~V Buldyrev,
  Hans~J Herrmann, and H~Eugene Stanley.
\newblock Finite-size scaling investigation of the liquid-liquid critical point
  in st2 water and its stability with respect to crystallization.
\newblock {\em The Journal of Chemical Physics}, 138(24):244506, 2013.

\bibitem{goswami2021thermodynamics}
Yagyik Goswami, Vishwas~V Vasisht, Daan Frenkel, Pablo~G Debenedetti, and
  Srikanth Sastry.
\newblock Thermodynamics and kinetics of crystallization in deeply supercooled
  stillinger--weber silicon.
\newblock {\em The Journal of Chemical Physics}, 155(19):194502, 2021.

\bibitem{ricci2019computational}
Francesco Ricci, Jeremy~C Palmer, Yagyik Goswami, Srikanth Sastry, C~Austen
  Angell, and Pablo~G Debenedetti.
\newblock A computational investigation of the thermodynamics of the
  stillinger-weber family of models at supercooled conditions.
\newblock {\em Molecular Physics}, pages 1--15, 2019.

\bibitem{kumar1992weighted}
Shankar Kumar, John~M Rosenberg, Djamal Bouzida, Robert~H Swendsen, and Peter~A
  Kollman.
\newblock The weighted histogram analysis method for free-energy calculations
  on biomolecules. i. the method.
\newblock {\em Journal of computational chemistry}, 13(8):1011--1021, 1992.

\bibitem{chodera2007use}
John~D Chodera, William~C Swope, Jed~W Pitera, Chaok Seok, and Ken~A Dill.
\newblock Use of the weighted histogram analysis method for the analysis of
  simulated and parallel tempering simulations.
\newblock {\em Journal of Chemical Theory and Computation}, 3(1):26--41, 2007.

\bibitem{debenedetti2020second}
Pablo~G Debenedetti, Francesco Sciortino, and G{\"u}l~H Zerze.
\newblock Second critical point in two realistic models of water.
\newblock {\em Science}, 369(6501):289--292, 2020.

\bibitem{maibaum2008comment}
Lutz Maibaum.
\newblock Comment on “elucidating the mechanism of nucleation near the
  gas-liquid spinodal”.
\newblock {\em Physical review letters}, 101(1):019601, 2008.

\bibitem{chakrabarty2008chakrabarty}
Suman Chakrabarty, Mantu Santra, and Biman Bagchi.
\newblock Chakrabarty, santra, and bagchi reply.
\newblock {\em Physical Review Letters}, 101(1):019602, 2008.

\bibitem{gao2012implementing}
Fuchang Gao and Lixing Han.
\newblock Implementing the nelder-mead simplex algorithm with adaptive
  parameters.
\newblock {\em Computational Optimization and Applications}, 51(1):259--277,
  2012.

\end{thebibliography}
\end{document}